\documentclass[fleqn,usenatbib]{mnras}

\usepackage{newtxtext,newtxmath}
\usepackage[T1]{fontenc}
\DeclareRobustCommand{\VAN}[3]{#2}
\let\VANthebibliography\thebibliography
\def\thebibliography{\DeclareRobustCommand{\VAN}[3]{##3}\VANthebibliography}

\usepackage{graphicx}
\usepackage{amsmath}

\def\inn{_{\rm in}}
\def\out{_{\rm out}}
\def\res{_{\rm res}}
\def\sec{_{\rm sec}}
\def\warp{_{\rm warp}}  
\newcommand{\HDone}{HD~110058}
\newcommand{\HDoneHIP}{HIP~61782}
\newcommand{\HDdec}{HD~15115}

\title[Vertical Structure of Debris Discs]{The vertical structure of debris discs and the role of disc gravity: A primer using a simplified model}

\author[A. A. Sefilian et al.]{Antranik A. Sefilian,$^{1}$
\thanks{E-mail: sefilian.antranik@gmail.com (51 Pegasi b Fellow)}
Kaitlin M. Kratter,$^{1}$
Mark C. Wyatt,$^{2}$
Cristobal Petrovich,$^{3}$
Philippe Th{\'e}bault,$^{4}$
\newauthor
Renu Malhotra$^{5}$
and
Virginie Faramaz-Gorka$^{1}$
\\
$^{1}$Department of Astronomy and Steward Observatory, University of Arizona, Tucson, AZ 85721, USA
\\
$^{2}$Institute of Astronomy, University of Cambridge, Madingley Road, Cambridge CB3 0HA, UK
\\
$^{3}$Department of Astronomy, Indiana University, Bloomington, IN 47405, USA
\\
$^{4}$LESIA-Observatoire de Paris, UPMC Univ. Paris 06, Univ. Paris-Diderot, France
\\
$^{5}$Lunar and Planetary Laboratory, The University of Arizona, Tucson, AZ 85721, USA
}

\date{Accepted 2025 September 5. Received 2025 July 29; in original form 2025 May 14}
\pubyear{\the\year{}}

\begin{document}
\label{firstpage}
\pagerange{\pageref{firstpage}--\pageref{lastpage}}
\maketitle

%%%%%%%%%%%%%%%%%%%%%%%%%%%%%%%%%%%
%%%     Abstract & Keywords     %%%
%%%%%%%%%%%%%%%%%%%%%%%%%%%%%%%%%%%

% Abstract
\begin{abstract} 
\noindent Debris discs provide valuable insights into the formation and evolution of exoplanetary systems. Their structures are commonly attributed to planetary perturbations, serving as probes of as-yet-undetected planets. However, most studies of planet-debris disc interactions ignore the disc's gravity, treating it as a collection of massless planetesimals. We develop a simplified analytical model as a primer to investigate how the vertical structure of a back-reacting debris disc responds to secular perturbations from an inner, inclined planet. Considering the disc's axisymmetric potential, we identify two dynamical regimes: planet-dominated and disc-dominated, which may coexist, separated by a secular-inclination resonance. In the planet-dominated regime ($M_d/m_p\ll1$), we recover the classical result: a transient warp propagates outward until the disc settles into a box-like structure centered around the planetary orbit's initial inclination $I_p(0)$, with a distance-independent aspect ratio $\mathcal{H}(R)\approx I_p(0)$. In contrast, in the disc-dominated regime ($M_d/m_p\gtrsim1$), the disc exhibits  dynamical rigidity, remaining thin and misaligned, with significantly suppressed inclinations and a sharply declining aspect ratio, $\mathcal{H}(R)\propto I_p(0)R^{-7/2}$. In the intermediate regime ($M_d/m_p\lesssim1$), the system exhibits a secular-inclination resonance, leading to long-lived, warp-like structures and a bimodal inclination distribution, containing both dynamically hot and cold populations. We provide analytic formulae describing these effects as a function of system parameters. We also find that the vertical density profile is intrinsically non-Gaussian and recommend fitting observations with non-zero slopes of $\mathcal{H}(R)$. Our results may be used to infer planetary parameters and debris disc masses based on observed warps and scale heights, as demonstrated for HD~110058 and $\beta$~Pic.
%%%%%%%%%
\end{abstract}


%%%%%%%%%%%%%%%%%%%%%%%%%%%%%%
\section{Introduction}
\label{sec:intro}
%%%%%%%%%%%%%%%%%%%%%%%%%%%%%%

%%%%%%%%%%%%%
\subsection{Motivation}
\label{sec:motiv}
%%%%%%%%%%%%%

To date, around 20 per cent of main sequence stars in the solar neighborhood are known to host debris disks \citep{montesinos2016, sibthorpe2018}. These circumstellar disks, representing remnants of planet-formation processes, are typically optically thin and  gas-poor \citep[][]{hughes2018review, wyatt19review}, although a few contain detectable amounts of gas \citep[e.g.,][]{Marino2016, Kral2020}. Debris disks are populated by solids ranging from {observed} micron-sized dust grains to {inferred} kilometer-sized or larger planetesimals, akin to the asteroid and Kuiper belts in the Solar System. Unlike the primordial interstellar dust in protoplanetary disks, debris disk dust is considered to be secondary in origin,  continuously replenished through collisional cascades among larger parent planetesimals \citep{wyatt08collisionsreview}. Thus, sustaining the levels of dust estimated from observations -- typically, $\sim 10^{-3} - 1 M_{\earth}$ \citep[][]{wyattdent2002, panic13, holland17,  reasons_matra} --  requires a massive reservoir of parent planetesimals. Depending on the assumed properties, such as size distribution and collisional behaviors, estimates for this reservoir can be as high as $\sim 10^3 M_{\earth}$ \citep[][]{krivovwyatt21}. As such, debris disks are often considered massive analogues of the Kuiper belt \citep{wyatt19review}.

In recent years, high-resolution observations of debris disks, particularly with ALMA and JWST, have revealed a variety of intricate substructures \citep[e.g.,][]{hughes2018review, wyatt19review, gaspar2023, reasons_matra}. Such substructures have been observed in the radial (e.g., asymmetries, gaps, and spirals), vertical (e.g., warps and scale heights), and azimuthal (e.g., clumps) distributions of dust populations. These observations span a range of wavelengths, including the millimeter range, which most closely traces the distribution of the underlying parent planetesimals \citep{Krivov2010, hughes2018review}. Motivated by planet-disk interaction studies applied to the Solar System \citep[see, e.g., review by][]{morby-review}, observed structures are often attributed to the gravitational effects of (invoked, yet-unseen) planets, along with non-gravitational forces such as radiation pressure and stellar winds \citep[e.g.,][]{wyattetal99, leechiang2016, Dong2020}. Accordingly, debris disks are treated as probes of planetary systems. They offer insight into planet formation and evolution, and may even reveal the existence of otherwise undetectable planets.

Vertical structures in particular, including warps and scale heights (or aspect ratios), are of central interest. Indeed,  warped structures -- characterized by a tilted inner disk extending into a flat outer disk -- are often interpreted as evidence of planets on inclined orbits \citep[e.g.,][]{augereau2001, Dawson2011, nesvold15, nesvold17,  Brady2023, farhat-sefilian-23, smallwoodbeta}. Part of the argument here is that debris disks are expected to retain the nearly flat configurations established during the gas-rich protoplanetary disk phase, in the absence of other perturbers. An iconic example is the $\beta$ Pic system, where the detection of a warped structure \citep{lagage94, Kalas95} led to the prediction \citep{Mouillet97, augereau2001} and subsequent direct imaging of the planet responsible for it \citep{lagrange2010}, $\beta$ Pic b.\footnote{It is noteworthy that $\beta$ Pic b is not completely aligned with the inner disk as expected based on theoretical considerations; for a detailed discussion, see \citet{Dawson2011}. Additionally, a second inner planet, $\beta$ Pic c, has been detected \citep{lagrange-betapic-c}; this, however, does not appear to significantly affect the warp \citep{Dong2020, smallwoodbeta}.} While warps are most readily accessible in disks with near edge-on configurations -- introducing observational biases against their detection -- a few systems have been identified to date: namely, AU Mic \citep{Wang2015}, HD 111520 \citep{draper2016, Crotts2022}, and \HDone\,\citep{Kasper2015, hales2022, Stasevic2023}.

Another observationally measurable vertical property of debris disks is their scale height $\mathfrak{h}_d$ and thus aspect ratio $\mathcal{H}$ $\equiv \mathfrak{h}_d / R $. Such measurements have been made for tens of debris disks, revealing a striking diversity, with aspect ratio values ranging from $\sim 0.01$ to $0.3$ \citep[e.g.,][]{grant18, hales2022, jonty23, Terrill2023, reasons_matra}. Interestingly, while most measurements assume a radius-independent aspect ratio (to simplify fitting procedures),  advanced (non-parametric) fitting methods have also revealed radius-dependent variations \citep{Han2022, Han25}. 
%%%%%%%
Additionally, observations of the AU Mic debris disk suggest a wavelength-dependent increase in aspect ratio \citep{Daley2019, johan22, vizgan2022}. While AU Mic is currently the only system with such evidence, it highlights the potential for wavelength-dependent aspect ratios, adding further complexity. 
%%%%%%%%
Finally, a recent study analyzing debris disks observed with ALMA suggests a tentative trend of a decreasing aspect ratio with increasing dust mass around A-type stars \citep{Han25}.

%%%%%%%%%%%%%
\subsection{Previous works}
\label{sec:prev_works}
%%%%%%%%%%%%%

Three key mechanisms have been explored in the literature to explain vertical structures in debris disks.  The first mechanism involves flybys, where the passage of a nearby star or massive object can induce tilts in the disk's vertical structure, as well as puff it up \citep{Kobayashi2001, lestrade2011, moore2020, batygin2020}.  The second mechanism involves gravitational stirring by large bodies embedded within the disk itself--such as Pluto-sized or larger embryos--where two-body encounters with these bodies dynamically excite the surrounding smaller planetesimals \citep[e.g.,][]{ida90, ida92,ida93,kokubo2012}. This mechanism can thicken the disk vertically \citep[e.g.,][]{marco2015, Daley2019, matra2019}, but when operating alone, it is not expected to induce warps. Recently, however, the combined effects of this mechanism and a planet located interior to the disk -- referred to as `mixed-stirring' -- have also been explored \citep{marco2023}.

The third mechanism, perhaps the most commonly studied, involves gravitational perturbations due to bound perturbers on inclined orbits, such as stellar companions or planets that are not embedded in the disk. Various configurations of this mechanism have been explored: (i) perturbers located interior to the disk \citep{wyattetal99, Mouillet97, pearcewyatt2014, Dong2020, Brady2023, Pedro2024}, (ii) exterior to the disk \citep{jilkova2015, nesvold17, Rodet2017}, or (iii) involving both interior and exterior perturbers \citep{batp9, Volk2017, farhat2021, farhat-sefilian-23, chai2024}. Finally, we note that, in principle, the three aforementioned mechanisms can influence the vertical distribution of debris across all sizes. This is in contrast to radiation pressure, which, when acting in combination with collisions, has been shown to impose a minimum `natural' aspect ratio of $\approx 0.05$ on dust particles susceptible to it, i.e., those of $\sim$micrometer in size \citep{thebault2009}.

%%%%%%%%%%%%%
\subsection{This work}
\label{sec:this_work_Q}
%%%%%%%%%%%%%

\begin{figure*}
%%%%%%%%
\centering
\includegraphics[width=0.75\textwidth]{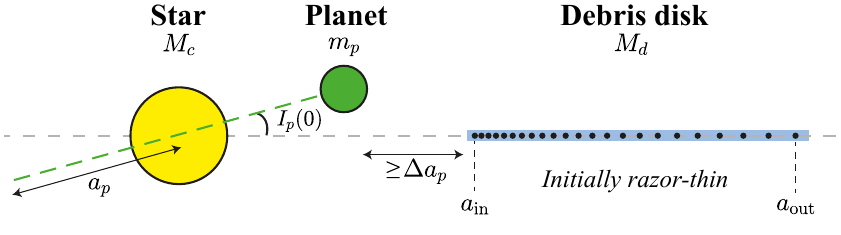}
%%%%%%%%
\vspace{-1.5em}
{\linespread{0.95}\selectfont \caption{An illustration depicting an edge-on view of the initial setup of the model system examined in this work. The system consists of a central star of mass $M_c$ which is orbited by a planet of mass $m_p$ and an external debris disk of mass $M_d$ (with $m_p, M_d \ll M_c$). The planetary orbit,  depicted by the dashed green line, is taken to be circular with a semimajor axis $a_p$ and is slightly inclined with respect to the disk plane (typically, $I_p(0) \lesssim 20^{\circ}$). The debris disk, extending from $a\inn$ to $a\out$, is circular and initially vertically razor-thin, lying in the equatorial plane of the star. The disk's inner edge is located at a minimum distance of $\Delta a_p$ away from the planetary semimajor axis (Equation (\ref{eq:MMR_condition_delta_ap})), and its initial surface density distribution is given by a truncated power-law profile (Equation (\ref{eq:Sigma_d})). The disk is represented as a collection of $N\gg 1$ massive planetesimals, each with a fixed, logarithmically-spaced semimajor axis $a_j$. As the system evolves, the inclinations ($I_j$) and longitudes of the ascending nodes ($\Omega_j$) of each planetesimal ($j = 1, \ldots, N$) and that of the planet ($I_p$ and $\Omega_p$) may vary due to gravitational interactions. For further details on the model system and the theoretical framework describing its long-term evolution, see Sections \ref{sec:modelsystem} and \ref{sec:analytical-model}.}
\label{fig:model-system}}
\end{figure*}

In this work, we focus on the role of planets in shaping the vertical structures of debris disks, considering a simple scenario: a single planet on a circular, inclined orbit interior to a flat debris disk; see Figure \ref{fig:model-system}. While much work has explored this setup, existing studies typically neglect the mass of the disk itself, i.e., treating planetesimals as massless particles. This simplification raises several questions, motivating the present work: Are planetary inferences based on  warps compromised by neglecting the disk's gravity, and if so, to what extent? Can massive debris disks `resist' planetary perturbations and remain thin? How does the disk's gravity affect its vertical density profile and aspect ratio? With an increasing number of debris disks being imaged at unprecedented spatial resolutions, these questions are timely as JWST is poised to test theories by detecting (or ruling out) the predicted planets \citep{carter-jwst}.

Previous studies have explored related questions in the context of coplanar, eccentric planet--debris disk interactions. These include analytical and numerical investigations into the formation of gapped (or double-ringed) structures \citep{pearce2015, Paper1, Paper2}, production of spiral patterns \citep{sefilian-phd}, as well as stirring processes and dust production  \citep{Sefilian2024}. However, investigations into misaligned planet--massive debris disk systems are more limited, typically carried out via direct $N$-body simulations \citep[][]{pearcewyatt2014, Pedro2024}. Thus, to maintain computational efficiency, these studies tend to be confined in terms of (i) physical parameters, such as planetesimal numbers $N$ and disk-to-planet mass ratios $M_d/m_p$, and (ii) disk gravity-related physics. For instance, \citet{pearcewyatt2014} considered a relatively large number of planetesimals ($N\sim 10^{3}-10^4$) in a relatively low-mass disk with $M_d/m_p \leq 0.1$ that back-reacts onto the inclined planet (i.e., ignoring self-gravity). On the other hand, \citet{Pedro2024} considered a smaller planetesimal count ($N\sim 10^2$) in a fully self-gravitating (but radially narrow) disk for a single value of $M_d/m_p = 0.5$. These studies also incorporate additional physics -- namely, eccentric planets, planet-planetesimal scattering, and/or self-stirring by large planetesimals -- which, while useful, render it more challenging to discern the specific effects due to the disk's gravity. 

%%%%%%%%%%%%%%%%%%%%%%%%%%%%%
%%%%%%%%%%%%%%%%%%%%%%%%%%%%%

In this paper, we analytically address this gap by investigating the long-term, secular interactions between a planet on a circular, inclined orbit and an external, massive debris disk. Our primary aim is to explore the role of disk gravity on the inclination distribution of planetesimals -- and, consequently, the vertical structure of debris disks -- across a wide range of  system parameters, including $M_d/m_p$. To flesh out the basic physics analytically, we focus on the disk’s back-reaction onto the planet (i.e., excluding the disk's self-gravity, as also done in \citet{pearcewyatt2014}). This paper focuses on the axisymmetric component of the disk potential, while the non-axisymmetric component will be addressed in an upcoming work. As we show below, even at this level of approximation, the disk's gravity plays a crucial role in shaping its vertical structure, with important implications  as to how  warps and aspect ratios are interpreted.

This paper is organized as follows. First, Section \ref{sec:modelsystem} outlines the model system adopted in this work, before  delineating the  theoretical framework for its long-term, secular evolution in Section \ref{sec:analytical-model}. Specializing to the case of an axisymmetric back-reacting disk potential, Section  \ref{sec:analytical_analysis} then characterizes the evolution of the planetary and planetesimal orbits over a broad range of parameters.  Using these analytical solutions, Section \ref{sec:vert_struc_num} presents the main results in terms of the evolution of the disk morphology, including cases where $M_d/m_p \lesssim 1$ and $M_d/m_p \gtrsim 1$. Section \ref{sec:discussion} discusses the main results and their implications for the interpretation of vertical structures (including warps, scale heights, and planetary inferences based on such features), before critically assessing the model's limitations and proposing future work. Our findings are summarized in Section \ref{sec:summary}. Technical details of some calculations are provided in Appendices \ref{sec:app_A}--\ref{app:disk_sigma_analysis}.

%%%%%%%%%%%%%%%%%%%%%%%%%%%%%%
\section{Model System}
\label{sec:modelsystem}
%%%%%%%%%%%%%%%%%%%%%%%%%%%%%%

We consider a model system composed of a central star of mass $M_c$ that is orbited by a single planet of mass $m_p$ and an external debris disk of mass $M_d$ (with $m_p, M_d \ll M_c$). For reference, a schematic of the model system and its initial configuration, as viewed edge-on, is depicted in Figure \ref{fig:model-system}.

We characterize the planetary orbit, assumed to be circular, by its semimajor axis $a_p$, inclination $I_p$, longitude of ascending node $\Omega_p$, and mean motion $n_p \approx \sqrt{G M_c /a_p^3}$. The planet is taken to be initially inclined with respect to the disk  (which lies in the equatorial plane of the star, more on this below), typically with $I_p(0) \lesssim 20^{\circ}$ and, without loss of generality, $\Omega_p(0) = 0$.\footnote{We limit $I_p(0)$ for two reasons: first, the linear Laplace-Lagrange secular theory we employ (Section \ref{sec:analytical-model}) is valid for small inclinations; second, to avoid scenarios where von Zeipel-Lidov-Kozai  effects \citep[][]{Lidov62, kozai62, Naoz2016,Ito19} cannot be ignored. The latter can modulate the eccentricities of the bodies involved, introducing further complexity.}

The debris disk, on the other hand, extending from $a\inn$ to $a\out$, consists of $N\gg 1$ planetesimals on circular orbits. Each planetesimal is characterized by its mass $m_j$, orbital semimajor axis $a_d = \{a_j\}$, inclination $I_d=\{I_j\}$, longitude of ascending node $\Omega_d = \{\Omega_j\}$, and mean motion $n_d = \{n_j\} \approx \sqrt{G M_c/a_j^3}$ (with $j=1, \ldots , N$). The disk is taken to be initially razor-thin with $I_j(0) = 0$ and $\Omega_j(0)$ that is uniformly drawn between $0$ and $2\pi$. The semimajor axes of the planetesimals are taken to be logarithmically spaced, with a geometric spacing ratio of $a_{j+1}/a_j = (a\out/a\inn)^{1/{(N-1)}}$. Based on this, the masses $m_j$ are assigned such that the disk's initial surface-density distribution follows a truncated power-law profile given by:
%%%%%%%%%%%%%%%%%%%%%%%
\begin{eqnarray}
    \Sigma_d^{t=0}(a_d) &=& \Sigma_0 \bigg(\frac{a\out}{a_d}\bigg)^p 
     ~~~~\textrm{for}  ~~~~ a\inn \leq a_d \leq a\out, 
     \label{eq:Sigma_d}
\end{eqnarray} 
%%%%%%%%%%%%%%%%%%%%%%%
and $\Sigma_d^{t=0}(a_d) = 0$ elsewhere. This is done using the relationship $dm_j/da_j=2\pi a_j \Sigma_d^{t=0}(a_j)$ \citep{statler99, irina18}. Thus, in the continuum limit ($N \rightarrow \infty$), the total disk mass $M_d = \sum_{j=1}^{N} m_j$ can be expressed as:
%%%%%%%%%%%%%%%%%%%%%%%
\begin{equation}
    M_d 
    = 2\pi \int_{a\inn}^{a\out} a_d \Sigma_d^{t=0}(a_d) da_d 
    = \frac{2\pi}{2-p} \Sigma_0 a\out^2 ( 1 - \delta^{p-2}) , 
    \label{eq:M_disc_Sigma0}
\end{equation}
%%%%%%%%%%%%%%%%%%%%%%%
where, for brevity, we have defined $\delta \equiv a\out/a\inn \geq 1$.

Here, we note in passing that  Laplace-Lagrange theory for secular inclination dynamics (as adopted in this work; Section~\ref{sec:analytical-model}) does not prescribe a natural reference plane. As a result, it is most meaningful to consider mutual inclinations, rendering the choice of a reference plane arbitrary. In this work, we adopt the stellar equatorial plane as our reference (Fig. \ref{fig:model-system}).

From hereon, unless otherwise stated, we adopt a Solar-mass  star ($M_c = 1 M_{\odot}$), and a fiducial disk model with $p=3/2$ composed of $N=5 \times 10^{3}$ planetesimals. The chosen value of $p$ is motivated by the slope of the Minimum Mass Solar Nebula \citep[MMSN;][]{hayashi}. We further set the disk's inner edge at $a\inn = 30$ au, ensuring that it is at a minimal distance of $\Delta a_p$ away from the planet:
%%%%%%%%%%%%%%%%%%%
\begin{equation}
    a\inn \geq a_p + \Delta a_p ~~ \mathrm{where} ~~  \Delta a_p \approx 1.3 \bigg(\frac{m_p}{M_c+m_p}\bigg)^{2/7} a_p .
    \label{eq:MMR_condition_delta_ap}
\end{equation}
%%%%%%%%%%%%%%%%%%%
Here, $\Delta a_p$ represents the half-width of the chaotic zone centered around the planetary orbit, arising from the overlap of first-order mean motion resonances \citep[][]{Wisdom1980, Duncan1989}. Particles within this region would be cleared out relatively quickly (i.e., $\sim$tens of orbital periods; see \citet{Morrison2015}), thus the condition on $a\inn$. Additionally, we set the disk's outermost edge at $a\out = 5 a\inn$ (i.e., $\delta = 5$); this choice is somewhat arbitrary, made to ensure a radially-broad disk.

This completes our description of the model system and its initial state. Finally, we note that the considered setup is very similar to those explored in \citet{Paper1, Paper2} and \citet{Sefilian2024}, but now for circular and mutually inclined -- rather than eccentric and coplanar -- configurations.

%%%%%%%%%%%%%%%%%%%%%%%%%%%%%%
\section{Analytical Treatment: Secular Laplace-Lagrange Theory}
\label{sec:analytical-model}
%%%%%%%%%%%%%%%%%%%%%%%%%%%%%%

Our aim is to investigate the long-term evolution of large, $\gtrsim$km-sized planetesimals with non-zero masses that are perturbed by a single, inclined planet interior to them. Given that large planetesimals would not be affected by non-gravitational forces (such as radiation pressure and stellar winds, \citet{burns79}), we focus purely on gravitational effects. In doing so, we account for gravitational perturbations not only due to the planet but also due to the massive planetesimals. However, both for simplicity and to highlight the important role of the disk's back-reaction onto the planet, its self-gravitational effects are ignored. In other words, within our framework, the planet perturbs the planetesimals and \textit{vice-versa}, but the planetesimals do not perturb themselves.

Generally, the disk's back-reacting potential consists of two components: axisymmetric and non-axisymmetric. To develop an analytical understanding, this paper focuses on the axisymmetric component, while the non-axisymmetric torque will be investigated in a future work. The implications of relaxing this limitation, as well as those related to the self-gravitational effects of the disk, are discussed in Section \ref{subsec:limit_future_work}. For completeness, however, we present here a general theory. To this end, we work within the framework of orbit-averaged, secular Laplace--Lagrange perturbation theory \citep{Plummer1918}, expanded to second order in inclinations. In this limit, the semimajor axis (or Keplerian energy) of each interacting body is conserved, while the inclinations and the ascending nodes may evolve over time. Additionally, since the planetary and planetesimal orbits are initially circular (Section \ref{sec:modelsystem}), no secular evolution ensues in the eccentricities, which, regardless of this, remain decoupled from the inclinations to second order \citep{mur99}. This decoupling allows the inclinations to be treated independently, as we outline next.

%%%%%%%%%%%%%%%%%%%%%%%%%%%%%%%%%%%%%%%%%%%
\subsection{The Disturbing Functions: Orbit-averaged}
\label{sec:dist_fn_d_p}
%%%%%%%%%%%%%%%%%%%%%%%%%%%%%%%%%%%%%%%%%%%

%%%%%%%%%%%%%%%%%%%%%%%%%%%%%%%%%%%%%%%%%%%
\subsubsection{Effects of the Planet on Planetesimals}
\label{sec:Rpj-sec}
%%%%%%%%%%%%%%%%%%%%%%%%%%%%%%%%%%%%%%%%%%%

To lowest order in inclinations, the secular dynamics of the $j$-th planetesimal due to an inclined planet can be described by the orbit-averaged disturbing function $R_{p,j}$ given by:
%%%%%%%%%%%%%%%%%%%%%
\begin{equation}
    R_{p,j} = n_j a_j^2 \bigg[ \frac{1}{2} A_p I_j^2 + B_p I_p I_j \cos(\Omega_p - \Omega_j)    \bigg] , 
    \label{eq:R_planet}
\end{equation}
%%%%%%%%%%%%%%%%%%%%%
where $j = 1, .., N$ \citep{mur99}. In Equation (\ref{eq:R_planet}), the coefficient $A_p$ represents the free precession rate of the planetesimal's longitude of ascending node (due to the axisymmetric component of the planetary potential), while $B_p$ governs its inclination excitation (due to the non-axisymmetric component of the planetary potential). The expression for $A_p$ is given by \citep{mur99}: 
%%%%%%%%%%%%%%%%%%%%%
\begin{eqnarray} 
    A_p &=& - \frac{1}{4} n_j \frac{m_p}{M_c} \frac{a_p}{a_j} b_{3/2}^{(1)}(a_p/a_j)   < 0  , 
     \label{eq:A_planet}
    \\
&\approx&     -0.39~{\rm {Myr}}^{-1}  \bigg(\frac{m_p}{1 M_J} \bigg) a_{p,20}^2 ~ a_{d,80}^{-7/2} ~ M_{c,1}^{-1/2} , 
    \nonumber
\end{eqnarray}
%%%%%%%%%%%%%%%%%%%%%
which, we note, apart from the minus sign, has the exact same expression as the free \textit{apsidal} precession rate induced by an \textit{eccentric} planet \citep[see, e.g.,][]{Sefilian2024}.\footnote{This is strictly true in the linear regime: in the non-linear regime, the finite inclinations and eccentricities of orbits would modify both the in-plane and out-of-plane precession rates \citep[e.g.,][]{laskar-boue}.} The expression for $B_p$, on the other hand, is simply:
%%%%%%%%%%%%%%%%%%%%%
\begin{eqnarray} 
    B_p &=& - A_p  > 0  .
    \label{eq:B_planet}
\end{eqnarray}
%%%%%%%%%%%%%%%%%%%%%
In Equation (\ref{eq:A_planet}), $b_s^{(m)}(\alpha)$ is the  Laplace coefficient given by
%%%%%%%%%%%%%%%%%%%%%
\begin{equation}
    b_{s}^{(m)}(\alpha) = \frac{2}{\pi} \int_{0}^{\pi} \cos(m\theta) \big[1+\alpha^2 - 2\alpha \cos\theta  \big]^{-s} d\theta , 
    \label{eq:bsm_normal}
\end{equation}
%%%%%%%%%%%%%%%%%%%%%
$M_J$ is Jupiter mass, and we have defined $a_{p,20} \equiv a_p/(20 ~{\rm au})$, $a_{d,80} \equiv a_d/(80 ~ {\rm au})$, and $M_{c,1} \equiv M_c/(1 M_{\odot})$. Note that for $\alpha \ll 1$, one has $b_{3/2}^{(1)}(\alpha) \approx 3 \alpha$; an approximation adopted in the numerical estimate of Equation (\ref{eq:A_planet}). Throughout this work, to optimize computations, we evaluate Laplace coefficients using their relationships with hypergeometric functions \citep{Izsak63}; see Appendix \ref{app:laplace_coeff_geom}.

%%%%%%%%%%%%%%%%%%%%%%%%%%%%%%%%%%%%%%%%%%%
\subsubsection{Effects of Massive Planetesimals on the Planet}
\label{subsec:Rdp_sect}
%%%%%%%%%%%%%%%%%%%%%%%%%%%%%%%%%%%%%%%%%%%

The back-reaction of the massive planetesimals on the planet can be described in a similar fashion to $R_{p,j}$ given by Equation (\ref{eq:R_planet}). That is, each planetesimal can be considered as a massive point-mass that perturbs the planet \citep[e.g.,][]{hahn2003, beust2014, SR19, petrovich19, Paper2}. Denoting the disturbing function of the planet due to the $j$-th planetesimal by $R_{j,p}$, the combined effect of all planetesimals $R_{d,p} = \sum_{j = 1}^{N} R_{j,p}$ can be written as:
%%%%%%%%%%%%%%%%%%%%%
\begin{equation}
    R_{d,p} =
    n_p a_p^2 \bigg[ \frac{1}{2} A_{d,p} I_p^2 + \sum_{j = 1}^{N} \delta B_{j,p} I_p I_j \cos(\Omega_p - \Omega_j)    \bigg] , 
    \label{eq:R_disc_planet}
\end{equation}
%%%%%%%%%%%%%%%%%%%%%
where the coefficients $A_{d,p}$ and $\delta B_{j,p}$ read as:
%%%%%%%%%%%%%%%%%%%%%
\begin{eqnarray} 
    A_{d,p} &=& \sum_{j = 1}^{N} A_{j,p} =  - \frac{1}{4} n_p \sum_{j = 1}^{N} \frac{m_j}{M_c} \frac{a_j}{a_p} b_{3/2}^{(1)}(a_j/a_p)   < 0  ,
    \label{eq:Adp}
    \\
    \delta B_{j,p} &=&  + \frac{1}{4} n_p  \frac{m_j}{M_c} \frac{a_j}{a_p} b_{3/2}^{(1)}(a_j/a_p) > 0  .  
    \label{eq:Bdp}
\end{eqnarray}
%%%%%%%%%%%%%%%%%%%%%
Here, $A_{d,p}$ represents the planet's free nodal precession rate due to all planetesimals, while $\delta B_{j,p}$ is a measure of the non-axisymmetric torque exerted by the $j$-th planetesimal.

For analytical purposes, it is also useful to consider the continuum limit of Equation (\ref{eq:Adp}), i.e., as $N \rightarrow \infty$. In that limit, $A_{d,p}$ takes the following form (see Appendix \ref{sec:app_A}):
%%%%%%%%%%%%%%%%%%%%%
\begin{eqnarray}
    A_{d,p} &=&  - \frac{3}{4} n_p \frac{2-p}{p+1} \frac{M_d}{M_c} \bigg(\frac{a_p}{a\out}\bigg)^3 \frac{\delta^{p+1}-1}{1-\delta^{p-2}}\phi_1^c  ,  
    \label{eq:Adp_Bdp_approx} 
    \\
    &\approx&  - 0.41  ~ \textrm{Myr}^{-1} 
        ~ \frac{\phi_1^c}{1.85}
    ~ \bigg( \frac{M_d}{30  M_{\earth}} \bigg)
    ~ \frac{a_{p,20}^{3/2} }{a_{\textrm{out}, 150}^{3}}
    ~  M_{c,1}^{-1/2} ,   
    \nonumber
\end{eqnarray}
%%%%%%%%%%%%%%%%%%%%% 
where the numerical estimate is obtained for the fiducial disk model ($p=3/2$,  $\delta = 5$), assuming $a_p/a\inn = 2/3$ so that $\phi_1^c \approx  1.85$. Here, $\phi_1^c$ is a dimensionless factor of order unity accounting for the contributions of planetesimals close to the planetary orbit (more on this below and in App. \ref{sec:app_A}). A similar expression can also be derived for the summation involving the $\delta B_{j,p}$  term in Equation (\ref{eq:R_disc_planet}),  provided that certain assumptions are made about $I_j$ and $\Delta \Omega \equiv \Omega_j - \Omega_p$, which may or may not depend on $a_j$. Such a calculation is presented in Appendix \ref{sec:app_A}; however, as this paper neglects the disk's non-axisymmetric torque, we do not elaborate further on its behavior.

Finally, we note that apart from the minus sign, $A_{d,p}$ of Equation (\ref{eq:Adp_Bdp_approx}) has the exact same expression as the planetary free \textit{apsidal} precession rate due to an external disk \citep[][]{Paper1}.\footnote{This is because the disk's potential acting on the planet resembles that of an external planet, which, to lowest order, induces \textit{free} precession  rates in apsidal angles and  ascending nodes in opposite directions \citep{mur99}. This also explains the change of sign noted in Section \ref{sec:Rpj-sec} for $A_p$ (Eq. \ref{eq:A_planet}).} Therefore, the coefficient $\phi_1^c$ in Equation (\ref{eq:Adp_Bdp_approx}) is the same dimensionless factor -- which is of order unity --  as given by Equation (A7) in \citet{Paper1}, originally obtained for a coplanar, eccentric disk. For completeness, we note that $\phi_1^c$ is an increasing function of $a_p/a\inn$ that also depends on $p$ and $\delta$, although very weakly. Indeed, regardless of ($p, \delta$), one has $\phi_1^c \approx 1$ for $a_p/a\inn \ll 1$, and $\phi_1^c \sim 10$ for $a_p \sim a\inn$; see  Eq. (\ref{eq:phi2_c_app}) and \citet[][Fig. {13(A)}]{Paper1}.

%%%%%%%%%%%%%%%%%%%%%%%%%%%%%%
\begin{figure}
%%%%%%%%
\centering
\includegraphics[width=\columnwidth]{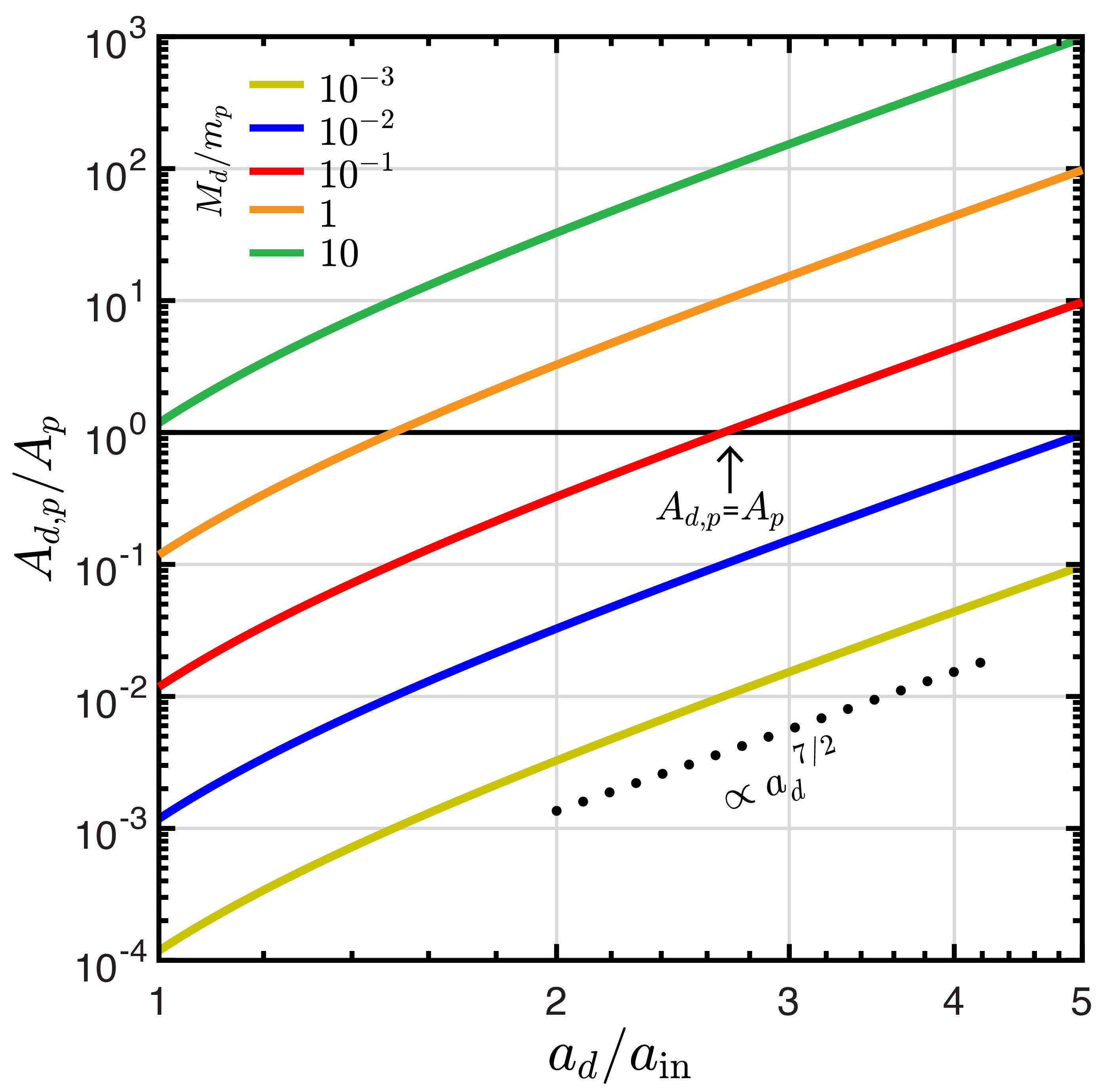}
%%%%%%%%
\vspace{-1.5em}
{\linespread{0.95}\selectfont \caption{The ratio $A_{d,p}/A_p$, i.e., the disk-induced planetary nodal recession rate ($A_{d,p}<0$; Eq. (\ref{eq:Adp})) relative to that of the planetesimals due to the planet ($A_p(a_d)<0$; Eq. (\ref{eq:A_planet})), as a function of planetesimal semimajor axis $a_d/a\inn$. Calculations are done for various disk-to-planet mass ratios, $M_d/m_p$ (shown by different colors), assuming the fiducial disk model ($p=3/2$, $a\inn = 30$ au, $\delta =5$; Section \ref{sec:modelsystem})  and a planet at $a_p = 20$ au. For $10^{-2} \lesssim M_d/m_p \lesssim 10$, the nodal recession rates are equal, $A_{d,p}/A_{p} = 1$, at a unique semimajor axis $a\res$. Interior to this, $|A_{d,p}| \lesssim |A_p|$, while exterior to it, $|A_{d,p}| \gtrsim |A_p|$. For more massive disks, i.e., $M_d/m_p \gtrsim 10$, the ratio $A_{d,p}/A_p \gtrsim 1$ across the entire disk. Conversely, for low-mass disks, i.e., $M_d/m_p \lesssim 10^{-2}$, the ratio $A_{d,p}/A_p \lesssim 1 $ at all semimajor axes. Note that $A_{d,p}/A_p \propto a_d^{7/2}$ for $a_d/a\inn \gg 1 $ (black dotted line; Eq. \ref{eq:Adp_over_Ap}). See the text (Sections \ref{sec:dist_fn_d_p} and \ref{sec:dyn_reg}) for more details.}
\label{fig:A_a}}
\end{figure}
%%%%%%%%%%%%%%%%%%%%%%%%%%%%%%%%%%

This completes the mathematical description of the disk's long-term effects on the planet. Before proceeding, we note two important points that are already clear at this stage. First, for the particular set of parameters in Equations (\ref{eq:A_planet}) and (\ref{eq:Adp_Bdp_approx}),  $A_{d,p} \approx A_p(a_d)$ at $a_d = 80$ au, i.e., a secular inclination resonance is established. Additionally, at larger distances from the star and/or for more massive disks (with other parameters unchanged),  the planet's free nodal precession rate can significantly exceed that of the planetesimals, $|A_{d,p}/A_p(a_d)| \gg 1$, in some or all regions of the disk. This is so even when $M_d/m_p \lesssim 1$ (contrary to naive expectations), as can be clearly seen in Figure \ref{fig:A_a}, where we plot the profiles of  $|A_{d,p}/A_p|$ as a function of planetesimal semimajor axis $a_d/a\inn$ for various values of  $M_d/m_p$. Calculations assume the fiducial parameters ($p = 3/2, a\inn = 30$ au, and $\delta =5$; Section \ref{sec:modelsystem}) and a planet at $a_p = 20$ au. Note that the condition of Equation (\ref{eq:MMR_condition_delta_ap}) would be satisfied if $m_p \lesssim 38 M_J$ for  $M_c = 1 M_{\odot}$, i.e., for all practical purposes. These observations have crucial implications for the system's evolution, including the disk morphology, which we will elaborate on in the subsequent sections.

%%%%%%%%%%%%%%%%%%%%%%%%%%%%%%%%%%%%%%%%%%%%%%%%%%%%%%%%%%%%%%%
\subsection{General Evolution Equations}
%%%%%%%%%%%%%%%%%%%%%%%%%%%%%%%%%%%%%%%%%%%%%%%%%%%%%%%%%%%%%%%

The disturbing functions given by Equations (\ref{eq:R_planet}) and (\ref{eq:R_disc_planet}) fully describe the secular inclination dynamics of a massive debris disk interacting with an interior planet (as long as the disk \textit{self}-gravity is ignored; see Section \ref{subsec:Rdp_sect}). The evolution of the planetary and planetesimal orbits can be determined with the aid of Lagrange's planetary equations \citep{mur99}. Namely, expressing the inclination vector of each body, $\mathbf{I} = (Q, P) = I (\cos \Omega, \sin \Omega)$, in terms of the complex Poincar\'e variable, 
%%%%%%%%%%%%%%%%%%%%%
\begin{equation}
    \zeta \equiv I  \cdot {\rm exp}(i \Omega) = Q + i P , 
\end{equation}
%%%%%%%%%%%%%%%%%%%%%
we find that Lagrange's planetary equations read as follows:
%%%%%%%%%%%%%%%%%%%%%
\begin{eqnarray}
   \dot{\zeta}_j &\equiv& \frac{d\zeta_j}{dt} =  i \bigg[ A_p(a_j) \zeta_j + B_p(a_j) \zeta_p     \bigg] 
   ,  
    \label{eq:zdot_j}
     \\
    \dot{\zeta}_p &\equiv& \frac{d\zeta_p}{dt} =   i \bigg[ A_{d,p} \zeta_p + \sum_{j = 1}^{N} \delta B_{j,p} \zeta_j     \bigg] 
    , 
    \label{eq:zdot_p}
\end{eqnarray}
%%%%%%%%%%%%%%%%%%%%%
where  Equation (\ref{eq:zdot_j}) applies for $j= 1,...,N$. These $N+1$ coupled ordinary differential equations can be further compacted into a matrix equation of the form $\dot{\zeta} = i \mathbf{B} \zeta$:
%%%%%%%%%%%%%%%%%%%%%
\begin{equation}
\frac{d}{dt}
\begin{bmatrix}
    {\zeta}_{p} \\
    {\zeta}_{1} \\
    {\zeta}_{2} \\
    \vdots \\
    {\zeta}_{N}
\end{bmatrix}
= i 
\begin{bmatrix}
    A_{d,p} & \delta B_{1,p}  
    & \delta B_{2,p} 
    &\dots & \delta B_{N,p} 
    \\
    B_p(a_1) & A_p(a_1) & 0 &\dots & 0 \\
     B_p(a_2) & 0 & A_p(a_2) &\ddots & \vdots \\
    \vdots & \vdots & \ddots & \ddots & 0
    \\
    B_p(a_N)  &0 & \dots & 0 & A_p(a_N)
\end{bmatrix}
\begin{bmatrix}
    {\zeta}_{p} \\
    {\zeta}_{1} \\
    {\zeta}_{2} \\
    \vdots \\
    {\zeta}_{N}
\end{bmatrix}
,
\label{eq:matrix_eom}
\end{equation}
%%%%%%%%%%%%%%%%%%%%%
where $a_1 = a\inn$ and $a_N = a\out$. Equation (\ref{eq:matrix_eom}) represents the master equation needed for our work, encapsulating the coupled secular inclination evolution of the planet and the back-reacting disk. In this work, we neglect the terms $\delta B_{j,p}$ in Equation (\ref{eq:matrix_eom}); full consideration of this non-axisymmetric component is deferred to future work (see, however, Section \ref{subsec:limit_future_work} and App. \ref{app:res_friction}). As we shall see,  this allows us to fully flesh out the basic effects of disk gravity analytically.

Finally, we note that in the case of a massless disk,  $A_{d,p} = \delta B_{j,p}  = 0$ for all $j=1,..,N$, and Equation (\ref{eq:matrix_eom}) reduces to the classical evolution equations of test-particles due to an internal, non-precessing planet on an inclined orbit \citep[][]{mur99}. Similarly, for $N=1$, one recovers the equations governing the secular interactions of two mutually inclined planets on circular orbits. Note that this is because when $N=1$, the planetesimal is formally treated as a single non-self-gravitating ring. To model a self-gravitating ring instead, one must consider $N > 1$ planetesimals sharing a fixed, single-valued semimajor axis and spatially soften the ring-ring interaction potential -- e.g., by modeling the planetesimals as Plummer spheres -- to properly account for their mutual interactions \citep[][]{Touma2002, hahn2003, SR19}.

%%%%%%%%%%%%%%%%%%%%%%%
\section{Secular behavior: simplified analytical results}
\label{sec:analytical_analysis}
%%%%%%%%%%%%%%%%%%%%%

We aim to elucidate the dynamical effects of the disk's back-reaction on the planet, providing insights into the vertical structure of massive debris disks. Although Equation (\ref{eq:matrix_eom})  does not admit an analytical closed-form solution for $M_d \neq 0$,\footnote{The difficulty arises due to the \textit{non-axisymmetric} component of the disk's potential which is proportional to $I_p I_d \cos\Delta\Omega$ and may vary over time (Equation \ref{eq:R_disc_planet}). In principle, an approximate analytical solution  could be attempted with an educated guess for the time-averaged $I_d(t)$ and $\Delta\Omega(t)$, but this is not obvious to us a priori (see also Appendix \ref{app:res_friction}).} progress can be made in a simplified case by considering only the \textit{axisymmetric} component of the disk's potential (i.e., setting $\delta B_{j,p} = 0$ for all $j$), as outlined below. At the outset, we emphasize that adopting this simplified framework does not imply that the neglected terms are unimportant. Rather, it allows us to isolate and analytically examine the most basic role of disk gravity within a tractable setting, avoiding the need for numerical simulations. Nevertheless, in Appendix \ref{app:res_friction}, we derive an analytical expression from first principles describing the effects of the neglected terms on the planet,  while their impact on the disk dynamics is discussed qualitatively in Section \ref{subsec:limit_future_work}.

In this simplified case with $\delta B_{j,p} =0$, assuming planetesimals are initiated on coplanar orbits  ($\zeta_j(0) = 0$), Equation (\ref{eq:matrix_eom}) can be solved analytically. Doing so, we  find that the planetary orbit evolves such that: 
%%%%%%%%%%%%%%%%%%%%%%%%%%%%
\begin{eqnarray}
    I_p(t) &=& I_p(0)  ,  
    \label{eq:Ip_sol}
    \\
    \Omega_p(t) &=& A_{d,p} t + \Omega_p(0) .  
    \label{eq:Omegap_sol}
\end{eqnarray}
%%%%%%%%%%%%%%%%%%%%%%%%%%%%
The evolution of planetesimal orbits, on the other hand, can be described by the sum of the free and forced inclination vectors, $\mathbf{I}_d(t) = \mathbf{I}_{d, \rm free}(t) + \mathbf{I}_{d, \rm forced}(t)$, so that:\footnote{{Assuming $\delta B_{j,p} =0$, the solutions given by Equations (\ref{eq:Id_sol})--(\ref{eq:I_forced_gen}) can be alternatively obtained via the disturbing function $R_{p,j}$ (Equation \ref{eq:R_planet}), upon switching to a frame co-precessing with the planet. This can be done by subtracting $\Psi A_{d,p}$ from Equation (\ref{eq:R_planet}), where $\Psi = n_j a_j^2 (1-\cos(I_j)) \approx n_j a_j^2 I_j^2 /2$ is the action conjugate to the angle $\Delta\Omega$} \citep{Goldstein1950}.} 
%%%%%%%%%%%%%%%%%%%%%%%%%%%%
\begin{eqnarray}
   & &  I_d(t) = 2 \bigg|I_{d, \rm{forced}}  \sin \bigg(\frac{A_p-A_{d,p}}{2}t \bigg) 
  \bigg|  , 
  \label{eq:Id_sol}
    \\
   & & \tan \Delta\Omega(t) =   \tan \bigg( \frac{A_p - A_{d,p}}{2}t + \frac{\pi}{2} \bigg)  ,  
   \label{eq:tan_DO}
\end{eqnarray}
%%%%%%%%%%%%%%%%%%%%%%%%%%%%
where $\Delta\Omega = \Omega_d - \Omega_p$ stays in the range $[-\pi, \pi]$. In Equation (\ref{eq:Id_sol}), the forced planetesimal inclination $I_{d, \rm{forced}}$ is:
%%%%%%%%%%%%%%%%%%%%%%%%%%%%
\begin{equation}
  I_{d, \rm forced}(a_d)  = I_{d, \rm free}(a_d) =I_p(0) \frac{A_p(a_d)}{A_p(a_d)- A_{d,p}}   ,
  \label{eq:I_forced_gen}
\end{equation}
%%%%%%%%%%%%%%%%%%%%%%%%%%%% 
where we have used the fact that $B_p = -A_p$ (Equation (\ref{eq:B_planet})). As expected, when $A_{d,p} \propto M_d = 0$, Equations (\ref{eq:Ip_sol})--(\ref{eq:I_forced_gen}) reduce to the well-known results for massless particles perturbed by an inner inclined planet \citep[e.g.,][]{Dawson2011}.

We note that $I_{d, \rm forced}  = I_{d, \rm free}$ due to the initial conditions (namely, $\zeta_j(0) =0$), and $I_{d, \rm forced}$ is time-independent as $I_p(t)$ remains constant. These points, while straightforward, will prove to be useful for the interpretation of our results, particularly as we incorporate the disk's full back-reaction in an upcoming work (see Appendix \ref{app:res_friction}). For this reason, we highlight that, by definition, Equation (\ref{eq:I_forced_gen}) can be interpreted as the time-averaged planetesimal inclinations (Equation \ref{eq:Id_sol}), which, from hereon, we denote as $\langle I_d(t) \rangle_t$.

%%%%%%%%%%%%%%%%%%%%%%%%%%%%%%
\begin{figure}
%%%%%%%%
\centering
\includegraphics[width=\columnwidth]{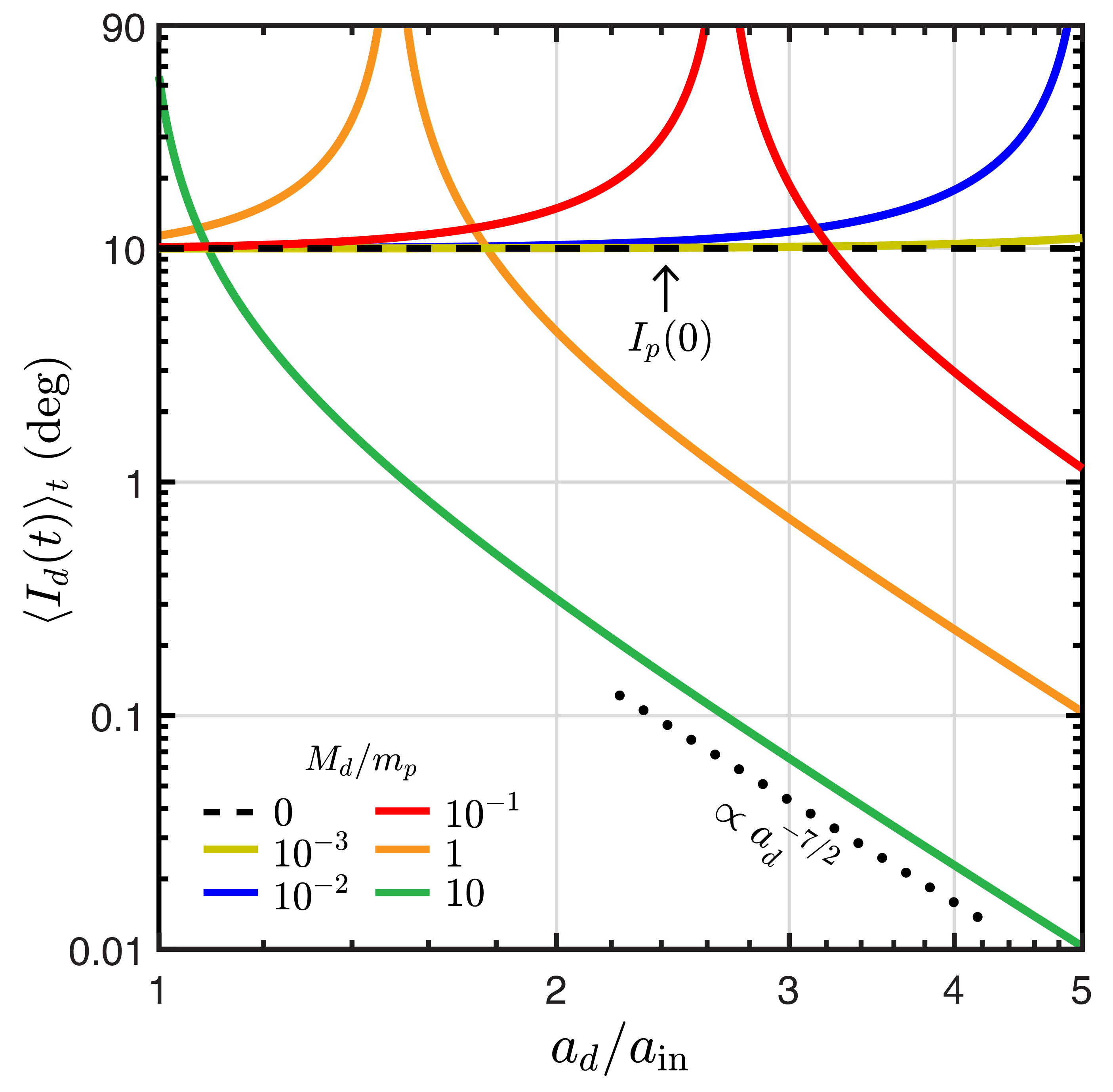}
%%%%%%%%
\vspace{-1.5em}
{\linespread{0.95}\selectfont \caption{Profiles of the time-averaged planetesimal inclinations $\langle I_{d}(t) \rangle_t$ as a function of their semimajor axes $a_d/a\inn$, computed using Equation (\ref{eq:I_forced_gen}) for different values of disk-to-planet mass ratios $M_d/m_p$. These profiles scale linearly with the initial planetary inclination, which here we have taken to be {$I_p(0) =10^{\circ}$}. All other system parameters are identical to those in Figure \ref{fig:A_a}. For reference, the black dashed line shows the results for a massless disk, i.e., $\langle I_{d}(t) \rangle_t = I_p(0)$ (see also Eq. (\ref{eq:Im_no_disk})). In contrast, the black dotted line illustrates the asymptotic behavior of the inclinations when suppressed by the disk gravity, i.e., $\langle I_{d}(t) \rangle_t \propto a_d^{-7/2}$ (see also Eq. (\ref{eq:Im_disk_sup_numest})). Note the occurrence of a secular inclination resonance for $10^{-2} \lesssim M_d/m_p \lesssim 10$, where $\langle I_{d}(t) \rangle_t \rightarrow \infty$ (see also Eq. (\ref{eq:lambda_If_gen})). See the text (Section \ref{sec:analytical_analysis}) for details.}
\label{fig:I_forced_a}}
\end{figure}
%%%%%%%%%%%%%%%%%%%%%%%%%%%%%%%%%%

For illustrative purposes,   profiles of $\langle I_d(t) \rangle_t$ as a function of planetesimal semimajor axis $a_d/a\inn$ are shown in Figure \ref{fig:I_forced_a}  for different values of $M_d/m_p$. These profiles,  computed using Equation (\ref{eq:I_forced_gen}), correspond to the same system parameters as in Figure \ref{fig:A_a}. The calculations assume an initial value of $I_p(0) = 10^{\circ}$ for the planetary orbit; this serves as a linear scaling factor since  $\langle I_d(t) \rangle_t \propto I_p(0)$ (Equation \ref{eq:I_forced_gen}). Figure \ref{fig:I_forced_a} reveals several noteworthy features, which we unpack below.

A prominent result in Figure \ref{fig:I_forced_a} is the divergence of planetesimal inclinations, $\langle I_d(t) \rangle_t \rightarrow \infty$, around a specific semimajor axis, $a_d =a\res$, across a broad range of disk masses, $10^{-2} \lesssim M_d/m_p \lesssim 10$. Looking at Figures \ref{fig:A_a} and \ref{fig:I_forced_a}, one can see that this happens when $A_{p} = A_{d,p}$, that is to say, when a \textit{secular inclination resonance} occurs.\footnote{{The possibility of establishing a secular inclination resonance in debris disks of non-zero masses  was noted in passing by \citet{Paper1}.}}  The resonance site $a\res$ is closer to the disk's inner edge than to its  outer edge for larger values of $M_d/m_p$, and vice versa. Furthermore, beyond this location, i.e., at $a_d \gtrsim a\res$ where $|A_{d,p}| \gtrsim |A_p|$, planetesimal inclinations are suppressed in magnitude compared to the case of a massless disk, in which case $\langle I_d(t)\rangle_t = I_p(0)$; see the black dashed line. This happens such that the inclinations roughly follow a power-law profile scaling as $\propto a_d^{-7/2}$; see the black dotted line. We refer to this case as the ``disk-dominated'' regime. On the other hand, interior to $a\res$,  where $ |A_p| \gtrsim |A_{d,p}|$, the inclinations converge to that expected from massless disk calculations with $\langle I_d(t)\rangle_t \rightarrow I_p(0)$. We refer to this case as the ``planet-dominated'' regime. Finally, disk gravity is negligible only for very low-mass disks, typically $M_d/m_p \lesssim 10^{-2}$, resulting in planetesimal inclinations consistent with the massless disk case.

In the remainder of this section, we analytically demonstrate that the behavior of the planetesimals falls into two regimes (Section \ref{sec:dyn_reg}) and describe the orbital evolution in each case (Section  \ref{subsec:orbit-evol-massive-less-disks}). 
%%%%%%%%%%%%
To aid interpretation, we provide numerical estimates for key equations using the fiducial disk parameters ($p = 3/2$ and $\delta = 5$; Section \ref{sec:modelsystem}).
%%%%%%%%%%%%%
Readers interested in the impact on the resultant disk vertical structure may proceed directly to Section \ref{sec:vert_struc_num}.

%%%%%%%%%%%%%%%%%%%%%%%%%%%%%%%%%%%%%%%%%%%%%%%%%
\subsection{Dynamical regimes}
\label{sec:dyn_reg}
%%%%%%%%%%%%%%%%%%%%%%%%%%%%%%%%%%%%%%%%%%%%%%%%%

Equations (\ref{eq:Ip_sol})--(\ref{eq:I_forced_gen}), along with Figures \ref{fig:A_a} and \ref{fig:I_forced_a}, demonstrate that the evolution of planetesimal orbits is governed by the nodal precession rate of the planetesimals relative to the planet, $A(a_d) \equiv A_p(a_d) - A_{d,p}$; see Equations (\ref{eq:A_planet}) and (\ref{eq:Adp_Bdp_approx}). We have identified two dynamical regimes:\footnote{These regimes are akin to those identified in coplanar setups \citep{rafikov_ptype, silsbeekepler, Paper1, Sefilian2024}.}
(i) a \textit{disk-dominated} regime, where $|A_{d,p}| \gtrsim |A_p|$ ($A \approx - A_{d,p}$), and (ii) a \textit{planet-dominated} regime, where $|A_p| \gtrsim |A_{d,p}|$ ($A \approx A_p$). In the extreme limit when  $|A_p| \gg |A_{d,p}|$, we recover the case of a massless disk. The transition between these two regimes occurs through a \textit{secular inclination resonance} at $a_d = a\res$, where $A_p(a\res) = A_{d,p}$ and $\langle I_d(t) \rangle_t \rightarrow \infty$.

To characterize this analytically,  we first analyze the ratio $A_{d,p}/A_{p}$. Using Equations (\ref{eq:A_planet}) and (\ref{eq:Adp_Bdp_approx}), we find that: 
%%%%%%%%%%%%%%%%%%%%%%%
\begin{eqnarray}
    \frac{A_{d,p}}{A_p} 
    &\approx& \frac{(2-p) \phi_1^c}{(p+1) C \delta^3} \frac{M_d}{m_p} \bigg(\frac{a_d}{a\inn}\bigg)^{7/2}  \bigg( \frac{a\inn}{a_p}\bigg)^{1/2} , 
    \label{eq:Adp_over_Ap}
    \\
   &\approx& 1.24
     ~ \frac{\phi_1^c  }{1.85}
     ~ \bigg( \frac{M_d/m_p}{0.1} \bigg)
     ~ \bigg(\frac{a_d/a\inn}{ 2.75}\bigg)^{7/2}
     ~ \bigg( \frac{a\inn/a_p}{1.5}\bigg)^{1/2} .
    \nonumber
\end{eqnarray}
%%%%%%%%%%%%%%%%%%%%%%%
In Equation (\ref{eq:Adp_over_Ap}), we have assumed  $a_p/a_d \ll 1$, so that $b_{3/2}^{(1)}(\alpha) \approx 3 \alpha$, and $C \equiv (1-\delta^{p-2})/(\delta^{p+1}-1)$. Note that for $\delta = a\out/a\inn \gg 1$, one has $C \delta^3 \rightarrow \delta^{2-p}$ for  $0  < p < 2$, i.e., for disks that contain more mass in the outer parts than in the inner parts (Equation \ref{eq:Sigma_d}). Accordingly, the ratio $A_{d,p}/A_p$ is positive and, more importantly, an increasing function of $a_d/a\inn$ for all astrophysically motivated values of $p$; see also Figure \ref{fig:A_a}. Note that $A_{d,p}/A_p$ depends on the planet and disk masses only through their ratio, $M_d/m_p$, and is independent of $M_c$.

%%%%%%%%%%%%%%%%%%%%%%%%%%%%%%%%%%
\subsubsection{Secular inclination resonance}
\label{subsubsec:sr}
%%%%%%%%%%%%%%%%%%%%%%%%%%%%%%%%%%

A secular inclination resonance occurs when the nodal precession rates of planetesimals and the planet match:
%%%%%%%%%%%%%%%%%%%%%%%
\begin{equation}
A(a\res) = 0 
\Rightarrow
A_{d,p} = A_p(a\res) ,
\label{eq:res_condition_gen_exp}
\end{equation}
%%%%%%%%%%%%%%%%%%%%%%%
see also Equation (\ref{eq:I_forced_gen}). Thus, by setting $A_{d,p}/A_p$, as given by Equation (\ref{eq:Adp_over_Ap}), equal to unity and solving for $a_d = a\res$, we find that the resonance location (for $a_p/a\res \lesssim 1$) is:
%%%%%%%%%%%%%%%%%%%%
\begin{eqnarray}
   \frac{a\res}{a\inn}  &\approx&  \bigg[ \frac{ (2-p)\phi_1^c  }{(p+1) C \delta^3}  \frac{M_d}{m_p} 
   \bigg( \frac{a\inn}{a_p}  \bigg)^{1/2}    \bigg]^{-2/7}, 
   \label{eq:res_loc}
    \\
    &\approx& 
    2.6  ~ \bigg( \frac{M_d/m_p}{0.1} \bigg)^{-2/7} \bigg( \frac{a_{\rm in}/a_p}{1.5} \bigg)^{-1/7} \bigg(\frac{\phi_1^c}{1.85}\bigg)^{-2/7} . 
    \nonumber
\end{eqnarray}
%%%%%%%%%%%%%%%%%%%% 
Since $A_{d,p}/A_p$ is an increasing function of $a_d$ (Equation \ref{eq:Adp_over_Ap}), it follows from Equation (\ref{eq:res_loc}) that when a secular resonance occurs within the disk, planetesimal dynamics will be planet-dominated interior to the resonance (i.e., for $a\inn \leq a_d \lesssim a\res$, where $A_{d,p}/A_p \lesssim 1$), and disk-dominated exterior to it (i.e., for $ a\res \lesssim a_d \leq a\out$, where $A_{d,p}/A_p \gtrsim 1$). We note that Equations (\ref{eq:Adp_over_Ap}) and (\ref{eq:res_loc}) well approximate the results shown in Figures \ref{fig:A_a} and  \ref{fig:I_forced_a}, respectively, except for $a_d \approx a\inn$, since the assumption $a_p \ll a_d$ used in deriving those equations no longer holds.

\begin{figure}
%%%%%%%%
\centering
\includegraphics[width=\columnwidth]{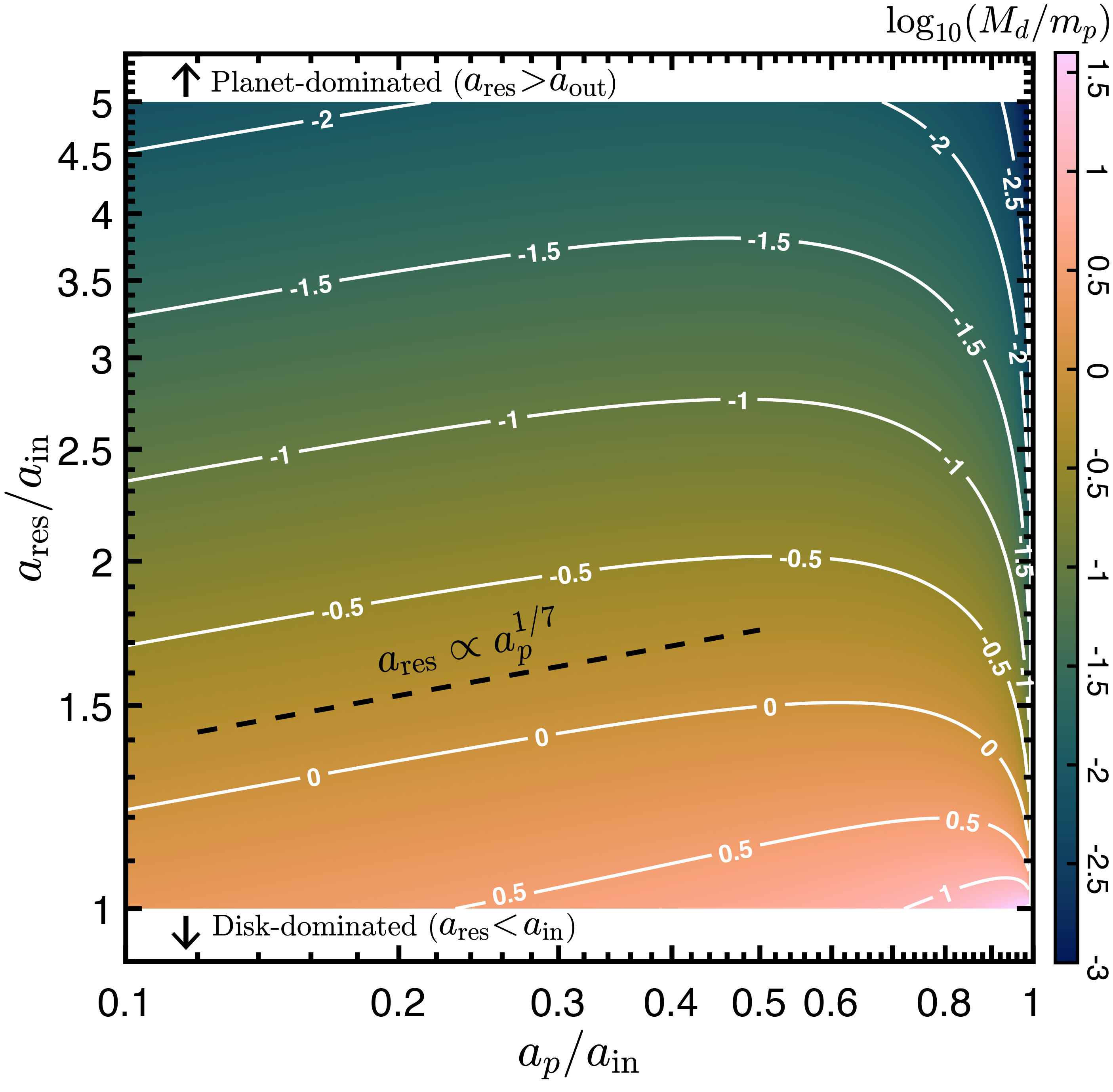}
%%%%%%%%
\vspace{-1.5em}
{\linespread{0.95}\selectfont \caption{The locations of secular-inclination resonances relative to the disk inner edge, $a\res/a\inn$, as functions of planetary semimajor axes $a_p/a\inn$ and disk-to-planet mass ratios $M_d/m_p$. Results are obtained by numerically solving the resonance condition (Equation \ref{eq:res_condition_gen_exp}) for a disk model with $p=3/2$ and $\delta = a\out/a\inn = 5$ (Section \ref{sec:modelsystem}). The white solid curves represent constant $M_d/m_p$ contours in $\log_{10}$ units (see the color bar). The black dashed line shows the scaling of $a\res$ with $a_p$ for fixed $M_d/m_p$ (Equation \ref{eq:res_loc}). Note the robustness of establishing a resonance across a wide range of $a_p/a_{\text{inn}}$ and $M_d/m_p$. When no resonances occur within the disk, planetesimal dynamics across the entire disk will be dominated either by the planet's potential (i.e., $M_d/m_p \lesssim 10^{-3}$) or by the disk's potential (i.e., $M_d/m_p \gtrsim 1$) alone. See the text (Section \ref{sec:dyn_reg}) for details. }
\label{fig:resonance_map}}
\end{figure}

Equation (\ref{eq:res_loc}) shows that a secular-inclination resonance can occur at some location within the disk, $a\inn \leq a\res \leq a\out$, for a broad range of system parameters. To illustrate this further, Figure \ref{fig:resonance_map} shows a contour plot of $M_d/m_p$ (measured in $\log_{10}$ units) in the $(a_p/a\inn, a\res/a\inn)$ space, derived by numerically solving the resonance condition (Equation \ref{eq:res_condition_gen_exp}) for the fiducial disk model.
%%%%%%%% 
Looking at Figure \ref{fig:resonance_map}, it is evident that for a given planet, one or no secular resonance occurs within the disk, depending on $M_d/m_p$. Indeed, for a given $a_p/a\inn \lesssim 1$, resonances are established provided that $10^{-3} \lesssim M_d/m_p \lesssim 10$. Moreover, Figure \ref{fig:resonance_map} shows that for a given $a_p/a\inn$, increasing $M_d/m_p$ shifts the resonance location closer to the disk's inner edge, and vice versa (see also Figures \ref{fig:A_a} and \ref{fig:I_forced_a}). Relatedly, for planets closer to the disk, a larger $M_d/m_p$ is required to maintain a given resonance location. These behaviors are well explained by Equation (\ref{eq:res_loc}), which accurately approximates the slopes of the constant $M_d/m_p$ contours -- namely, $a\res \propto a_p^{1/7}$ -- as illustrated by the black dashed line in Figure \ref{fig:resonance_map}. As expected, the contours deviate from this scaling as $a_p, a\res \rightarrow a\inn$, at which point $\phi_1^c \rightarrow \infty$ and $b_{3/2}^{(1)}(\alpha)$ deviates from the $\approx 3 \alpha$ approximation (see Equation \ref{eq:res_loc}).

%%%%%%%%%%%%%%%%%%%%%%%%%%%%%%
%%%%%%%%%%%%%%%%%%%%%%%%%%%%%%%%%%

%%%%%%%%%%%%%%%%%%%%%%%%%%%%%%%%%%%%%%%%%%%%
%%%%%%%%%%%%%%%%%%%%%%%%%%%%%%%%%%%%%%%%%%%%
\begin{table*}
\centering
\caption{Parameters of the planet--debris disk systems modeled in Sections \ref{sec:analytical_analysis} and \ref{sec:vert_struc_num}. 
\label{table:models}}
%%%%%%%%%%%%%%%%%%%%%%%%%
\begin{tabular}{lccccccccc}
\hline
\hline
Model 
&  $M_d/m_p$
& $m_p $     
& $a_p$ 
& Regime  
& $I_p(0)$  
& $T_{\dot{\Omega}_p}$ 
& $T_{\rm sec}(a\out)$
& Other timescales 
\\
%%%%%%%%%%%%%%
\hline
%%%%%%%%%%%%%%
\texttt{A} 
&  $0$
& $1 M_J$ 
& $20$ {\rm au}
&  Planet-dominated: entire disk   
& $10^{\circ}$ 
& $\infty$
& $139.6$  Myr 
& $T_{{\rm sec}}(80~{\rm au}) \approx 14.2$  Myr 
\\
\texttt{B} 
&  $0.1$
&  \dots 
& \dots  
& Secular resonance: $a\res \approx 80.2$ au 
& \dots  
& $14.3$  Myr
& $16 $ Myr
& $\tau\res \approx 20.5$  Myr 
\\
\texttt{C} 
& $10$ 
&  $10 M_{\earth}$ 
& \dots  
& Disk-dominated: entire disk 
& \dots  
& $4.6$ Myr
&  $ \approx T_{\dot{\Omega}_p}$
& $T_{\rm sec}(a\inn) \approx 29.5$ Myr
\\
\hline
\end{tabular}
%%%%%%%%%%%%%%%%%%%%%%%%%
\begin{flushleft}
\textbf{Notes.}
{The combinations of $M_d/m_p$, $m_p$, and $a_p$ (Columns 2--4)  are chosen to cover all dynamical regimes (Column 5) identified in Section \ref{sec:dyn_reg}. Column 6 shows the planet's initial orbital inclination, with an ascending node $\Omega_p(0) =0$ that precesses with a period listed in Column 7 (Eq. \ref{eq:planetary_node_rate}). Columns 8 and 9 list the secular oscillation timescale at $a\out = 150$ au and the resonance (if any) or secular timescale at other locations, respectively (Eqs. (\ref{eq:Tsec_planetesimals}), (\ref {eq:tau_res_growth})). All other parameters are assigned their fiducial values (Section \ref{sec:modelsystem}), satisfying Equation (\ref{eq:MMR_condition_delta_ap}).}
\end{flushleft}
\end{table*}
%%%%%%%%%%%%%%%%%%%%%%%%%%%%%%%%%%%%%%%%%%%%
%%%%%%%%%%%%%%%%%%%%%%%%%%%%%%%%%%%%%%%%%%%%

%%%%%%%%%%%%%%%%%%%%%%%%%%%%%%%%%%
\subsubsection{Disk-dominated and planet-dominated regimes}
%%%%%%%%%%%%%%%%%%%%%%%%%%%%%%%%%%

A corollary of the results in Section \ref{subsubsec:sr} is that for values of $M_d/m_p$ where no resonance occurs within the disk, planetesimals across the entire disk would be in either the disk-dominated or planet-dominated regime, as indicated in Figure \ref{fig:resonance_map}. This behavior can also be seen in Figure \ref{fig:A_a}. Indeed, for relatively high-mass disks, planetesimal dynamics at all semimajor axes will be dominated by the disk gravity. This happens when $A_{d,p}/A_p \geq 1 $ at $a_d \geq a\inn$ (recall that $A_{d,p}/A_p \propto a_d^{7/2}$),  or  equivalently, $a\res \leq a\inn$, requiring disk masses of (Equations \ref{eq:Adp_over_Ap} and \ref{eq:res_loc}):
%%%%%%%%%%%%%%%%%%%%%%
\begin{eqnarray}
   \frac{M_d}{m_p}\bigg|_{d}   &\geq& 
   \frac{(p+1) C \delta^3}{ (2-p)\phi_1^c  } \sqrt{\frac{a_p}{a\inn}}
   \approx {2.8} ~ \frac{1.85}{\phi_1^c}
   \sqrt{ \frac{1.5}{a\inn/a_p} } ,
   \label{eq:Mdmp_disk_dominated}
\end{eqnarray}
%%%%%%%%%%%%%%%%%%%%%%
see also Figure \ref{fig:resonance_map}. In other words, this is so for all values of $M_d \sim m_p$ and larger for a given planet, noting that the numerical factor in Equation (\ref{eq:Mdmp_disk_dominated}) represents a lower limit for $a_p\sim a\inn$ (where $b_{3/2}^{(1)}(\alpha)$ deviates from $\approx 3 \alpha$). Alternatively, adopting a representative disk mass of $M_d = 100 M_{\earth}$ (noting the upper bound of $\sim 10^{3}M_{\earth}$ according to \citet{krivovwyatt21}), the disk-dominated regime holds for planets less massive than 
%%%%%%%%%%%%%%%%%%%%%%
\begin{equation}
   {m_p} \leq 2.1  M_N ~\frac{\phi_1^c}{1.85}    
    \sqrt{\frac{a\inn/a_p}{1.5}} \frac{M_d}{100 M_{\earth}}  ,   
    \label{eq:Mdmp_dd_dominated}
\end{equation}
%%%%%%%%%%%%%%%%%%%%%%
where $M_N$ is Neptune mass.

%%%%%%%%%%%%%%%%%%%%%%%%%%%%%%%%
\begin{figure*}
%%%%%%%%%%%%
\begin{minipage}{2.1\columnwidth}
\includegraphics[width=\columnwidth]{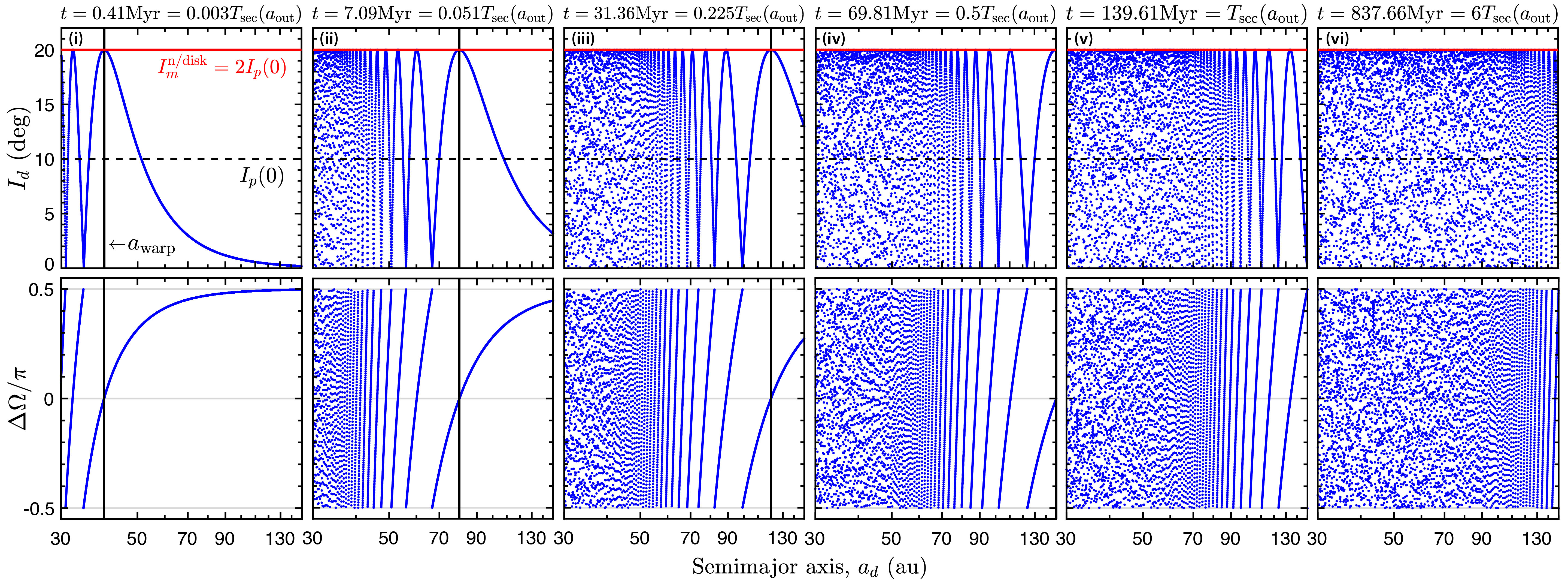}
\par \smallskip
\centering{(a) \texttt{Model A}  ($M_d = 0$). Dynamical regime: planet-dominated across the entire disk.}
\end{minipage}
\\[0.5ex]
%%%%%%%%%%%%
\begin{minipage}{2.1\columnwidth}
\includegraphics[width=\columnwidth]{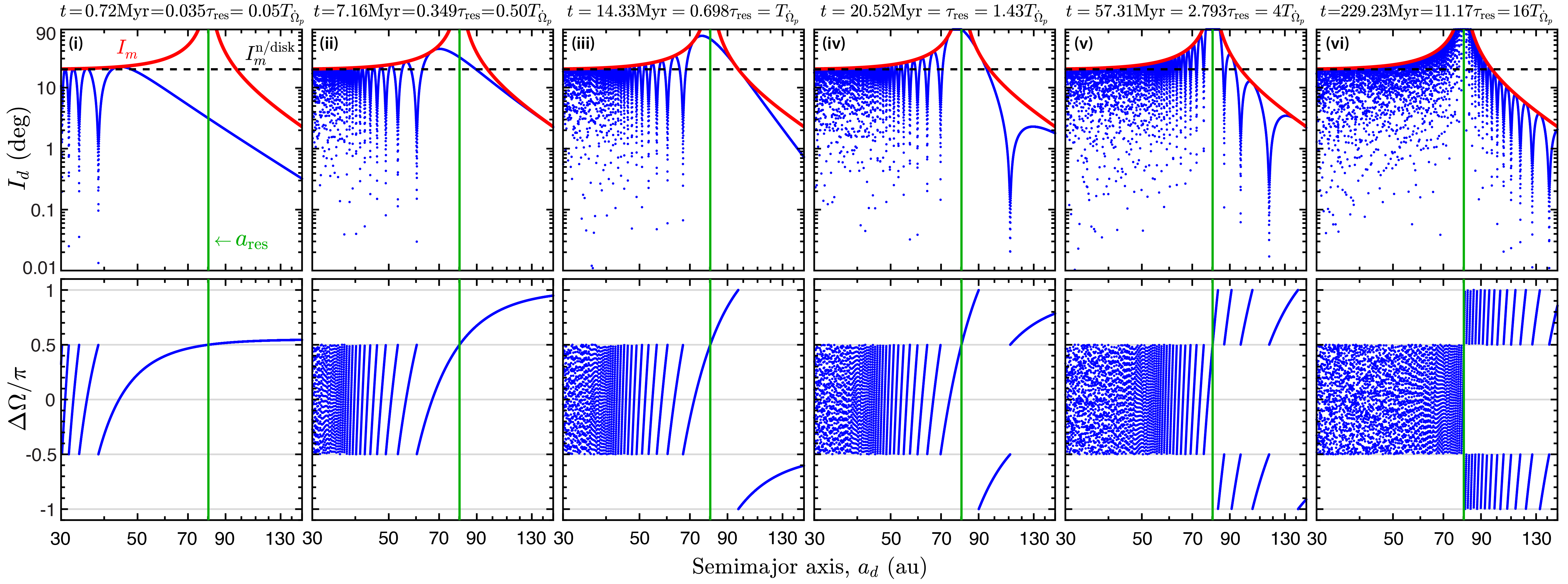}
\par \smallskip
\centering{(b) \texttt{Model B} ($M_d/m_p = 0.1$). Dynamical regime: secular resonance at $a\res \approx 80.2$ au, planet (disk)-dominated interior (exterior) to $a\res$.}     
\end{minipage}
\\[0.5ex]
%%%%%%%%%%%%
\begin{minipage}{2.1\columnwidth}
\includegraphics[width=\columnwidth]{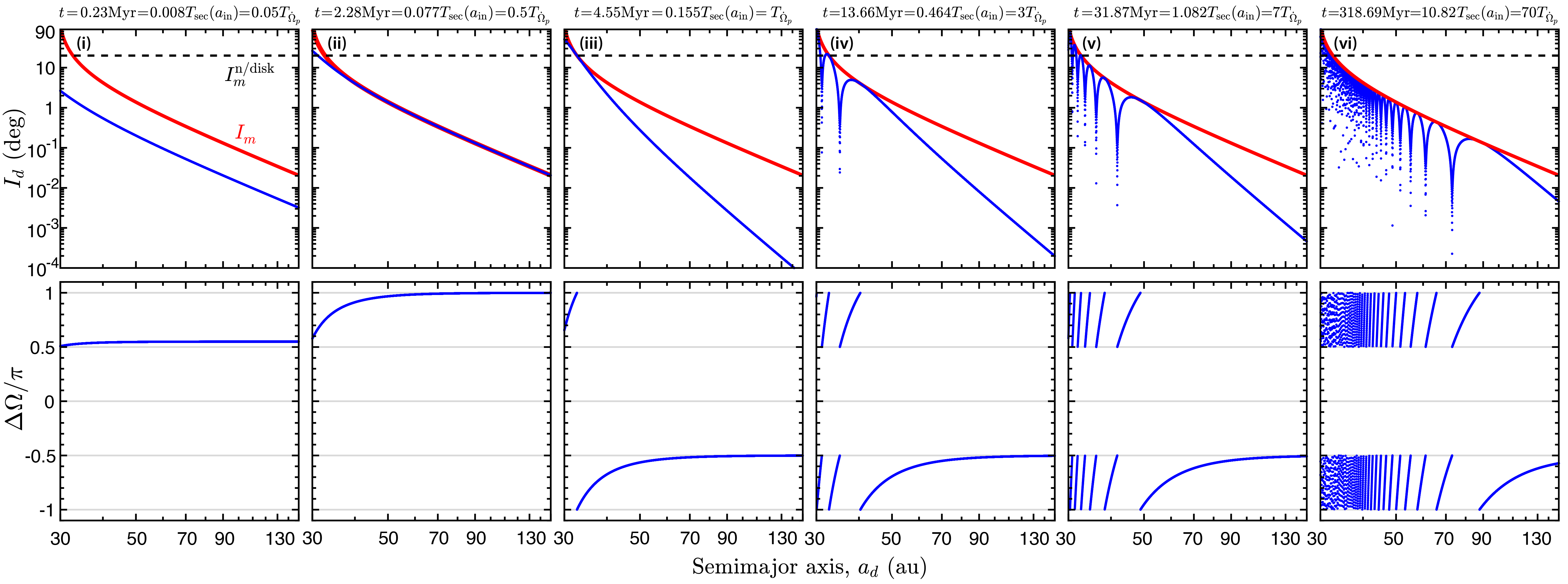}
\par \smallskip
\centering{(c) \texttt{Model C} ($M_d/m_p = 10$). Dynamical regime: disk-dominated across the entire disk.}
\end{minipage}
\\[0.5ex]
%%%%%%%%%%%%
\vspace{-0.75em}
%%%%%%%%%%%%
{\linespread{0.95}\selectfont \caption{The radial profiles of planetesimal inclinations $I_d$  and longitudes of ascending node $\Delta\Omega = \Omega_d - \Omega_p$ (top and bottom rows in each panel, respectively) at different snapshot times (left to right). Panels (a), (b), and (c) correspond to different planet--debris disk systems (Table \ref{table:models}), as indicated in subcaptions. Snapshot times are provided in absolute terms $t$ and relative to relevant secular timescales (Section \ref{subsec:orbit-evol-massive-less-disks}). Each model includes $N = 5,000$ planetesimals initialized on coplanar orbits ($I_d(0) = 0$) and a planet with $I_p(0) = 10^{\circ}$ and $\Omega_p(0) =0 $. In each panel, the red curves show the maximum inclinations, $I_m = 2 | I_{d, {\rm forced}} | $ (Eq. \ref{eq:I_forced_gen}). For reference, the dashed black lines in panels (b) and (c) show the maximum inclinations in the absence of disk gravity,  i.e., $I_m^{{\rm n/disk}} = 2 I_p(0)$ (Eq. \ref{eq:Im_no_disk}). In panel (a), the vertical black line (when present) marks the warp location $a\warp$ (Eq. \ref{eq:awarp_explicit}). In panel (b), the vertical green line marks the secular inclination resonance location $a\res$ (Eq. \ref{eq:res_loc}). Note the different y-axis scales for $I_d$ and $\Delta\Omega$ in each of panels (a), (b), and (c). See the text (Section \ref{sec:analytical_analysis}) for  details. } 
%%%%%%%%%%%%
\label{fig:master_orb_el}}
\end{figure*}
%%%%%%%%%%%%%%%%%%%%%%%%%%%%%%%%

Alternatively, for relatively low-mass disks, the disk's gravitational effects are negligible, and planetesimal dynamics at all semimajor axes are dominated by the planetary gravity alone (see also Figure \ref{fig:A_a}). This occurs when $A_{d,p}/A_p \leq 1$ at $a_d \leq a\out$,  or equivalently, when $a\res \geq a\out$, which corresponds to disk masses given by (Equations \ref{eq:Adp_over_Ap} and \ref{eq:res_loc}):
%%%%%%%%%%%%%%%%%%%%%%
\begin{eqnarray}
   \frac{M_d}{m_p}\bigg|_{n/d}  &\leq& \delta^{-7/2}  \frac{M_d}{m_p}\bigg|_{d}  
   \approx  10^{-2} ~  \frac{1.85}{\phi_1^c}  \sqrt{ \frac{1.5}{a\inn/a_p} } , 
   \label{eq:Mdmp_pl_dom}
\end{eqnarray}
%%%%%%%%%%%%%%%%%%%%%%
see also Figure \ref{fig:resonance_map}. In other words, the disk's back-reaction on the planet can be safely ignored only for relatively low disk-to-planet mass ratios. For instance, for a Jupiter-mass planet at $a_p = 20$ au with $a\inn = 30$ au, the condition of Equation (\ref{eq:Mdmp_pl_dom}) translates to as low as $M_d \lesssim 3 M_{\earth}$. For all other disk masses, $M_d/m_p|_{n/d}  \lesssim M_d/m_p \lesssim M_d/m_p|_{d} $, the system will feature a secular resonance within the disk (see Section \ref{subsubsec:sr} and Figure \ref{fig:resonance_map}).

%%%%%%%%%%%%%%%%%%%%%%%%%%%%%%%%%%%%%%%%%%%%%%%%%
\subsection{Orbital evolution in massive and massless disks}
\label{subsec:orbit-evol-massive-less-disks}
%%%%%%%%%%%%%%%%%%%%%%%%%%%%%%%%%%%%%%%%%%%%%%%%%

We now analyze the orbital evolution of the system, as described by Equations (\ref{eq:Ip_sol})--(\ref{eq:I_forced_gen}). For illustrative purposes, Figure \ref{fig:master_orb_el} presents the radial profiles of instantaneous inclinations $I_d(t)$ and longitudes of ascending nodes $\Delta\Omega(t)$ (measured relative to the planet) of planetesimals at various times, as indicated at the top of each column. These profiles are computed using Equations (\ref{eq:Ip_sol})--(\ref{eq:I_forced_gen}), which do not require any direct integration. Results are shown for three systems,\footnote{When a secular resonance occurs,  we limit the inclinations to $I_d(t) = 90^{\circ}$ since the action variable is $J \propto (1 - \cos I_d)$ \citep{Goldstein1950}. This bounds the system's angular momentum and prevents unphysical divergence of $I_d$.} with parameters summarized in Table \ref{table:models}, chosen to cover all dynamical regimes identified in Section \ref{sec:dyn_reg}. The primary difference among these systems is the value of $M_d/m_p$: namely, $M_d = 0$ in \texttt{Model A} (first two rows; entirely planet-dominated), $M_d/m_p = 0.1$ in \texttt{Model B} (second two rows; featuring a secular resonance), and $M_d/m_p = 10$ in \texttt{Model C} (last two rows; entirely disk-dominated). These same model systems are considered in detail later in Section \ref{sec:vert_struc_num}, where we analyze their morphological evolution.

%%%%%%%%%%%%%%%%%%%%%%%%%%%%
\subsubsection{Secular evolution of the planetary orbit}
\label{subsec:planet_evol_analysis}
%%%%%%%%%%%%%%%%%%%%%%%%%%%%

According to Equations (\ref{eq:Ip_sol}) and (\ref{eq:Omegap_sol}), the evolution of the planetary orbit due to the axisymmetric potential of a massive debris disk is trivial: it maintains a constant inclination, $I_p(t) = I_p(0)$, while its ascending node regresses linearly in time, $\dot{\Omega}_p \equiv d\Omega_p/dt = A_{d,p} < 0$ (Equation \ref{eq:Adp_Bdp_approx}). This is characterized with a period of:
%%%%%%%%%%%%%%%%%%%%
\begin{equation}
    T_{\dot{\Omega}_p} = \frac{2\pi} {|A_{d,p}| }
    \approx
    {15.3}~ \textrm{Myr} ~ \frac{1.85}{\phi_1^c} ~ \bigg( \frac{30  M_{\earth}} {M_d}\bigg) ~ 
    \frac{a_{\textrm{out}, 150}^{3}}{a_{p,20}^{3/2} }
    ~  M_{c,1}^{1/2}
    .
    \label{eq:planetary_node_rate}
\end{equation}
%%%%%%%%%%%%%%%%%%%% 
Note that for a massless disk, $A_{d,p} =0$ and the planetary orbit does not evolve, that is, $I_p(t) = I_p(0)$ and $\Omega_p(t) = \Omega_p(0)$. Before moving on, we emphasize that when the full back-reaction of the disk is taken into account, the planetary inclination may evolve over time (e.g., decay exponentially) and the planet's nodal precession period may deviate by a factor of a few from that given by Equation (\ref{eq:planetary_node_rate}); see Appendix \ref{app:res_friction} and Section \ref{subsec:limit_future_work} for further discussion.

%%%%%%%%%%%%%%%%%%%%%%%%%%%%
\subsubsection{Secular evolution of the planetesimal orbits}
%%%%%%%%%%%%%%%%%%%%%%%%%%%%

In conjunction with planetary precession, planetesimal orbits undergo both nodal precession and inclination oscillations. The inclinations oscillate between the initial value of $0$ and a maximum of $I_m(a_d) = 2 | I_{d, \rm{forced}}|  = 2 \langle I_d(t) \rangle_t$; see  Equations (\ref{eq:Id_sol})--(\ref{eq:I_forced_gen}) and Figure \ref{fig:master_orb_el}. The oscillation period is
%%%%%%%%%%%%%%%%%%%%
\begin{equation}
    T_{\rm sec}(a_d) =   \frac{2 \pi}{ |A(a_d)| } \equiv 
    \frac{2 \pi} { |A_p(a_d) - A_{d,p}| } , 
    \label{eq:Tsec_planetesimals}
\end{equation}
%%%%%%%%%%%%%%%%%%%%
as would be expected in a reference frame co-precessing with the planetary node. However, depending on whether planetesimals are in the disk-dominated or planet-dominated regime (Section \ref{sec:dyn_reg}), their orbits evolve in distinct ways.

\noindent $\bullet$ \textbf{Planet-dominated regime} ($|A_{d,p}/A_p| \lesssim 1$): In this regime, the maxima of planetesimal inclinations are 
%%%%%%%%%%%%%%%%%%%%
 \begin{equation}
 I_m \rightarrow 
 I_m^{{\rm n/disk}}
 %   2 I_{d, \rm forced}^{n/disk} 
  %  =  2 I_{d, \rm free}^{n/disk} 
  = 2 I_p(0) , 
    \label{eq:Im_no_disk}
\end{equation}
%%%%%%%%%%%%%%%%%%%%
irrespective of their semimajor axes; see Equation (\ref{eq:I_forced_gen}) and Figure \ref{fig:master_orb_el}(a). 
Thus, when time-averaged, each planetesimal lies within the plane of the planetary orbit, $\langle I_d(t) \rangle_t^{{\rm n/disk}} = I_p(0)$, and the disk's angular momentum vector is aligned with that of the planet. This corresponds to the well-studied limiting case of  a \textit{massless} disk \citep[e.g.,][]{Dawson2011}. In the course of oscillations, planetesimal inclinations are maximized when their orbits are nodally aligned with the planet, $\Delta \Omega(t) = 0$, and, since the forced inclination is positive, $\Delta \Omega(t)$ remains within the range $[-\pi/2, \pi/2]$ -- see also Figure \ref{fig:master_orb_el}(a). Finally, the oscillations occur with a period of:
%%%%%%%%%%%%%%%%%%%%
\begin{equation}
 T_{{\rm sec}} \rightarrow
    T_{{\rm sec}}^{{\rm n/disk}}(a_d) \equiv \frac{2 \pi}{|A_p|} \approx 
    {16.1} ~ {\rm {Myr}}  \bigg(\frac{1 M_J}{m_p} \bigg)  ~ \frac{a_{d,80}^{7/2}}{a_{p,20}^{2}} ~ M_{c,1}^{1/2} ,    
    \label{eq:Tsec_no_disk}
\end{equation}
%%%%%%%%%%%%%%%%%%%%
which is an \textit{increasing} function of planetesimal semimajor axis $a_d$. The behavior described here can also be seen in the inner part of the disk shown in Figure \ref{fig:master_orb_el}(b).

\noindent $\bullet$ \textbf{Disk-dominated regime} ($|A_{d,p}/A_p| \gtrsim 1$): In this regime, the maximum inclinations of the planetesimals are
%%%%%%%%%%%%%%%%%%%%%
\begin{equation}
I_m \rightarrow  I_m^{{\rm disk}}(a_d)  = 2 I_p(0) \bigg|  \frac{ A_p(a_d)}{A_{d,p}}  \bigg| ,
\label{eq:Im_disk}  
\end{equation}
%%%%%%%%%%%%%%%%%%%%%
which, for the fiducial disk parameters gives
%%%%%%%%%%%%%%%%%%%%%
\begin{equation}
    I_m^{{\rm disk}}(a_d) \approx 
    %0.09
    9 \times 10^{-2} 
    ~  I_m^{{\rm n/disk}} \bigg( \frac{1.85}{\phi_1^c  }\bigg) \bigg( \frac{m_p}{M_d} \bigg)  
     \frac{a_{p,20}^{1/2}}{a_{d,80}^{7/2}} , 
  \label{eq:Im_disk_sup_numest}
\end{equation}
%%%%%%%%%%%%%%%%%%%%%
see Equations (\ref{eq:I_forced_gen}) and (\ref{eq:Adp_over_Ap}). Comparing $I_m^{{\rm disk}}$ with $I_m^{{\rm n/disk}}$, two key differences emerge: first, planetesimal inclinations in the disk-dominated regime decrease with distance ($\propto a_d^{-7/2}$); second, their amplitudes can be much smaller than those in the planet-dominated regime, with $ I_m^{{\rm disk}} \lesssim  I_m^{{\rm n/disk}}$ -- see also Figures  \ref{fig:I_forced_a} and \ref{fig:master_orb_el}(b), (c). Relatedly, when time-averaged, the angular momentum vectors of planetesimals will be misaligned with that of the planet, with the degree of misalignment increasing as the planet-planetesimal separation grows. Finally, since  the forced inclination is negative in this regime,  $\Delta \Omega(t)$ oscillations are confined to the range $\pm[\pi/2, \pi]$, and inclinations are maximized when planetesimals are nodally anti-aligned with the planet, $\Delta \Omega = \pi $; see also Figures \ref{fig:master_orb_el}(b), (c). Such oscillations occur with a period of (Equation (\ref{eq:Tsec_planetesimals})):
%%%%%%%%%%%%%%%%%%%%
\begin{equation}
 T_{{\rm sec}} \rightarrow    T_{{\rm sec}}^{{\rm disk}}
   \equiv
    \frac{2 \pi}{|A_{d,p}|} = T_{\dot{\Omega}_p} ,  
    \label{eq:Tsec_with_disk}
\end{equation}
%%%%%%%%%%%%%%%%%%%%
where we have assumed that $|A_{d,p}/A_p| \gg 1$. Note that, unlike $T_{{\rm sec}}^{{\rm n/disk}}$ (Equation \ref{eq:Tsec_no_disk}),  $T_{{\rm sec}}^{{\rm disk}}$ is independent of planetesimal semimajor axis (see also Equation  (\ref{eq:planetary_node_rate})). We clarify, however, that this does not necessarily imply that the planetesimal orbits remain strictly nodally aligned, as $A_p$ is small (compared to $A_{d,p}$) but nonzero in the disc-dominated regime. This subtle contribution accounts for the differential evolution of $\Delta \Omega(a_d)$ observed in the relevant regions of Figures \ref{fig:master_orb_el}(b) and (c).

\noindent $\bullet$ \textbf{Secular resonance} ($A_{d,p}=A_p$): At the resonance location $a\res$ (if present), the relative nodal precession rate goes to zero, causing the free and forced inclinations to diverge (Equation \ref{eq:I_forced_gen}). When this happens, planetesimal inclinations at (and around) $a\res$ grow linearly with time, such that (Equation \ref{eq:Id_sol}): 
%%%%%%%%%%%%%%%%%%%%
\begin{equation}
    I_d(t)\big|_{a\res} \approx I_p(0) |A_p(a\res)| ~ t   =  
     I_p(0) |A_{d,p}| ~ t , 
    \label{eq:I_res_time}
\end{equation} 
%%%%%%%%%%%%%%%%%%%%
while their orbits remain locked to the planetary node such that $\Delta\Omega(t) = \pi/2$; see also Figure \ref{fig:master_orb_el}(b). Thus, the timescale $\tau\res$ for the inclinations to grow from their initial value of $0$ to $\pi/2$, i.e., $I_d(t)/90^{\circ} = t / \tau\res$, can be approximated as: 
%%%%%%%%%%%%%%%%%%
\begin{eqnarray}
\tau\res &=& 
\frac{1}{4} \frac{  T_{\dot{\Omega}_p}}{ I_p(0)} = 
\frac{1}{4} \frac{  T_{{\rm sec}}^{{\rm n/disk}}(a\res) }{ I_p(0)} ,
\label{eq:tau_res_growth}
\\
&\approx&   23 ~ {\rm {Myr}} \frac{10^{\circ}}{I_p(0)} \bigg(\frac{1 M_J}{m_p} \bigg) \frac{ a_{{\rm res},80}^{7/2}}{a_{p,20}^{2}} ~ M_{c,1}^{1/2} , 
\nonumber
\end{eqnarray}
%%%%%%%%%%%%%%%%%%
where $a_{{\rm res}, 80} \equiv a\res/(80 ~ {\rm au})$; see also Figure \ref{fig:master_orb_el}(b). Finally, we note that the divergent behavior of planetesimal inclinations around $a\res$ can be approximated using the fact that $A_p(a\res) - A_p(a_d) \approx (a\res - a_d) dA_p/da_d|_{a\res}$. Inserting this relationship into Equation (\ref{eq:I_forced_gen}), it is easy to show that:
%%%%%%%%%%%%%%%%%%
\begin{equation}
I_m(a_d)  
%I_{d, {\rm forced}}(a_d) 
\approx  2 \lambda \frac{a\res}{|a\res - a_d|} 
~~~~ \mathrm{as} ~~ a_d \rightarrow a\res , 
\label{eq:lambda_If_gen}
\end{equation}
%%%%%%%%%%%%%%%%%%
see also Figures \ref{fig:I_forced_a} and \ref{fig:master_orb_el}(b), where $\lambda>0$ is a scaling factor given by
%%%%%%%%%%%%%%%%%%
\begin{equation}
\lambda = I_p(0)
\bigg|\frac{A_p(a_d)}{dA_p/d\log a_d}\bigg|_{a\res} \approx \frac{2}{7} I_p(0) ,
\label{eq:lambda_If}
\end{equation}
%%%%%%%%%%%%%%%%%%
where we have assumed $a_p \ll a_d$ so that $dA_p/d \log a_d \approx - (7/2) A_p$ (Equation \ref{eq:A_planet}). This concludes our analytical investigation of the secular orbital evolution of planetesimals.

%%%%%%%%%%%%%%%%%%%%%%%%%%%%%
\section{Evolution of the disk Morphology }
\label{sec:vert_struc_num}
%%%%%%%%%%%%%%%%%%%%%%%%%%%%%

%%%%%%%%%%%%%%%%%%%%%%%%%%%%%%
\begin{figure*}
%%%%%%%%
\centering
\includegraphics[width=\textwidth]{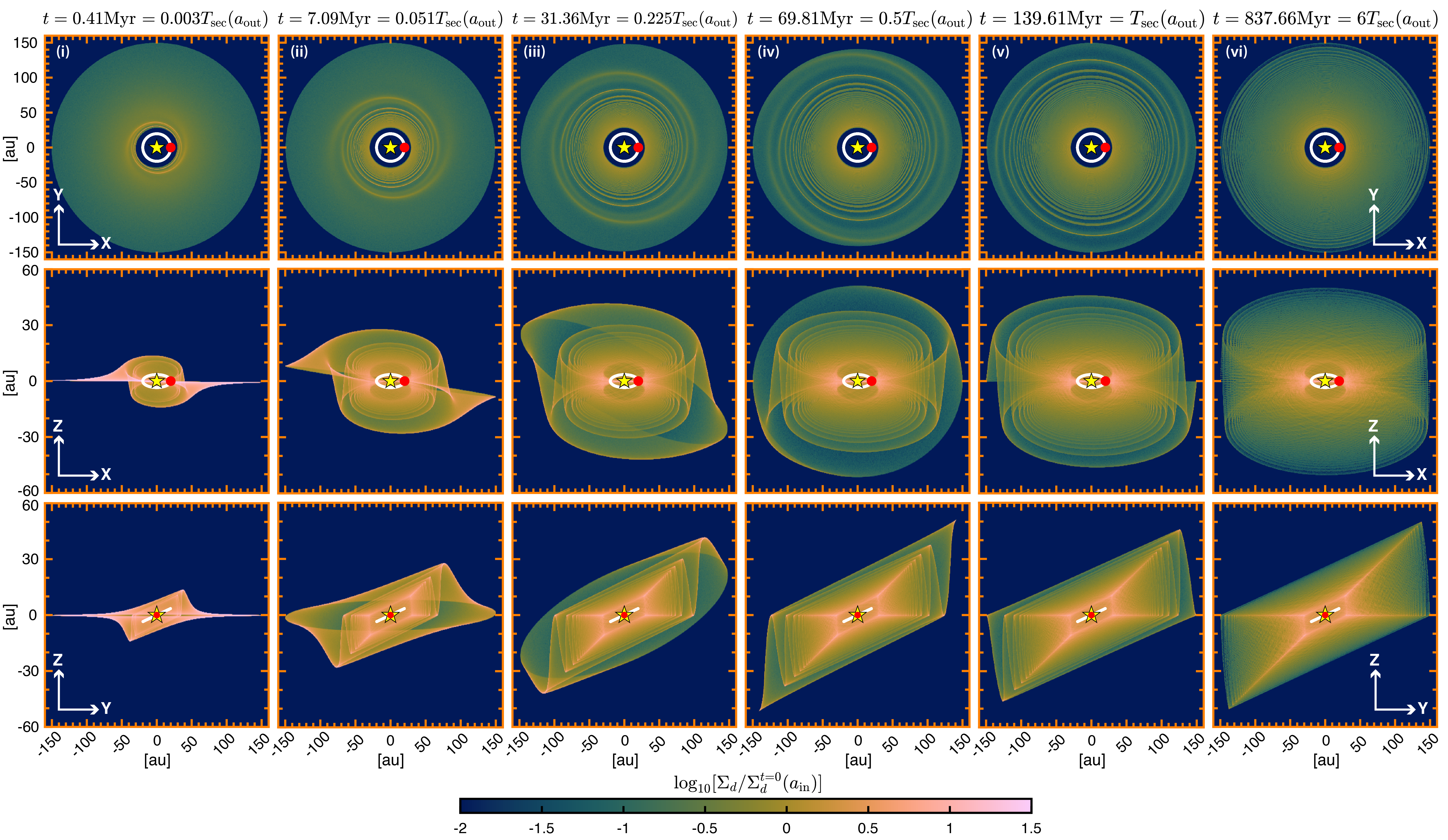}
%%%%%%%%
\vspace{-1.5em}
{\linespread{0.95}\selectfont \caption{Snapshots showing the time evolution of the surface density $\Sigma_d$ of the debris disk in \texttt{Model A}  (i.e., $M_d=0$; Table \ref{table:models}),  viewed from three orientations: face-on (top row), end-on (center row), and edge-on (bottom row). The snapshots correspond to the same times $t$ as those in Figure \ref{fig:master_orb_el}(a), as indicated above each column (left to right). In each panel, the stellar position is marked by the yellow star, while the planet's orbit and its (ascending) nodal position are shown by the white line and red circle, respectively. To highlight fine details,  surface densities are plotted on a logarithmic scale (as indicated by the color bar), with all panels normalized by $\Sigma_d^{t=0}(a\inn)$. The density maps derive from $N =  5,000$ parent planetesimals, each  with $N_{np} = 10,000$ particles per orbit, populating the $3$-D volume divided into equally sized cubic cells, each with a volume of $(0.3 ~{\mathrm{au}})^3$. Calculations are done using the analytically derived distribution of orbital elements (rather than direct $N$-body simulations; see Appendix \ref{app:map_construction} for technical details). It is evident that at early times (panels (i)), the planet launches a leading bending wave at the disk's inner edge, $a\inn$, which appears as a warp when viewed edge-on and as a double-armed spiral when viewed face-on. The location of the warp, where the disk reaches its maximum vertical height from the $(X,Y)$ plane, moves outward over time (panels (i)--(iv)). The warp feature fades away upon reaching the disk's outer edge $a\out$ (panel (iv)); which occurs after half a secular precession period at $a\out$, i.e., $t = 0.5 T_{\rm sec}(a\out)$. Beyond this time (panels (v), (vi)), the entire disk becomes fully inclined, forming a thick, box-like structure characterized by an ``X''-shaped density pattern that is symmetrical about the planetary orbital plane. In other words, the disk and the planet are aligned. At late times, the disk's aspect ratio is distance-independent, given by $\mathcal{H} \approx I_p(0)$. See the text (Section \ref{subsec:massless_case_Md0}) for further details.}
\label{fig:master_Md0_maps}}
\end{figure*}
%%%%%%%%%%%%%%%%%%%%%%%%%%%%%%%%%%

Having analyzed the evolution of the planetesimal and planetary orbits in each dynamical regime (Section \ref{sec:analytical_analysis}), we now examine how the combined planet-disk gravity influences the disk's morphology and its evolution over time.

To facilitate the presentation, we  consider the same three model systems analyzed in Section \ref{sec:analytical_analysis}, the parameters and relevant secular timescales of which are summarized in Table \ref{table:models}. We begin by considering \texttt{Model A} (Section \ref{subsec:massless_case_Md0}), in which the disk is massless ($M_d =0$).
%%%
This serves as a reference case for comparison with the results for massive disks, starting with $M_d/m_p \lesssim 1$ (\texttt{Model B}; Section \ref{subsec:small_Mdmp_sim})  and then $M_d/m_p \gtrsim 1$ (\texttt{Model C}; Section \ref{subsec:large_Mdmp_sim}). To this end, we convert the distributions of orbital elements shown in Figure \ref{fig:master_orb_el} -- which, we recall, were obtained using  the simplified analytical solutions given by Equations (\ref{eq:Ip_sol})--(\ref{eq:I_forced_gen}) -- into distributions of disk surface density $\Sigma_d(t)$. To provide a comprehensive view, density  maps are then projected onto three planes: namely, 
(i) $X-Y$ (face-on view, with the observer looking along the $Z$-axis), 
(ii) $X-Z$ (end-on view,  with the observer looking along the $Y$-axis), 
and
(iii) $Y-Z$ (edge-on view,  with the observer looking along the $X$-axis). 
%face-on ($X,Y$), edge-on ($Y,Z$), and end-on ($X,Z$). 
Technical details of this procedure are provided in Appendix \ref{app:map_construction}. 
For clarity, we note that the disk initially lies in the $X-Y$ plane ($Z=0$), with the positive $X$-axis pointing towards the planet's initial ascending node.
We close this section by examining the effects of varying system parameters relative to their fiducial values (Section \ref{subsec:variations}).

From here on, unless otherwise specified, we define the disk's scale height, $\mathfrak{h}_d$, as the square root of the mass-weighted second moment of the planetesimal's vertical distribution about their mean height (see Appendix \ref{app:scale_height} for details).
The corresponding local aspect ratio is then given by $\mathcal{H}(R) = \mathfrak{h}_d/R$, where $R = \sqrt{X^2 + Y^2}$ is the cylindrical radius.  This definition is adopted because the shape of the vertical density is not known a priori.

%%%%%%%%%%%%%%%%%%%%%%%%%%%%%
\subsection{Massless Disks \texorpdfstring{($M_d/m_p =0$)}{Md/mp=0}}
\label{subsec:massless_case_Md0}
%%%%%%%%%%%%%%%%%%%%%%%%%%%%%

We begin by examining the case of a planet interacting with a massless debris disk, $M_d = 0$ (\texttt{Model A}; Table \ref{table:models}). By definition, planetesimal dynamics in this case will be planet-dominated (Section \ref{sec:dyn_reg}). The main results are summarized in Figure \ref{fig:master_Md0_maps}, which shows maps of the  disk surface density, $\Sigma_d(t)$, at the same snapshot times as those in Figure \ref{fig:master_orb_el}(a).\footnote{Although disk gravity is ignored in \texttt{Model A}, we assign $M_d \neq 0$ (following $p = 3/2$; Eq. \ref{eq:Sigma_d}) to enable the plotting of the disk's surface density.} The snapshot times are also given in terms of the secular oscillation timescale at the disk's outer edge, $T_{\rm sec}^{\rm{n/disk}}(a\out)$; see Eq.  (\ref{eq:Tsec_no_disk}). For reference, Figure \ref{fig:master_Md0_maps} also shows the planet's orbit and its nodal position, which, since $M_d =0$, remain fixed over time (Section \ref{subsec:planet_evol_analysis}). This scenario is studied in the literature \citep{Mouillet97, wyattetal99, Dawson2011, pearcewyatt2014, nesvold15, Brady2023, Stasevic2023}; nevertheless, we describe below two key evolutionary stages governing the disk structure.

%%%%%%%%%%%%%%%%%%%%%%%%%%%%%%%%%%%%%%%%%%%%
\noindent
\textbf{Stage 1} $[0 \leq t < 0.5 T_{\rm sec}^{{\rm n/disk}}(a\out)]$:
%%%%%%%%%%%%%%%%%%%%%%%%%%%%%%%%%%%%%%%%%%%%
At early times, the disk rapidly departs from its initial axisymmetric, razor-thin state by developing a spiral bending wave (Fig. \ref{fig:master_Md0_maps}(i)). The bending wave,  launched at the disk's inner edge  ($a\inn = 30$ au), travels outward as a leading pattern, i.e., rotating in the same direction as the planetesimals’ motion. In the top-down images, the bending wave appears as a double-armed spiral, while in the edge-on images, it manifests as vertical oscillations (more on this later). The leading nature of the wave follows from the fact that $d\Omega_d/da_d >0$ at all times (see, e.g., Figure \ref{fig:master_orb_el}(a)).
The wave's propagation rate is determined by the planet's mass and semimajor axis, causing it to propagate more slowly as it extends to larger radii.
For instance, the wave travels from $a\inn$ to $\sim 80$ au in $\approx 7$ Myr, and it takes an additional $\approx 24$ Myr to 
extend to $\sim 120$ au (Figs. \ref{fig:master_Md0_maps}(i)--(iii)).
This is  because $|A_p(a_d)| \propto a_d^{-7/2}$, rendering  $T_{\rm sec}^{{\rm n/disk}}(a_d)$  longer at larger distances; see Eq. (\ref{eq:Tsec_no_disk}).

As the waves travel outward, they wrap around the star, becoming more tightly wound closer to the planet than at larger distances; see, e.g., Figs. \ref{fig:master_Md0_maps}(i)--(iv). The outermost portion of the bending wave is associated with planetesimals at $a_d = a_{\rm warp}$ that are in the midst of their first oscillation cycle about the planetary orbit: that is, attaining their maximum inclinations for the first time with $I_m(a_d) = 2 I_p(0)$ and $\Delta \Omega = 0$; see Figure \ref{fig:master_orb_el}(a). Interior to this region, the waves are tightly wound, making them difficult to discern. This is because planetesimals have completed more than one precession cycle, and their inclinations and longitudes of ascending nodes are relatively well-mixed within the ranges $[0, 2 I_p(0)]$ and $[-\pi/2, \pi/2]$, respectively (Figure \ref{fig:master_orb_el}(a)). Nevertheless, some windings remain identifiable, particularly just inside the outermost portion of the waves. In the top-down images, these appear as a double-armed spiral: a structure that is closely reproduced by the nodal positions of planetesimal orbits (both ascending and descending, i.e., $[X,Y, Z] = \pm a_d [\cos \Omega_d, \sin\Omega_d, 0]$; Equation (\ref{eq:XYZ-planetesimals})), along with overdense regions due to  planetesimals that have not yet phase-mixed. A complimentary view of this can be seen in the vertical oscillations apparent in e.g. the edge-on images.

At $a_d = a_{\rm warp}$, the disk attains its maximum vertical extent $|z_{\rm warp}|$ from the initial midplane, featuring a warped structure, as best observed in the edge-on images (Figs. \ref{fig:master_Md0_maps}(i)--(iv)). The maximum height $|z_{\rm warp}|$ from the $X-Y$ plane is:
%%%%%%%%%%%%%%%%%%%%
\begin{equation}
    |z_{\rm warp}| =  a_{\rm warp} \sin[2 I_p(0)] , 
   \label{eq:zwarp}
\end{equation}
%%%%%%%%%%%%%%%%%%%%
as expected from basic geometrical considerations.\footnote{Eq. (\ref{eq:zwarp}) can be derived by plugging \mbox{$I = 2I_p(0)$}, \mbox{$\Omega = 0$}, and  \mbox{$f = \pm \pi/2$} into Eq. (\ref{eq:XYZ-planetesimals}), yielding {$(x, y, z)_{\rm warp} =  a_{\rm warp} (0, \pm \cos[2I_p(0)], \pm \sin[2I_p(0)])$}. Thus, $z\warp = R\warp  \tan[2 I_p(0)]$ where $R\warp = \sqrt{ x\warp^2+y\warp^2}$ \citep[see also][]{Dawson2011}.} 
This happens at the warp production time, $\tau_{\rm warp}$, corresponding to half a secular precession cycle at $a_{\rm warp}$ (Eq. \ref{eq:Tsec_no_disk}):
%%%%%%%%%%%%%%%%%%%%
\begin{equation}
    \tau_{\rm warp} =  \frac{1}{2} T_{\rm sec}^{{\rm n/disk}}(a_{\rm warp}) =   \frac{\pi}{|A_p(a_{\rm warp})|} ,
    \label{eq:tau_warp_ppdom}
\end{equation}
%%%%%%%%%%%%%%%%%%%%
so that $I_d(a_{\rm warp}) = I_m $ for the first time. Thus,
%%%%%%%%%%%%%%%%%%%%
\begin{equation}
    {a_{\rm warp}} \approx 80 ~ {\rm au} \bigg(\frac{\tau_{\rm warp}}{8 {\rm Myr}} \cdot \frac{m_p}{1 M_J}  \bigg)^{2/7} {a_{p, 20}^{4/7}} M_{c,1}^{-1/7}  ,
    \label{eq:awarp_explicit}
\end{equation}
%%%%%%%%%%%%%%%%%%%%
where we have assumed $a_p \ll a_{\rm warp}$. Equation (\ref{eq:awarp_explicit}) well reproduces the warp locations seen in Figure \ref{fig:master_Md0_maps}: indeed, the warp is located at $a_{\rm warp} = 40$ au by $\sim 0.4$ Myr, and at $120$ au by $\sim 31$ Myr (see also Figure \ref{fig:master_orb_el}(a)).

Finally, the warp location corresponds to an inflection point in the disk's vertical structure, marking the transition between the interior region ($a\inn \lesssim a_d \lesssim a_{\rm warp}$), where the disk is inclined and aligned with the planetary orbital plane, and the exterior region ($ a_{\rm warp}  \lesssim a_d \lesssim a\out$), where planetesimals remain close to their initial low inclinations and are thus misaligned with the planetary orbit. As best viewed in the edge-on images, the transition occurs smoothly, rather than abruptly, ensuring a continuous change in the disk's vertical profile between the two regions.

%%%%%%%%%%%%%%%%%%%%%%%%%%%%%%
\begin{figure}
%%%%%%%%
\centering
\includegraphics[width=\columnwidth]{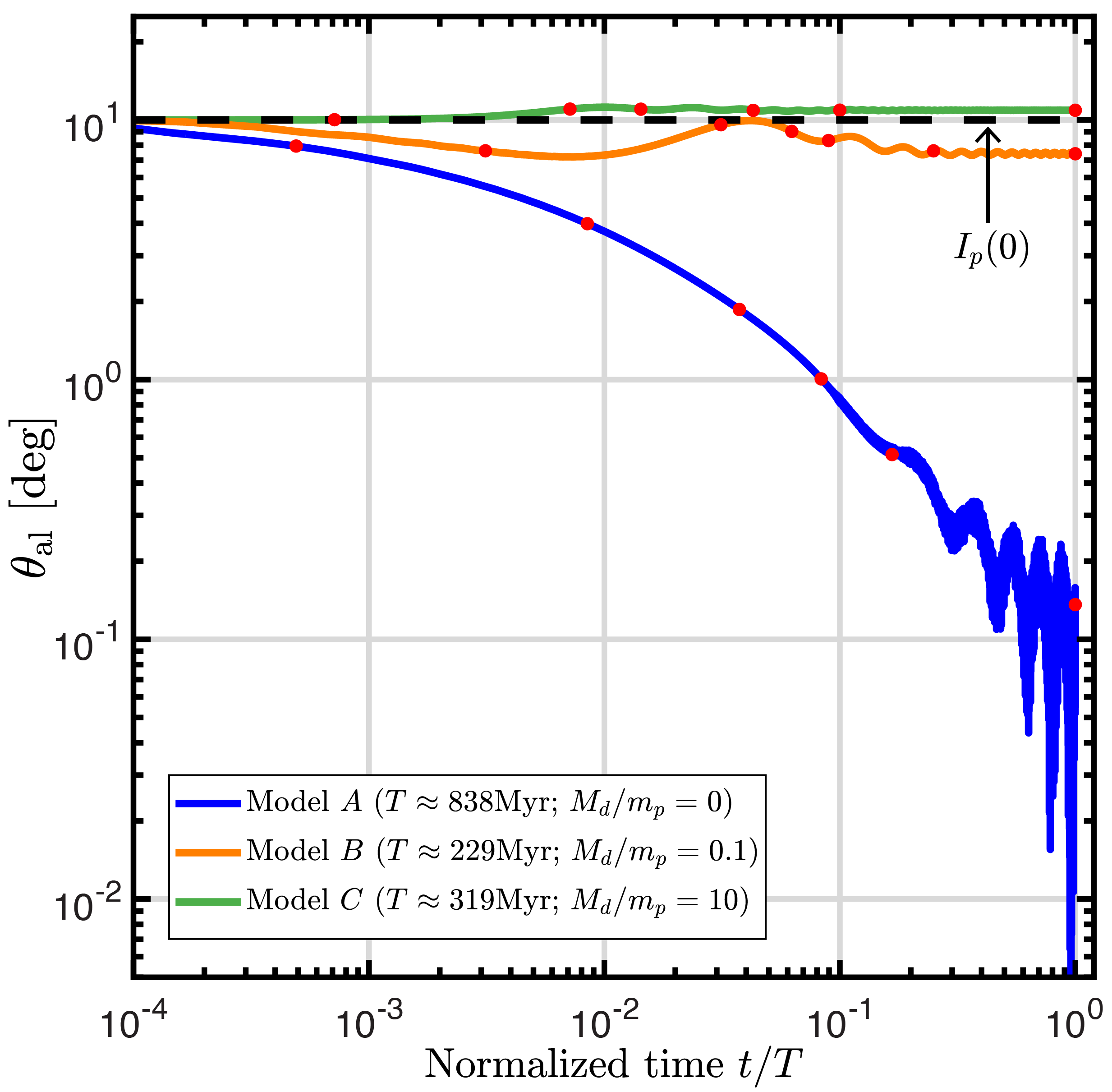}
%%%%%%%%
\vspace{-1.5em}
{\linespread{0.95}\selectfont \caption{The time evolution of the misalignment  between the planet and the disk, $\theta_{{\rm al}}$, computed in degrees using their angular momentum vectors (Eq. \ref{eq:theta_al_app_D}). Results for models \texttt{A} ($M_d = 0$; blue), \texttt{B} ($M_d/m_p =0.1$; orange), and \texttt{C} ($M_d/m_p = 10$; green) are shown, with time $t/T$ normalized by the corresponding last snapshot time  in Figure \ref{fig:master_orb_el}, as indicated in the legend. For reference, the red dots along each curve represent the snapshot times in Figure \ref{fig:master_orb_el}. Note that for massless disks, the planet and the disk align quite rapidly, with $\theta_{{\rm al}}(t)\rightarrow 0$. In contrast, when the disk is massive, the planet cannot torque the disk onto its plane, and the system nearly maintains the initial misalignment, $\theta_{\rm al}(t)\approx I_p(0) $. See the text (Sections \ref{subsec:massless_case_Md0}-- \ref{subsec:large_Mdmp_sim}) for details.}
%%%%%%%%%%%%%%%%
\label{fig:misalignment_ABC}}
\end{figure}
%%%%%%%%%%%%%%%%%%%%%%%%%%%%%%%%%%

%%%%%%%%%%%%%%%%%%%%%%%%
\noindent \textbf{Stage 2} $[t \geq 0.5 T_{\rm sec}^{{\rm n/disk}}(a\out)]$: 
%%%%%%%%%%%%%%%%%%%%%%%%
The warp, however, is a transient phenomenon. Indeed, at later times, i.e., when $t = 0.5 T_{\rm sec}^{{\rm n/disk}}(a\out)$, the warp location reaches the disk's outer edge, $a_{\rm warp} = a\out$ (Figure \ref{fig:master_Md0_maps}(iv)). Subsequently, when all planetesimals have completed at least one full precession cycle, i.e., when $t \gtrsim T_{\rm sec}^{{\rm n/disk}}(a\out)$, the warp structure disappears entirely (Figures \ref{fig:master_Md0_maps}(v), (vi)).

%%%%%%%%%%%%%%%%%%%%%%%%%%%%%%
\begin{figure}
%%%%%%%%
\centering
\includegraphics[width=\columnwidth]{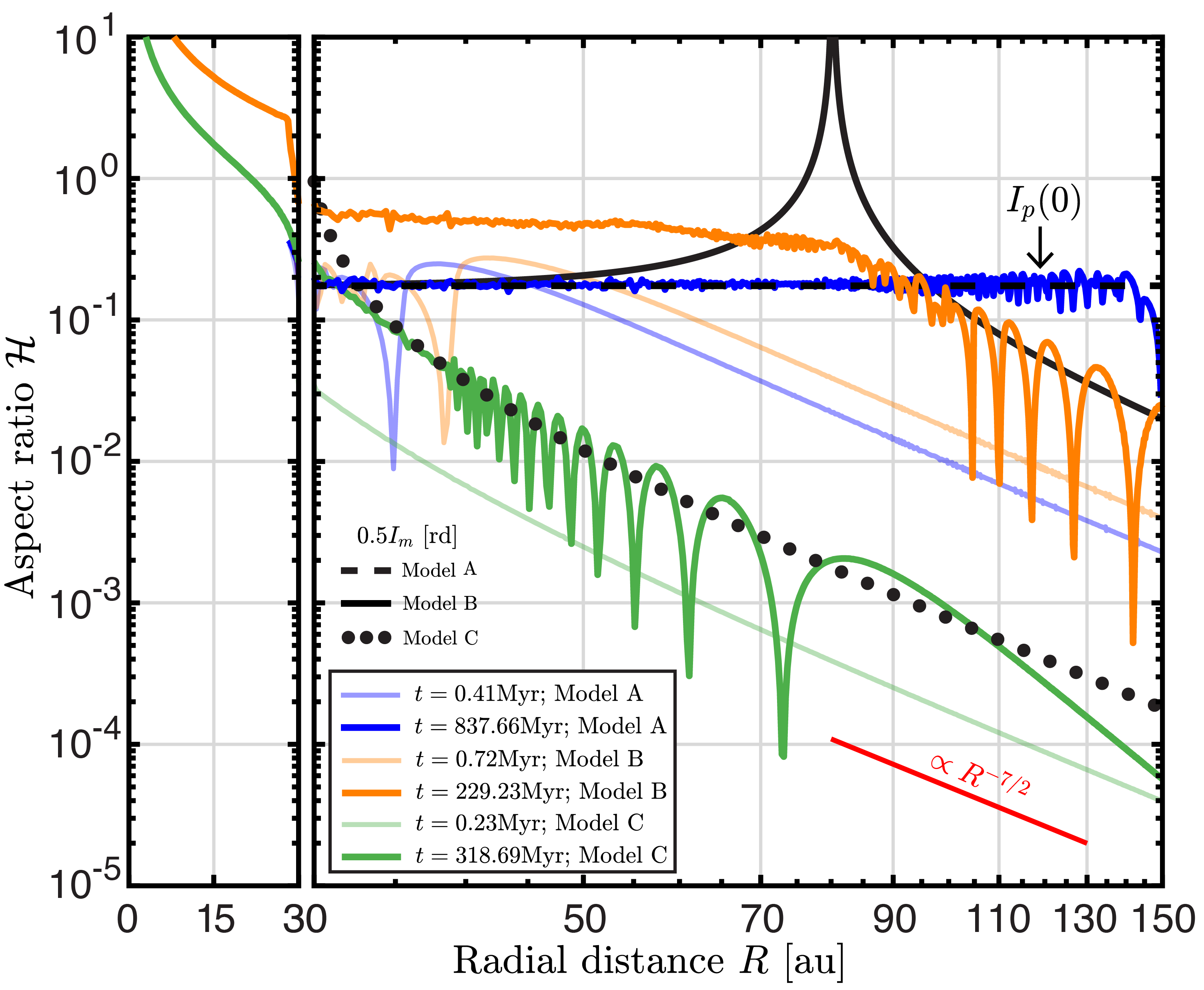}
%%%%%%%%
\vspace{-1.5em}
{\linespread{0.95}\selectfont \caption{Profiles of the disk aspect ratio, $\mathcal{H} = \mathfrak{h}_d/R$, as a function of cylindrical radius $R = \sqrt{X^2+Y^2}$, computed from the mass-weighted second moment of the vertical positions of planetesimals (Eq. \ref{eq:H_appD2}). Results are shown for models \texttt{A} ($M_d = 0$; blue), \texttt{B} ($M_d/m_p = 0.1$; orange), and \texttt{C} ($M_d/m_p = 10$; green), with low and high opacity representing the first and last snapshots in Figure \ref{fig:master_orb_el}, respectively. It is evident that when $M_d = 0$,  $\mathcal{H} \rightarrow I_p(0)$ at late times, irrespective of $R$. In massive disks,  $\mathcal{H}\propto I_p(0)$ interior to the resonance location (if any), and $\mathcal{H}/ I_p(0) \propto R^{-7/2}$ outside. Results at $t\rightarrow \infty$ are well approximated by  $\mathcal{H} \approx I_m(a_d)/2$ (Equations \ref{eq:I_forced_gen} and \ref{eq:aspect_ratio_analytical_H}), shown using different black curves (see the legend). Note that results for Models \texttt{B} and \texttt{C} interior to $R \lesssim a\inn = 30$ au are unrepresentative due to a tail of high-$I_d$ orbits approaching $R\rightarrow0$ (involving  $\lesssim 3 \%$ of planetesimals with $Z>R$; Appendix \ref{app:scale_height}). See the text (Sections \ref{sec:vert_struc_num} and \ref{subsec:aspect_discussion_Sec}) for details.}
%%%%%%%%%%%%%%%%%%%%
\label{fig:scaleheight_ABC}}
\end{figure}
%%%%%%%%%%%%%%%%%%%%%%%%%%%%%%%%%%

When this occurs, the disk settles into a quasi-equilibrium state, characterized by a fully inclined, thick, box-like structure that is symmetric about the planetary orbit. This is because, at such late times, planetesimal orbits become relatively well-mixed  such that $0 \leq I_d(t) \leq 2 I_p(0)$ and $ |\Delta \Omega| \leq \pi/2$ at all $a_d$ (Figure \ref{fig:master_orb_el}(a)). This implies that the total angular momentum vector of the disk is aligned with that of the planet, both tilted by $I_p(0)$ relative to the initial disk midplane. This is illustrated by the blue curve in Figure \ref{fig:misalignment_ABC}, which shows the time evolution of $\theta_{{\rm al}}$, the angle between the planet's and the disk's angular momentum vectors (Appendix \ref{app:ang_mom}). It is evident that $\theta_{{\rm al}}$ decreases from an initial value of $I_p(0)$ to $0$ over time. Relatedly, the disk acquires a vertical scale height, $\mathfrak{h}_d$, or equivalently an aspect ratio, $\mathcal{H} \equiv \mathfrak{h}_d/R$, set by the planet. Namely, the scale height increases linearly with radius, $\mathfrak{h}_d \propto R$, so that the aspect ratio is a distance-independent constant given by $\mathcal{H}   \approx I_p(0)$. Figure \ref{fig:scaleheight_ABC} illustrates the radial dependence of the aspect ratio as a function of time, as measured from the density maps (see Appendix \ref{app:scale_height} for technical details).

As this process unfolds, the disk as a whole  exhibits an ``X''-shaped density pattern in the $(Y,Z)$ plane, with enhanced densities symmetrically distributed around the planet's orbit. Namely, the densities are maximized along the lines $Z = 0$ and $Z / Y = \tan[2 I_p(0)]$, which appears as a ``fork''-like structure. The origin of this pattern can be understood as follows. The evolution of planetesimal inclinations is akin to that of a simple harmonic oscillator, causing them to linger longer near the peaks -- that is, around $I_d(t) \approx  2 I_p(0)$ with $\Omega_d(t) \approx 0$, where $dI_d(a_d)/dt \rightarrow 0$ (Eq. \ref{eq:Id_sol}). Consequently, when such planetesimals at different semimajor axes are viewed edge-on, their overlapping positions -- an effect absent in the end-on view -- create an overdense line. By contrast, the overdense $Z=0$ line in $(Y,Z)$, and absence thereof in $(X,Z)$, can be explained by the low-$I_d$ planetesimals with $\Omega_d \approx \pm \pi/2$ crossing through $Z=0$ along $Y \approx \pm a_d$ and $X \approx 0$. Note that the out-of-plane arm of the ``X'' corresponds to the warp trajectory over time.

%%%%%%%%%%%%%%%%%%%%%%%%%%%%%%
\begin{figure*}
\centering
%%%%%%%%
\includegraphics[width=\textwidth]{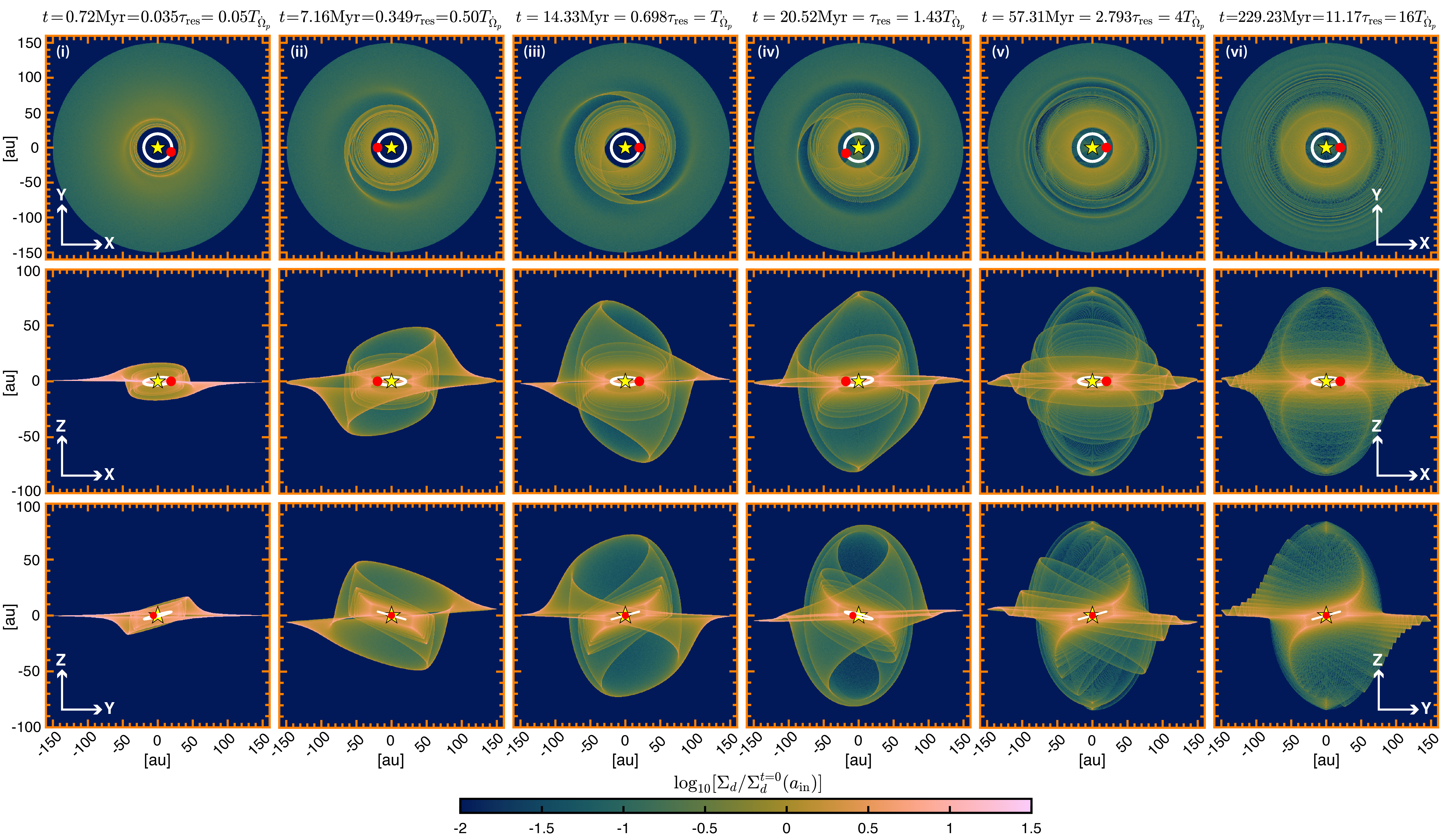}
%%%%%%%%
\vspace{-1.5em}
{\linespread{0.95}\selectfont \caption{Similar to Figure \ref{fig:master_Md0_maps}, but for \texttt{Model B} (Table \ref{table:models}). \texttt{Model B} is identical to \texttt{Model A}, except that its debris disk has a non-zero mass of $M_d/m_p =0.1$. Density maps are generated using the same method (Appendix \ref{app:map_construction}) and  parameters as in Figure \ref{fig:master_Md0_maps}, including grid cell volume, $N$, $N_{np}$, and the normalization constant. The snapshots (left to right) correspond to the same times $t$ as in Figure \ref{fig:master_orb_el}(b). All other notations and symbols are the same as in  Figures  \ref{fig:master_orb_el}(b) and \ref{fig:master_Md0_maps}. It is evident that the disk's back-reaction on the planet significantly affects its resultant morphology. At early times (panels (i),(ii)), a bending wave  launched at the disk's inner edge propagates outward; this appears as a warp (two-armed spiral) when viewed edge-on (top-down). The region interior to the warp is tilted and corotates with the planetary node, while remaining aligned and symmetric around the planet’s orbit. By the time the planet completes one precession cycle (panel (iii)), the warp extends out to $a\res \approx 80.2$ au, where it stalls at later times. The disk's inner part develops a curly ``X''-pattern, whose out-of-plane arm does not connect smoothly with the outer parts, thus the appearance of a ``broken'' disk. Beyond this time (panels (iv)--(vi)), the disk structure remains practically invariant, coprecessing with the planetary node. In terms of angular momentum, the disk maintains the initial misalignment with the planet, while its vertical aspect ratio develops a radial dependence (unlike when $M_d =0$). See the text (Section \ref{subsec:small_Mdmp_sim}) for further details.}
\label{fig:master_ModelB_maps}}
\end{figure*}
%%%%%%%%%%%%%%%%%%%%%%%%%%%%%%%%%%

Finally, we note that if the system were evolved for a longer time, e.g., for tens of $T_{\rm sec}^{{\rm n/disk}}(a\out)$, planetesimal orbits would become even more phase-mixed than in panel (vi) of Figure \ref{fig:master_orb_el}(a). Thus, in this limit, the surface density in the outermost regions would become more uniform, such that the waves which are (hardly) visible in Figure \ref{fig:master_Md0_maps}(vi) -- where $t = 6  T_{\rm sec}^{{\rm n/disk}}(a\out)$ -- disappear entirely.

In closing, we point out that the behavior described thus far for massless disks generally holds true for systems with small, non-zero disk masses such that $ M_d/m_p \lesssim M_d/m_p\big|_{n/d}$; see Equation (\ref{eq:Mdmp_pl_dom}).

%%%%%%%%%%%%%%%%%%%%%%%%%%%%%
\subsection{Disks Less Massive than the Planet \texorpdfstring{($M_d/m_p \lesssim 1$)}{Md/mp<1}}
\label{subsec:small_Mdmp_sim}
%%%%%%%%%%%%%%%%%%%%%%%%%%%%%

We now consider a scenario in which the disk is less massive than the planet, $M_d/m_p \lesssim 1$ (\texttt{Model B}; Table \ref{table:models}). Apart from the small but non-zero disk mass, $M_d/m_p =0.1$, \texttt{Model B} is otherwise identical to \texttt{Model A} (Section \ref{subsec:massless_case_Md0}). The system features a secular  resonance at $a\res \approx 80.2$ au, with planetesimal dynamics dominated by the planetary and disk gravity interior and exterior to $a\res$, respectively (Section \ref{sec:dyn_reg}). Figure \ref{fig:master_ModelB_maps} shows snapshots of the disk's surface density corresponding to the same times shown in Figure \ref{fig:master_orb_el}(b). Below, we outline three evolutionary stages of the disk's morphology, occurring on timescales measured relative to $T_{\dot{\Omega}_p} \approx 14.3$ Myr --  the period over which the planet's ascending node regresses linearly due to the disk (Eq. \ref{eq:planetary_node_rate}).

%%%%%%%%%%%%%%%%%%%%%%%%%%%%%%%%%%%%%%%%%%%%
\noindent
\textbf{Stage 1} $[0 \leq t \lesssim  T_{\dot{\Omega}_p}]$:
%%%%%%%%%%%%%%%%%%%%%%%%%%%%%%%%%%%%%%%%%%%%
At early times, i.e., before the planet completes one full nodal precession cycle, the disk's morphological evolution qualitatively resembles that of a massless disk. Indeed, as seen in Figures \ref{fig:master_ModelB_maps}(i) and (ii), a  spiral bending wave is launched at $a\inn$, which propagates outward as a leading pattern while gradually wrapping around the star.  This manifests as a  double-armed spiral in the top-down images and as a warp in the edge-on views, consistent with ``Stage 1'' evolution for $M_d = 0$ disks (Section \ref{subsec:massless_case_Md0}). However, this description is valid only within a certain radial range and is subject to minor modifications, as outlined below.

First, the bending waves -- and thus the warp --  extend only out to a radius of  $\approx a\res$, rather than reaching the disk's outer edge. This is essentially because the planet's gravity dominates planetesimal dynamics only within this region. Additionally, the warp propagation time is  modulated by the disk-induced planetary precession, causing it to increase relative to the massless disk case as $a\warp \rightarrow a\res$. Namely, the warp production time at $a\warp \lesssim a\res$ is  (Equation \ref{eq:Tsec_planetesimals}):
%%%%%%%%%%%%%%%%%%%%
\begin{equation}
    \tau\warp = \frac{1}{2} T\sec(a\warp) = \frac{\pi}{| A_p(a\warp) - A_{d,p} | }, 
\label{eq:tau_warp_ppdom_modified_w_Md}
\end{equation}
%%%%%%%%%%%%%%%%%%%%
which should be compared with Equation (\ref{eq:tau_warp_ppdom}). Conversely, since $|A_p| \gtrsim |A_{d,p}|$ in this region (Figure \ref{fig:A_a}), the warp forms slightly closer to the star than predicted by Equation (\ref{eq:awarp_explicit}) at a given time; see, e.g., panels (ii) in Figures \ref{fig:master_Md0_maps} and \ref{fig:master_ModelB_maps}. However, the maximum vertical extent  at the warp location remains  well approximated by Equation (\ref{eq:zwarp}), once $I_m^{\rm n/disk} =2I_p(0)$ is replaced with $I_m(a_d)$ (Figure \ref{fig:master_orb_el}(b)). Note that $I_d(a\res) \geq I_m^{{\rm n/disk}} $ (Equation \ref{eq:I_res_time}) at times exceeding
%%%%%%%%%%%%%%%%%%%%
\begin{equation}
    t =  \frac{1}{\pi} T_{\dot{\Omega}_p} 
    %\approx 0.32 T_{\dot{\Omega}_p} 
    \equiv \frac{2}{|A_p(a\res)|} , 
    \label{eq:t_significant_inc_res}
\end{equation}
%%%%%%%%%%%%%%%%%%%% 
which is shorter than $\tau\res$ of Equation (\ref{eq:tau_res_growth}) for $I_p(0)\leq 45^{\circ}$, as is the case in this study (Section \ref{sec:modelsystem}).

Second, the region interior to the warp becomes inclined and aligned with the planet's orbit -- exhibiting an ``X''-shaped structure -- just as in massless disks. However, unlike in that case, the structure corotates with the planetary node while maintaining its symmetry around the planetary orbit. This is because, although  planetesimals in this region  precess faster than the planet, their inclinations are maximized when their orbits are nodally aligned with the planet ($\Delta\Omega  =0$); see Figure \ref{fig:master_orb_el}(b). This is why, for instance, in Figure \ref{fig:master_ModelB_maps}(ii), where the planet has completed half a precession cycle, the inner inclined region  lies along $Z/Y = - \tan[ I_p(0)]$, rather than  $Z/Y =  \tan[I_p(0)]$ as in the $M_d =0$ case.

%%%%%%%%%%%%%%%%%%%%%%%%%%%%%%%%%%%%%%%%%%%%
\noindent
\textbf{Stage 2} $[t \sim T_{\dot{\Omega}_p}]$:
%%%%%%%%%%%%%%%%%%%%%%%%%%%%%%%%%%%%%%%%%%%%
By the time the planet has completed one full precession cycle, the warp  extends to $a\res$, and the disk's morphology exhibits significant deviations from that of a massless disk; see Figure \ref{fig:master_ModelB_maps}(iii). First, the inner tilted region of the disk no longer transitions smoothly into the outer parts but instead does so rather abruptly. This creates the appearance of a ``broken'' disk, in which the region interior to the resonance appears separated from the outer parts. This can be understood by noting that planetesimal inclinations are maximized when nodally aligned with the planet in the inner region, but anti-aligned in the outer region (Figure \ref{fig:master_orb_el}(b)). This disconnect is most evident around the resonance, where inclinations are higher, especially compared to the inclinations toward the disk's outer edge (recall that $I_d(a_d) \propto a_d^{-7/2}$ at $a_d \gtrsim a\res$; Equation \ref{eq:Im_disk_sup_numest}).

Second, the density structure in the inner region deviates from the typical ``X''-shaped morphology, with the out-of-plane arms  appearing curved and bent away from the $Z = 0$ plane when viewed edge-on. These curved overdense regions extend toward larger values of $|Z|$ with increasing distance from the star. This can be understood by noting that $I_m(a_d)$ is distance-dependent (unlike when $M_d = 0$), increasing as $a_d \rightarrow a\res$ (Figure~\ref{fig:master_orb_el}(b)). We note that this structure is reminiscent of, although not necessarily related to, the so-called ``cat's tail'' feature observed in $\beta$ Pic \citep{isa2024beta}.

Finally, when viewed top-down, the disk exhibits a depletion around $a\res$, with a two-armed spiral extending beyond that radius. This naturally results from the fact that $I_d(a_d \approx a\res) \rightarrow 90^{\circ}$, while the planetesimal ascending nodes are not yet phase-mixed at $a_d \gtrsim a\res$ (Figure \ref{fig:master_orb_el}(b)).

%%%%%%%%%%%%%%%%%%%%%%%%%%%%%%%%%%%%%%%%%%%%
\noindent
\textbf{Stage 3} $[t \gtrsim T_{\dot{\Omega}_p}]$: 
%%%%%%%%%%%%%%%%%%%%%%%%%%%%%%%%%%%%%%%%%%%%
Further into the evolution, the disk settles into and maintains a quasi-equilibrium state, remaining largely unaffected by the continued growth of inclinations at the resonance; see panels (iv)--(vi) in Figure \ref{fig:master_ModelB_maps}. Indeed, the disk structure -- including the curly ``X''-pattern -- remains invariant, coprecessing with the planet's ascending node as a rigid body. As this occurs, the broken, warp-like feature stalls near $a_d \approx a\res$, persisting as a long-lived feature.

Additionally, interior to the resonance, at least one secular period has elapsed for the planetesimals, leading to a more uniform surface density -- best seen in the top-down views. In contrast, exterior to the resonance, where secular timescales are relatively longer, spiral patterns emerge just beyond the resonance. These spirals gradually wind up, becoming increasingly difficult to discern at radii near $a\out$, where planetesimal inclinations are smaller (see also Figure \ref{fig:master_orb_el}(b)).

Furthermore, from an angular momentum perspective, the disk remains misaligned with the planet throughout the evolution. This is illustrated in Figure \ref{fig:misalignment_ABC}, which shows that  $\theta_{\rm al}(t)$ stays close to its initial value of $I_p(0)$   within $\approx 2^{\circ}$. While this might seem counterintuitive -- since the disk's inner part is torqued into alignment with the planetary orbit -- it can be reconciled by recognizing that most of the angular momentum is carried by the outer parts of the disk where planetesimal inclinations are suppressed due to the disk's back-reaction (see Section \ref{subsec:planet_evol_analysis} and Figure \ref{fig:master_orb_el}(b)).

Finally, as shown in Figure \ref{fig:scaleheight_ABC}, the disk's vertical aspect ratio $\mathcal{H}(R)$ exhibits two distinct behaviors. Indeed, interior to $R \approx a\res$, the disk becomes vertically thick at late times, with $\mathcal{H}(R)$ approximately constant but enhanced by a factor of $\sim 2 - 3$ relative to the massless disk case (in which case, we recall,   $\mathcal{H}(R) \approx I_p(0)$, more on this below). Beyond the resonance, $R \gtrsim a\res$, the aspect ratio decreases with radius, following $\mathcal{H}(R) \propto R^{-7/2}$, and becomes about an order of magnitude smaller than $I_p(0)$ near the disk's outer edge.

We find that the enhanced aspect ratio in the inner disk parts ($R \lesssim a\res$), compared to massless disks, is due to a dynamically hot population of planetesimals originating near the secular resonance (for details, see Section \ref{sec:cold_hot_rayleigh}). Their vertical trajectories, $Z_{\rm hot}(R)$, can be described using Equation (\ref{eq:XYZ-planetesimals}) evaluated at $a_d \approx a\res$, so that: 
%%%%%%%%%%%%%%%%%%%%%%%
\begin{equation}
    Z_{\rm hot}(R) \approx \pm \sqrt{a\res^2 - R^2} ~~~ \mathrm{for} ~~~ 0 \leq R \leq a\res . 
    \label{eq:Z_hot_res}
\end{equation}
%%%%%%%%%%%%%%%%%%%%%%%
We have verified that artificially removing these planetesimals restores the inner disk aspect ratio to $\approx I_p(0)$, without affecting the profile at $R \gtrsim  a\res$. This also explains why the enhancement is absent at early times, i.e., when $I_d(a\res) \lesssim I_m^{{\rm n/disk}}$; see Figure \ref{fig:scaleheight_ABC}. We note that this hot population coexists with a colder component; since our definition of $\mathcal{H}(R)$ (Appendix \ref{app:scale_height}) does not distinguish between them, it tends to overestimate the local aspect ratio.

In closing, we note that the behavior described herein remains generally applicable to  systems with $M_d/m_p|_{n/d}  \lesssim M_d/m_p \lesssim M_d/m_p|_{d}$; see Equations (\ref{eq:Mdmp_disk_dominated}) and (\ref{eq:Mdmp_pl_dom}).

%%%%%%%%%%%%%%%%%%%%%%%%%%%%%%
\begin{figure*}
\centering
%%%%%%%%
\includegraphics[width=\textwidth]{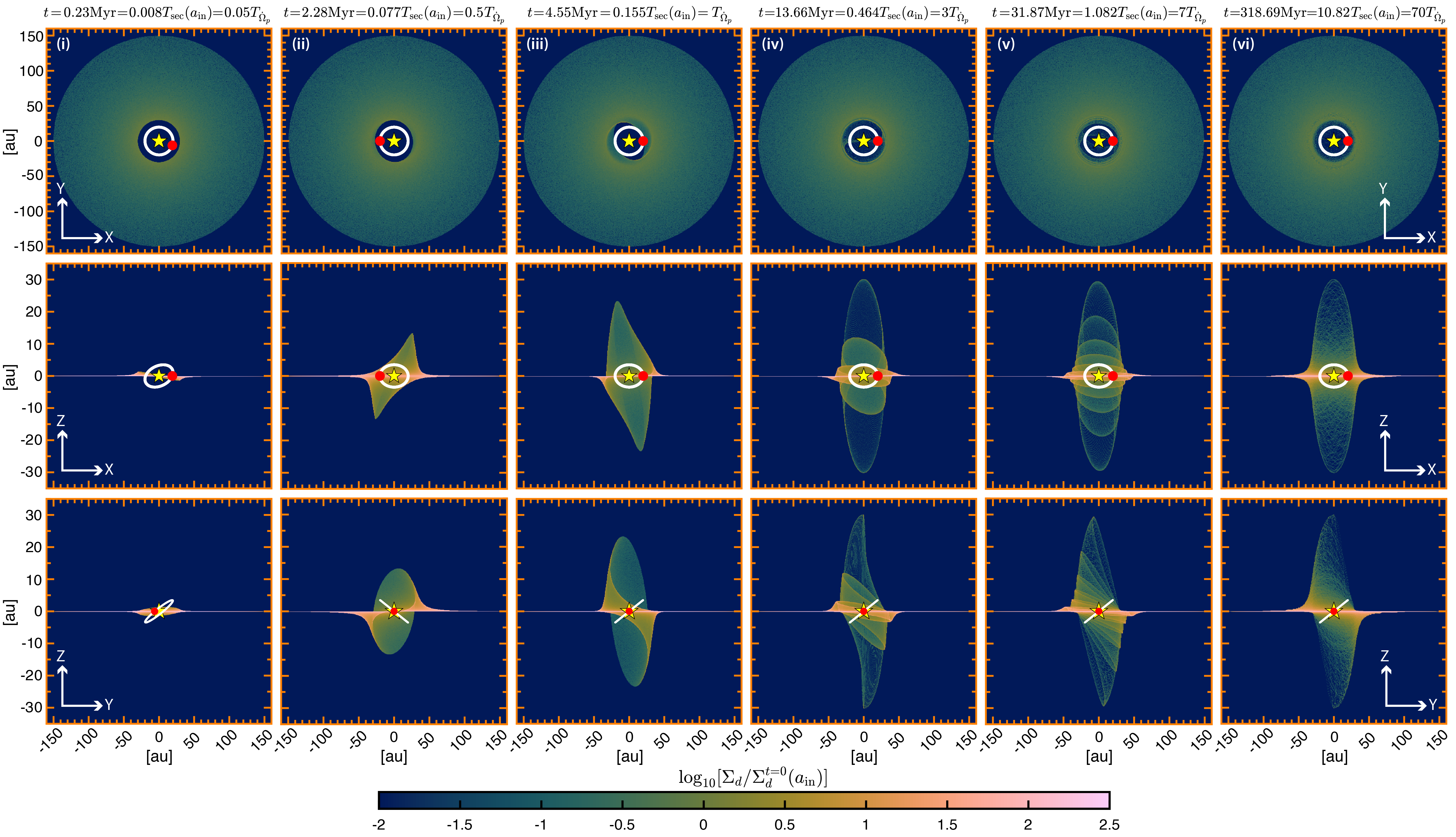}
%%%%%%%%
\vspace{-1.5em}
{\linespread{0.95}\selectfont 
\caption{Similar to Figure \ref{fig:master_Md0_maps}, but for \texttt{Model C} (Table \ref{table:models}). \texttt{Model C} is identical to \texttt{Model A}, except that its planetary mass is reduced to $m_p =10 M_{\earth}$ to allow for a massive debris disk with $M_d/m_p = 10$. Density maps are generated using the same method as in Figure \ref{fig:master_Md0_maps}, but with $(0.1~{\rm au})^3$ cubic cells and a broader logarithmic scale on the color bar to better resolve the density.
All other notations and symbols are the same as in Figures \ref{fig:master_Md0_maps} and \ref{fig:master_orb_el}(c). 
%%%%%%%%%%%%%%%%%%%
One can see that the disk's back-reaction on the planet plays an important role in shaping its structure.
At early times (panels (i)--(iii)), the disk vertically puffs up in the inner regions, developing a near-polar component that precesses over time.
Once the planet completes more than one full precession cycle ($t \gtrsim T_{\dot{\Omega}_p}$; panels (iv)--(vi)), a warp propagates outward, characterized by a maximum vertical extent that decreases with increasing distance.
At late times (panel (vi)), the disk settles into a configuration
dominated by a vertically nearly razor-thin component centered around 
$Z = 0$, along with a tenuous polar component in the inner regions. 
Overall, the disk maintains its initial misalignment with the planet, while exhibiting a decreasing vertical aspect ratio.
See the text (Section \ref{subsec:large_Mdmp_sim}) for details. 
} 
\label{fig:master_ModelC_maps}}
\end{figure*}
%%%%%%%%%%%%%%%%%%%%%%%%%%%%%%%%%%

%%%%%%%%%%%%%%%%%%%%%%%%%%%%%
\subsection{Disks More Massive than the Planet \texorpdfstring{($M_d/m_p \gtrsim 1$)}{Md/mp>1}}
\label{subsec:large_Mdmp_sim}
%%%%%%%%%%%%%%%%%%%%%%%%%%%%%

We now consider a system similar to those in Sections \ref{subsec:massless_case_Md0} and \ref{subsec:small_Mdmp_sim}, but with a debris disk  more massive than the planet; namely, $M_d/m_p = 10$ (\texttt{Model C}; Table \ref{table:models}). This mass ratio  results in planetesimal dynamics at all distances being disk-dominated (Section \ref{sec:dyn_reg}). Note that, compared to models \texttt{A} and \texttt{B}, the planet's mass is reduced to $m_p = 10 M_{\earth}$ to avoid unreasonably large values of $M_d$. However, this adjustment does not affect the system's dynamical regime, which is determined by the mass ratio $M_d/m_p$ (rather than the individual masses); it only affects the evolution timescales (see Section \ref{sec:var_Md_mp_results}). The main results are summarized in Figure \ref{fig:master_ModelC_maps}, which shows maps of the  disk surface density at the same snapshot times as those in Figure \ref{fig:master_orb_el}(c).  Below we describe the evolution of the disk's morphology over two stages, demarcated by the planetary nodal precession period $T_{\dot{\Omega}_p} \approx 4.6$ Myr (Table \ref{table:models}).

%%%%%%%%%%%%%%%%%%%%%%%%%%%%%%%%%%%%%%%%%%%%
\noindent
\textbf{Stage 1} $[0 \leq t \lesssim  T_{\dot{\Omega}_p}]$:
%%%%%%%%%%%%%%%%%%%%%%%%%%%%%%%%%%%%%%%%%%%%
At early times, the disk departs from its initial razor-thin configuration by developing a warp-like feature near its inner edge at $a\inn = 30$ au (Figure~\ref{fig:master_ModelC_maps}(i)). This feature, however, is not directly analogous to the warp that develops in massless disks (Section \ref{subsec:massless_case_Md0}). The key difference is that the location of the warp does not propagate outward over time. Instead, the maximum vertical displacement from the initial midplane gradually increases while remaining at an in-plane radius of $\approx a\inn$, such that $|Z_{\rm max}| \sim \sin[I_d(a\inn)]$ (Figures \ref{fig:master_ModelC_maps}(i)--(iii)). This effect is amplified as $I_d(a\inn) \rightarrow 90^{\circ}$ in this specific setup (Figure \ref{fig:master_orb_el}(c)), although similar behavior would occur even if $I_d(a\inn)$ were smaller.\footnote{For instance, for a system with a larger $M_d/m_p$ and/or a smaller $a_p/a\inn$, such that $A_{d,p}/A_p$ at $a_d \approx a\inn$ is relatively larger than unity, rather than comparable to as in \texttt{Model C}; see the green curve in Figure \ref{fig:A_a}.} Additionally, this structure precesses as a whole while maintaining its symmetry around the planetary orbit.

This  behavior and its differences from $M_d =0$ cases can be understood by noting that when $t \lesssim T_{\dot{\Omega}_p}$, the radial profile of planetesimal inclinations behaves like a global ``breathing'' mode. That is, the entire $I_d(a_d)$ profile oscillates vertically in an almost coherent manner, without developing nodes or small-scale oscillations  along $a_d$; see Figure \ref{fig:master_orb_el}(c). Additionally, since the secular precession timescale at $a_d \gtrsim  a\inn$ is approximately distance-independent, $T\sec(a_d) \rightarrow T_{\dot{\Omega}_p}$ (Equation \ref{eq:Tsec_with_disk}), but $T\sec(a\inn) \gtrsim T_{\dot{\Omega}_p}$ (Table \ref{table:models}), planetesimals across the disk appear nodally aligned with each other. This coherence also explains the absence of spiral features  in the disk when viewed top-down at early times (Figures \ref{fig:master_ModelC_maps}(i)--(ii)).

Finally, at  $a_d\gtrsim a\inn$, the disk remains mostly undisturbed, preserving its flat structure. This is because, although planetesimals may have already attained their maximum inclinations, these decrease steeply with distance and remain small, $I_m(a_d) \propto a_d^{-7/2} \ll 1^{\circ}$, with $\Delta\Omega(a_d) \approx$  $\rm{const}$; see, e.g., panel (ii) in Figures \ref{fig:master_orb_el}(c) and   \ref{fig:master_ModelC_maps}.

%%%%%%%%%%%%%%%%%%%%%%%%%%%%%%%%%%%%%%%%%%%%
\noindent
\textbf{Stage 2} $[t \gtrsim  T_{\dot{\Omega}_p}]$:
%%%%%%%%%%%%%%%%%%%%%%%%%%%%%%%%%%%%%%%%%%%%
Once the planet has completed more than one precession cycle, the warp-like feature begins to propagate outward, akin to the scenario with $M_d = 0$ (Figures \ref{fig:master_ModelC_maps}(iv)--(vi)). This occurs as the planetesimal inclination profile $I_d(a_d)$ develops oscillations along $a_d$, and the ascending nodes $\Delta \Omega(a_d)$ become phase-mixed in the inner disk regions; see also Figure \ref{fig:master_orb_el}(c). This description also explains the appearance of spiral patterns in the disk when viewed top-down, particularly near $a\inn$, where inclinations are larger.

The warp is more clearly visible in Figures \ref{fig:master_ModelC_maps}(iv) and (v), where it is located at $a\warp \approx 40$ and $48$ au, respectively. These correspond to the largest semimajor axes where the inclinations are maximized at a given time (Figure \ref{fig:master_orb_el}(c)). The warp's maximum vertical extent is well described by Equation (\ref{eq:zwarp}), with the substitution $I_m^{{\rm n/disk}} = 2 I_p(0) \rightarrow I_m^{\rm disk}(a\warp)$ (Equation \ref{eq:Im_disk}).  Unlike when $M_d = 0$, the line connecting the out-of-plane peaks of the warp is not aligned with the (precessing) planet but is instead anti-aligned. This is because inclinations are maximized when $\Delta \Omega(t) = \pi$ in the disk-dominated regime (Figure \ref{fig:master_orb_el}(c)). Note that at larger distances, the decreasing maxima of planetesimal inclinations cause the warp to peak at smaller values of $|Z|$. For example, by $t = 70 T_{\dot{\Omega}_p}$, the warp is located around $\approx 88$ au, with a vertical height as small as $\approx 0.2$ au, making it difficult to discern in Figure \ref{fig:master_ModelC_maps}(vi).

The evolution outlined above takes place while the inner disk regions continue to exhibit a near-polar component at small radii, as discussed in ``Stage 1''. However, this component is tenuous in surface density, typically lower by more than two orders of magnitude compared to the disk's midplane around $Z \approx 0$. As a result, in terms of angular momentum, the disk as a whole maintains its initial misalignment with the planet. This can also be seen in Figure \ref{fig:misalignment_ABC}, which shows that $\theta_{\rm al}(t) \approx I_p(0)$ at all times to within $\lesssim 1^{\circ}$.

Finally, in terms of its vertical structure, the disk not only remains nearly razor-thin but also develops a radial dependence in its aspect ratio $\mathcal{H}(R)$. This can be seen in  Figure \ref{fig:scaleheight_ABC}, which shows that  $\mathcal{H}(R)$ decreases with radius following $\propto R^{-7/2}$ with absolute values at least an order of magnitude or two smaller than $I_p(0)$ (as would otherwise be the case in massless disks). Physically, this means that the disk becomes progressively thinner at larger distances, thereby showing stronger resistance to planetary perturbations.

In closing, we note that the behavior described herein remains generally applicable to other planet-debris disk systems provided that $M_d/m_p \gtrsim M_d/m_p|_d$; see Equation (\ref{eq:Mdmp_disk_dominated}).

%%%%%%%%%%%%%%%%%%%%%%%%%%%%%
\subsection{Variations on a theme}
\label{subsec:variations}
%%%%%%%%%%%%%%%%%%%%%%%%%%%%%

We close this section by commenting on the generality of the results presented thus far. While quantitative differences may arise depending on the system parameters, certain scaling rules can account for such variations, as discussed below.

%%%%%%%%%%%%%%%%%%%%%%%%%%%%%%
\subsubsection{Variation with disk-to-planet mass ratio, \texorpdfstring{$M_d/m_p$}{Md/mp}}
\label{sec:var_Md_mp_results}
%%%%%%%%%%%%%%%%%%%%%%%%%%%%%%

We begin by considering how variations in the planet and disk masses affect the system's evolution, with all other parameters held constant. To this end, we  recall  from  Sections \ref{sec:analytical-model} and \ref{sec:analytical_analysis} that the individual masses $M_d$ and $m_p$ act as linear scaling factors for the  nodal precession rates  (Equations \ref{eq:A_planet} and \ref{eq:Adp_Bdp_approx}), while their ratio $M_d/m_p$ determines the dynamical regime (Equation \ref{eq:Adp_over_Ap}) and, consequently, the behavior of planetesimal inclinations (Equation  \ref{eq:I_forced_gen}). Accordingly, uniformly scaling both $m_p$ and $M_d$ -- while keeping the ratio $M_d/m_p$ constant -- does not affect the details of the system's dynamical and morphological end-states. The only effect of such scaling would be a change in the evolution timescales: namely, increasing both masses simultaneously results in a shorter secular timescale, and vice versa. This scaling remains valid as long as the Laplace--Lagrange description is applicable, i.e., for $m_p, M_d \ll M_c$, which is generally the case. Finally, we note that varying $M_d/m_p$ -- while all else is kept the same -- results in different dynamical regimes (and strengths thereof), as already explored in Sections \ref{subsec:massless_case_Md0}--\ref{subsec:large_Mdmp_sim} and illustrated in Figure \ref{fig:resonance_map}.

%%%%%%%%%%%%%%%%%%%%%%%%%%%%%%
\subsubsection{Variation with the initial disk mass distribution, \texorpdfstring{$p$}{p}}
\label{subsec:var_p}
%%%%%%%%%%%%%%%%%%%%%%%%%%%%%%

We next consider the effects of varying the initial mass distribution within the disk via $p$ in Equation (\ref{eq:Sigma_d}). While our fiducial choice of $p=1.5$ is motivated by the MMSN \citep{hayashi}, it is possible that debris disks are ``born'' with different values.\footnote{Observational data on $p$ is limited and primarily stems from (sub-)mm observations, which may be affected by, e.g., collisions \citep[][]{krivov2006}.} Kinematically, varying $p$ may introduce subtle changes in the disk's projected density maps. Dynamically, however, its impact could be more significant, as $p$ determines the planet's nodal precession rate $A_{d,p}$ (Equation \ref{eq:Adp_Bdp_approx}):  a higher concentration of mass near the planet results in faster precession. Thus, variations in $p$ affect the relative strength of perturbations, $A_{d,p}/A_p$  (Equation \ref{eq:Adp_over_Ap}), directly influencing the critical mass ratio $M_d/m_p$ that dictates whether the disk- or planet-dominated regime applies, as well as the corresponding disk regions (see Equation \ref{eq:res_loc}).

Generally, the disk contains more mass in its outer regions than its inner regions for $p<2$, and vice versa. Indeed, if we define the disk's midpoint as the geometric mean of its boundaries, $\bar{a}_d=\sqrt{a\inn a\out}$, then $M_d(a_d < \bar{a}_d) = M_d(a_d > \bar{a}_d) \delta^{(p-2)/2}$. Thus, considering decaying profiles of $\Sigma_d^{t=0}(a_d) \propto a_d^{-p}$ (i.e., $p \geq 0$), steeper (flatter) initial surface density profiles lead to faster (slower) planetary nodal precession rates: indeed, we find that regardless of $a_p/a\inn$, $|A_{d,p}|$ is approximately twice as small (large) for $p=0.5$ ($p=2.5$) compared to $p=1.5$; see also Equation (\ref{eq:Adp_Bdp_approx}). Accordingly, when $p$ is varied relative to the fiducial value,  reinstating the same dynamical regime and strength at a given location within the disk requires adjusting $M_d/m_p$ by a factor of  $\sim 1-2$: namely, decreasing it for larger $p$,  and vice versa. Otherwise, if $M_d/m_p$ is held constant, increasing (decreasing)  $p$ shifts the resonance location inward (outward). For instance, $a\res$  changes by at most $\approx 20\%$ from its nominal value  in \texttt{Model B} as $p$ is varied between $0.5$ and $2.5$ (see also Equation \ref{eq:res_loc}).

%%%%%%%%%%%%%%%%%%%%%%%%%%%%%%
\subsubsection{Variation with the disk's radial extent, \texorpdfstring{$\delta$}{delta}}
\label{subsec:var_delta}
%%%%%%%%%%%%%%%%%%%%%%%%%%%%%%

We now consider the effects of varying the disk's initial radial extent $\delta \equiv a\out/a\inn$, while all else is held constant.  Similar to variations in $p$ (Section \ref{subsec:var_p}), changes in $\delta$  directly affect the planet's nodal precession rate $A_{d,p}$, thereby, the system's  dynamical regime/evolution. This is to be expected, as radially narrower disks  concentrate more mass near the planet, leading to  larger $|A_{d,p}|$, and vice versa; see, e.g., Equation (\ref{eq:Adp_Bdp_approx}). Thus, the disk's back-reaction effects may become more significant and easier to induce for $\delta\rightarrow 1$. This is evident in Equation (\ref{eq:Mdmp_disk_dominated}), according to which:
%%%%%%%%%%%%%%%%%
\begin{equation}
  \frac{M_d}{m_p}\bigg|_d 
  \propto 
  \frac{1-\delta^{p-2}}{\delta^{p+1}-1} \delta^3 ,    
  \label{eq:mdmp_delta_sec}
\end{equation}
%%%%%%%%%%%%%%%%%
where we have ignored the weak dependence of $\phi_1^c$ on $\delta$ (App. \ref{sec:app_A}). Equation (\ref{eq:mdmp_delta_sec}) shows that for smaller values of $\delta$, the minimum disk-to-planet mass ratio -- beyond which all planetesimals would be in the disk-dominated regime -- decreases, and vice versa. For instance, in otherwise identical systems to those in Table \ref{table:models} but with $\delta = 1.1$ (akin to known narrow debris disks, \citet{reasons_matra}), the minimum $M_d/m_p|_d\approx 1$ instead of $\approx 8$; a reduction by about an order of magnitude. This behavior holds true for all values of $p$ and $a_p/a\inn$.

Additionally, since $M_d/m_p|_d \propto \sqrt{a_p/a\inn}$ (Equation \ref{eq:Mdmp_disk_dominated}), it follows that radially-narrow debris disks  could behave as in \texttt{Model C} (Section \ref{subsec:large_Mdmp_sim}) even when $M_d/m_p \lesssim 1$, provided that $a_p \ll a\inn$. For instance, for a disk with $\delta = 1.1$ and $p=1.5$, Equation (\ref{eq:Mdmp_disk_dominated}) yields $M_d/m_p|_d \approx 0.5 $ for $a_p/a\inn = 1/6$. In such cases, from a morphological perspective, the measurability of the disk's aspect ratio slope (i.e., $\mathcal{H} \propto R^{-7/2}$) would depend on its radial extent, let alone observational considerations (e.g., resolution). Nevertheless, the disk would exhibit notably small aspect ratios, while remaining misaligned with the planet (Figures \ref{fig:misalignment_ABC} and \ref{fig:scaleheight_ABC}).

%%%%%%%%%%%%%%%
\subsubsection{Variation of the planetary semimajor axis, \texorpdfstring{$a_p$}{ap}}
%%%%%%%%%%%%%%%

We now examine the effects of varying the planetary semimajor axis. To this end, we note that changing $a_p$ changes the value of $M_d/m_p$ needed to maintain the same dynamical regime and strength at a given distance. This relationship is illustrated, e.g., in Figure \ref{fig:resonance_map}: for $a_p/a\inn \lesssim 1$,  maintaining a given $a\res/a\inn$ with a larger $a_p/a\inn$ requires a larger $M_d/m_p$ (and vice versa). The departure from this scaling happens when $a_p/a\inn \rightarrow 1$ and $\phi_1^c \sim (1-a_p/a\inn)^{-1}$ diverges (Appendix \ref{sec:app_A}), reversing the requirement. Quantitatively, maintaining this balance requires  (see Equations (\ref{eq:Adp_over_Ap})--(\ref{eq:res_loc})):
%%%%%%%%%
\begin{equation}
    \phi_1^c(a_p/a\inn)  \frac{M_d}{m_p} \sqrt{\frac{a\inn}{a_p}} =  {\rm const}.
    \label{eq:condition_ap_vari}
\end{equation}
%%%%%%%%%
Accordingly, as long as Equation (\ref{eq:condition_ap_vari}) holds, the profiles of planetesimal inclinations and the secular resonance locations, if any,  remain unaffected (Equations (\ref{eq:res_loc}) and (\ref{eq:Im_disk_sup_numest})), leaving the disk structure unchanged. While subtle, the same applies to the widths of resonances, $w\res$: if we naively define $w\res$ as the range in semimajor axis where planetesimal inclinations around $a\res$ exceed their maximum amplitudes in a massless disk, $I_m(a_d) \geq I_m^{{\rm n/disk}} = 2 I_p(0)$, then
%%%%%%%%%%%%%%%%
\begin{equation}
    w\res \approx \frac{4}{7} a\res ,  
    \label{eq:w_res}
\end{equation} 
%%%%%%%%%%%%%%%%
where we have made use of Equations (\ref{eq:lambda_If_gen}) and (\ref{eq:lambda_If}). Thus, varying $a_p$ in this way affects only the secular timescales: the evolution is faster when the planet is closer to the disk than to the star, and vice versa (see Equations (\ref{eq:planetary_node_rate}), (\ref{eq:Tsec_planetesimals}), and (\ref{eq:Tsec_no_disk})). We note in passing that since $w\res \propto a\res$ and, according to Equation (\ref{eq:res_loc}), $a\res \propto (M_d/m_p)^{-2/7} $ for a fixed $a_p$, Equation (\ref{eq:w_res}) well explains the larger resonance widths (and locations) for smaller $M_d/m_p$ seen in Figure \ref{fig:I_forced_a}.

Finally, when $a_p/a\inn$ is changed without modifying any other parameter, the further the planet is from the disk, the easier it is for the latter to dominate the dynamics; see Figure \ref{fig:resonance_map} and Equation (\ref{eq:Adp_over_Ap}). For instance, decreasing $a_p$ for a fixed $M_d/m_p$ pushes the resonance inward, such that $a\res \propto a_p^{1/7}$ (see   Equation \ref{eq:res_loc} and Figure \ref{fig:resonance_map}).

%%%%%%%%%%%%%%%
\subsubsection{Variation of the planetary inclination, \texorpdfstring{$I_p(0)$}{Ip(0)}}
%%%%%%%%%%%%%%%

Finally, we consider the effects of varying the planet's initial inclination $I_p(0)$. Given the linear nature of the employed theoretical model (Section \ref{sec:analytical-model}), the value of $I_p(0)$ does not affect the nodal precession rates \citep[for a non-linear treatment, see, e.g.,][]{laskar-boue}; instead, it only controls the oscillation amplitudes of planetesimal inclinations (Equation \ref{eq:I_forced_gen}). Therefore, the secular evolution of otherwise identical systems, differing only in the value of $I_p(0)$, would follow the same qualitative dynamical regime (Section \ref{sec:dyn_reg}), with quantitative differences in terms of the disk morphology.

First, initially more inclined planets lead to vertically thicker disks, and vice versa, without affecting the overall planet-disk (mis)alignment. This is to be expected since $I_p(0)$ acts as a linear scaling factor for the planetesimal inclinations, $I_m(a_d) \propto I_p(0)$ (Section \ref{sec:analytical_analysis}). We have verified this by varying $I_p(0)$ within $1^{\circ} - 10^{\circ}$ in each of models \texttt{A}, \texttt{B}, and \texttt{C} (Table \ref{table:models}); the resulting curves for $\mathcal{H}(R)$  at $t \rightarrow \infty$, once normalized by $I_p(0)$, closely overlap with each other (apart from minor-to-no noise due to the random assignment of mean anomalies; Appendix \ref{app:map_construction}). The only exception occurs interior to the resonance location, if present: namely, $\mathcal{H}(R)/I_p(0)$ is more enhanced at $R \lesssim a\res$ for lower $I_p(0)$. This occurs due to the fact that the inclination dispersion difference between resonant and non-resonant planetesimals at $a\inn \lesssim R \lesssim a\res$ increases, compromising our definition of $\mathcal{H}(R)$ (Appendix \ref{app:scale_height}), which is not well-suited for bimodal distributions of $Z(R)$. To verify this, we removed the resonant planetesimals from the calculation (although not shown in Figure \ref{fig:scaleheight_ABC}), finding that  $\mathcal{H}(R) \rightarrow I_p(0)$ at  $R \lesssim a\res$ for all $I_p(0)$. As we show later in Section \ref{subsec:aspect_discussion_Sec}, the fact that  $\mathcal{H} \propto I_m(a_d) \propto I_p(0)$ can be derived analytically from first principles (see Equation \ref{eq:aspect_ratio_analytical_H}).

Second, and relatedly, the maximum height at the location of the warp (if present) is larger for planets that are initially more inclined, and vice versa (Equation (\ref{eq:zwarp})). Third, when the disk features a secular resonance, higher initial planet inclinations cause  $I_d(a\res) \rightarrow 90^{\circ}$ faster ($\tau\res \propto 1/I_p(0)$; Eq. \ref{eq:tau_res_growth}), but this does not affect the overall disk structure. This is because the ``broken" morphology, described in Section \ref{subsec:small_Mdmp_sim}, emerges once planetesimals around $a\res$ attain inclinations that are larger than those of their surroundings: according to Equation (\ref{eq:t_significant_inc_res}),  this occurs at $t \gtrsim T_{\dot{\Omega}_p} / \pi$, independently of $I_p(0)$. Finally, varying $I_p(0)$ leads to subtle differences  in the structure of the bending waves, e.g., when viewed face-on. Namely, with increasing $I_p(0)$, the bending waves appear more open and prominent, essentially due to the higher forced planetesimal inclinations.

%%%%%%%%%%%%%%%%%%%%%%%%%%%%%%%%%%%%%
\section{Discussion}
\label{sec:discussion}
%%%%%%%%%%%%%%%%%%%%%%%%%%%%%%%%%%%%%%%%%%%%%%%%%

Debris disk structures, including vertical features such as warps and scale heights, are often used to infer the presence and characteristics of unseen planetary companions. With the exception of $\beta$ Pictoris (Section \ref{sec:intro}), planets inferred solely from debris disk structures have yet to be detected. However, with JWST's exceptional sensitivity \citep[e.g.,][]{carter-jwst}, many of these planetary predictions are being  and will be tested, potentially expanding the album of exoplanets in the near future. It is, however, important to recognize that any planetary inference is only as good as the assumptions underlying the models employed. Given estimates of debris disk masses -- ranging from a few Earth  to tens of Neptune masses \citep{krivovwyatt21} -- models that neglect the role of disk gravity may lead to compromised conclusions. This served as the motivation for the present work, urging us to analytically investigate the  inclination dynamics of massive disks perturbed by inclined planets.

To address the questions raised in Section \ref{sec:this_work_Q} regarding the extent to which planetary inferences are compromised by the disk mass, we first discuss the implications of our findings for interpreting warped structures (\S \ref{sec:disc1_warps}), followed by vertical scale heights and density profiles (\S \ref{sec:scale_height_discussion}). Finally, we discuss the implications of our study for constraining the total masses of debris disks (\S \ref{sec:debrisdiskmass_problem_sec}), before reviewing the key assumptions and limitations of our model (\S \ref{subsec:limit_future_work}).

We preface this section by emphasizing that our goal is not to provide definitive predictions for planet and disk parameters in any specific observed system. Rather, our aim is to demonstrate that even at the level of simplicity considered here -- i.e., accounting only for the axisymmetric back-reaction of the disk on the planet, while neglecting its non-axisymmetric counterpart and the disk's self-gravity -- the resulting dynamics already alters the interpretation of debris disk observations, including planetary inferences, compared to models that assume massless disks. Thus, the quantitative predictions presented below should be viewed as illustrative rather than definitive. They serve as both a guide and motivation for future, more comprehensive studies that account for the full gravitational potential of the disk, whether via direct N-body integrators or specialized orbit-averaged codes \citep[e.g.,][]{hahn2003, Touma2009, Paper2}.

%%%%%%%%%%%%%%%%%%%%%%%%%%
\subsection{Implications for warp features}
\label{sec:disc1_warps}
%%%%%%%%%%%%%%%%%%%%%%%%%%

Results of Sections \ref{sec:analytical_analysis} and \ref{sec:vert_struc_num} indicate that, across a broad range of parameters, a planet--debris disk system may feature a secular inclination resonance, with implications for shaping warped structures (c.f. Figures \ref{fig:master_Md0_maps} and \ref{fig:master_ModelB_maps}). This is so even when  $M_d/m_p \lesssim 1$, contrary to naive expectations. Below, we apply our results to known warped debris disks to illustrate the implications of our findings.

%%%%%%%%%%%%%%%%%%%%%%%%%%%%%%%%%%%%%%%
\subsubsection{A practical guide}
\label{sec:how_to_use_warp}
%%%%%%%%%%%%%%%%%%%%%%%%%%%%%%%%%%%%%%%

We first provide a brief guide on how to use our results for interpreting warped structures. To this end, we note that:
\\
$\bullet$ For massless disks, the warp is transient.  Thus, for a given system age $t_{\rm age}$ and warp location $a\warp$, solving the condition $0.5 T\sec^{{\rm n/disk}}(a\warp) = t_{\rm age}$ (Equation \ref{eq:tau_warp_ppdom}) yields a family of solutions relating $m_p$ and $a_p$;  see also Equation (\ref{eq:awarp_explicit}).
\\
$\bullet$
For massive disks, two families of solutions exist:
\\ 
\indent \textbf{(i)} In ``{Family 1}'',  the observed warp is presumed to be stationary with time, i.e., its location coincides with the resonance location. Thus, setting $a\res = a\warp$ in Equation (\ref{eq:res_loc})  yields a relationship between $M_d/m_p$ and $a_p/a\inn$, the results of which are further subjected to the condition $ t_{\rm age} \geq  T_{\dot{\Omega}_p} /\pi$ given by Equation (\ref{eq:t_significant_inc_res}). The latter is to ensure that planetesimal inclinations at $a\res$ are excited to a ``significant'' degree within the system's age. Note that this solution family corresponds to the warps observed in the later stages in Section \ref{subsec:small_Mdmp_sim}; namely, Stages 2 and 3.
\\
\indent \textbf{(ii)} In ``{Family 2}'', the observed warp is presumed to be evolving outward in time. The system may or may not feature a secular resonance within the disk, but the warp has not yet propagated to that distance, i.e., $a\warp < a\res$. The solutions in this case are obtained by solving $0.5 T\sec(a\warp) = t_{\rm age}$ (Equation \ref{eq:tau_warp_ppdom_modified_w_Md}), subject to the condition that $|A_p(a\warp)/A_{d,p}| > 1$ (Equation \ref{eq:Adp_over_Ap}). This solution family can be thought of as the generalized version of that obtained in massless disk calculations. In terms of Section \ref{subsec:small_Mdmp_sim}, it corresponds to the propagating warp during Stage 1.

%%%%%%%%%%%%%%%%%%%%%%%%%%
\subsubsection{Application to HD 110058}
\label{sec:HD110058}
%%%%%%%%%%%%%%%%%%%%%%%%%%

We first consider  \HDone~(\HDoneHIP), which, in some ways, is reminiscent of $\beta$ Pictoris. This system, located at a distance of $130\pm 1$ pc \citep{GAIADR2}, is part of the Lower Centaurus-Crux (LCC) association, with an estimated age of $15 \pm 3$ Myr \citep{Pecaut2016}. It features a late A-type star with a dynamical mass of $\approx 1.84 \pm 0.16 M_{\odot}$ \citep{hales2022}, and a near edge-on debris disk but no known planets \citep[][]{meshkat2015, brandt2021, Stasevic2023}. The disk has been observed by multiple instruments \citep{Kasper2015, Esposito2020, Stasevic2023, Ren2023}, including  ALMA \citep{Lieman16ALMA, hales2022}, according to which the mm-sized dust grains  populate the region between $\sim 18$ au and $70$ au. Interestingly, the disk appears to be vertically thick \citep[$\mathcal{H} \approx 0.13$--$0.28$;][]{hales2022} and warped, though the exact warp location is not well constrained, with values reported between $40$ and $60$ au \citep{Kasper2015, hales2022, Stasevic2023}.

\begin{figure}
%%%%%%%%
\centering
\includegraphics[width=\columnwidth]{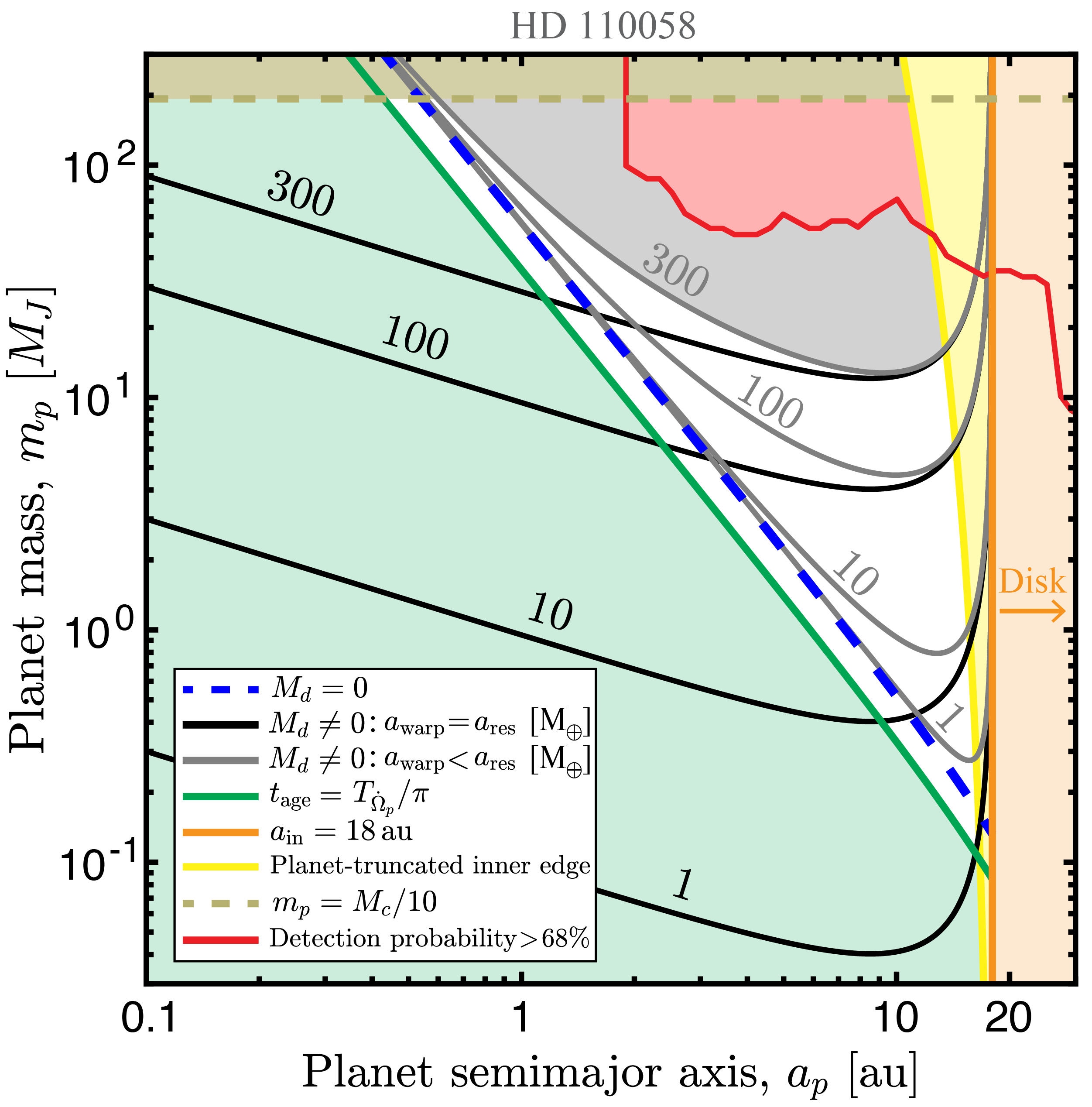} 
%%%%%%%%
\vspace{-1.5em}
{\linespread{0.95}\selectfont 
\caption{The mass of a hypothetical planet $m_p$ as a function of its semimajor axis $a_p$ capable of producing a warp in the \HDone~debris disk at $a_{\rm warp} = 50$ au. The solution with a massless disk  is depicted by the blue dashed curve ($M_d =0$).  For the massive disks, two families of solutions are presented (assuming $p=3/2$; Eq. (\ref{eq:Sigma_d}) and Section \ref{sec:how_to_use_warp}): the black and gray contours correspond to ``Family 1'' (i.e., $a\warp = a\res$) and ``Family 2'' (i.e., $a\warp < a\res$), respectively, with different contours representing different values of $M_d$ (in $M_{\earth}$; as indicated). Color-shaded regions are excluded: (i) the gray region represents where the disk would be too massive; (ii) the green region corresponds to inclination excitation timescales at the resonance that are longer than the system age; (iii) the yellow region indicates where the planet would have carved an inner disk edge larger than the one observed (Eq. \ref{eq:MMR_condition_delta_ap}); and (iv) the olive region represents where $m_p \geq M_c/10$. The remaining white (unshaded) area represents the allowed parameter space where all of the above conditions are satisfied. Note that the red region is where the presence of companions is ruled out with $>68\%$ confidence (based on direct imaging and Gaia astrometry; \citet{Stasevic2023}). See the text (\S \ref{sec:HD110058}) for details. 
\label{fig:HD_110058}}}
\end{figure}

This naturally raises the question: How do the inferred properties of the planet, as a function of disk mass, compare with predictions from massless disk calculations? This is addressed in Figure \ref{fig:HD_110058}, which shows the planetary mass required to sculpt a warp at $50$ au -- the average of the reported range -- as a function of its semimajor axis for different disk masses (assuming $p=3/2$; Eq. \ref{eq:Sigma_d}). The ``allowed'' range of planet-disk parameters is represented by the white (unshaded) region. The color-shaded regions are excluded based on various factors, including timescale arguments, MMR overlap considerations, and reasonable planet and disk mass constraints, as explained in the caption.

Figure \ref{fig:HD_110058} shows that when the disk is assumed to be massless, $M_d = 0$, the possible solutions lie along a single curve (shown in dashed blue), which scales  as $m_p \propto a_p^{-2}$; see also Equation (\ref{eq:awarp_explicit}). By contrast, when the disk has a non-zero mass, $M_d \neq 0$, the allowed parameter space expands significantly, forming an extended wedge-shaped area. Indeed, we find that, provided the disk mass is $1 M_{\earth} \lesssim M_d \lesssim 300 M_{\earth}$, a companion with a semimajor of $\sim 0.4 - 16$ au and a mass between $\sim 0.1$ and $190 M_J$ can produce the observed warp in \HDone~within the system's age.\footnote{The upper limit of $M_d = 300 M_{\earth}$ is somewhat arbitrary. This, however, corresponds to a maximum planetesimal size of $\sim 15$ km, assuming the observed dust ($\approx 0.08 M_{\earth}$; \citet{hales2022}) is produced in a steady-state collisional cascade; see \citet[][eq. 6]{krivovwyatt21} and Section \ref{sec:debrisdiskmass_problem_sec}.} This holds for both the solutions of ``Family 1'' ($ a\warp= a\res$; black contours) and those of ``Family 2'' ($a\warp < a\res$; gray contours). Note that for $a_p \lesssim a\inn$, the Family 1 solutions scale as $m_p \propto a_p^{-1/2}$  (Equation \ref{eq:res_loc}), while the Family 2 solutions asymptote to the massless disk solutions.

%%%%%%%%%%%%%%%%%%%%%%%%%%%%%%
%%%%%%%%%%%%%%%%%%%%%%%%%%%%%%%%%%

A corollary that follows from  Figure \ref{fig:HD_110058} is that for disks with $M_d \neq 0$, the inferred planetary masses are generally larger than those predicted by massless-disk models. Indeed, the wedge-shaped region of possible solutions roughly extends from the diagonal corresponding to the solutions with $M_d =0$ to the disk's inner edge. This observation is a general one: for a given $a_p$, the minimum planetary mass $m_p$ derived from the $t_{\rm age} = T_{\dot{\Omega}_p}/\pi$ condition for massive disks will always be a factor of $2/\pi$ smaller than that inferred with $M_d =0$; see Equations (\ref{eq:tau_warp_ppdom}) and (\ref{eq:t_significant_inc_res}). Thus, planetary mass inferences based on massless disk models, for this or any other system, can and should be considered as a lower bound for $m_p$ at a given $a_p$, the upper bound of which can be limited by constraints from disk masses and traditional planet-hunting techniques, as done in Figure \ref{fig:HD_110058}.

The results of Figure \ref{fig:HD_110058} are promising,  in the sense that  (planetary) companions responsible for the warp in \HDone~could be more massive than currently thought (especially at $\sim 10$ au separations); therefore, potentially detectable in future observations. Currently, direct imaging and proper motion anomaly have ruled out giant planets outside the \HDone~disk (namely, with $m_p \gtrsim 8 M_J$ at $ \gtrsim 50$ au; \citet{brandt2021, Wahhaj2013, meshkat2015, Kervella2022}), and there is a likelihood of $< 68\%$ of detecting any companion interior to the disk with a mass smaller than $\sim 50 M_J$ \citep[][see the red-shaded region in Figure \ref{fig:HD_110058}]{Stasevic2023}.

The inferences in Figure \ref{fig:HD_110058} hold true independently of $I_p(0)$. Can the disk's aspect ratio $\mathcal{H}$ help constrain it? The short answer is: in principle, yes, but  for \HDone, further data is required. This is because, when $M_d \neq 0$,  ``Family 2'' solutions (i.e., $a\warp < a\res$) yield aspect ratios  similar to massless disk models, whereas  ``Family 1'' solutions (i.e., $a\warp = a\res$) produce a bimodal $Z(R)$ distribution interior to $a\warp$ and $\mathcal{H} \propto R^{-7/2}$ outside it (Figure \ref{fig:scaleheight_ABC}). Thus, without direct $\mathcal{H}(R)$ constraints -- typically assumed constant in observational fittings -- we can only ascertain that the initial misalignment  cannot exceed that expected from massless disk models, which for \HDone~corresponds to $\sigma_i/2 \approx 6^{\circ}-12^{\circ}$. Here,  $\sigma_i \approx \sqrt{2} \mathcal{H}$ is the inclination dispersion obtained from the measured aspect ratio, assuming a vertical Gaussian distribution \citep{matra2019, hales2022}. Aside from planet detection, a definitive answer requires either (i) evidence of a decreasing aspect ratio beyond the warp, e.g., by fitting $\mathcal{H}(R)$ non-parametrically \citep[][]{Han25}, which will be difficult to measure since the warp in  \HDone~occurs near $a\out$; or (ii) the presence of a bimodal planetesimal distribution interior to the warp, a possibility which has not yet been tested (Section \ref{sec:cold_hot_rayleigh}).

Finally, we point out that a future detection of a planet within the allowed region of Figure \ref{fig:HD_110058} would provide an indirect measure (or constraint, in the case of non-detection) of the total debris disk mass, assuming no other significant perturbers (e.g., additional planets).
%%%%%%%
This is particularly appealing given the uncertainties in estimating total masses of debris disks using theoretical collisional models (see Section \ref{sec:debrisdiskmass_problem_sec} and \citet{krivovwyatt21}). 
%%%%%%%
Additionally, if such a planet is found to be lying along or close to the solution with $M_d =0$ in Figure \ref{fig:HD_110058}, this does not necessarily imply a massless disk, as solutions with massive disks either intersect or asymptote to that line. We next explore this possibility for $\beta$ Pic.

%%%%%%%%%%%%%%%%%%%%%%%%%%%%%%%%%%
\subsubsection{The case of \texorpdfstring{$\beta$}{beta} Pictoris}
\label{sec:beta_pic_luckyQ}
%%%%%%%%%%%%%%%%%%%%%%%%%%%%%%%%%%

The discovery of  $\beta$ Pic b -- with parameters apparently consistent with those predicted by massless disk models -- is often celebrated as a major success in ascribing debris disk structures to planets (Section \ref{sec:motiv}). However, several features of the $\beta$ Pic debris disk, such as its inner edge location, aspect ratio, and mutual inclination with respect to the planet, remain difficult to explain \citep{Currie11, Dawson2011, Apai15,  Maxwell15, matra2019, beust25}. Thus, a natural question arises: does the seemingly successful prediction of $\beta$ Pic b confirm the validity of the massless disk models, or if it is merely a coincidence, what implications can be drawn about the disk mass?

\begin{figure}
%%%%%%%%
\centering
\includegraphics[width=\columnwidth]{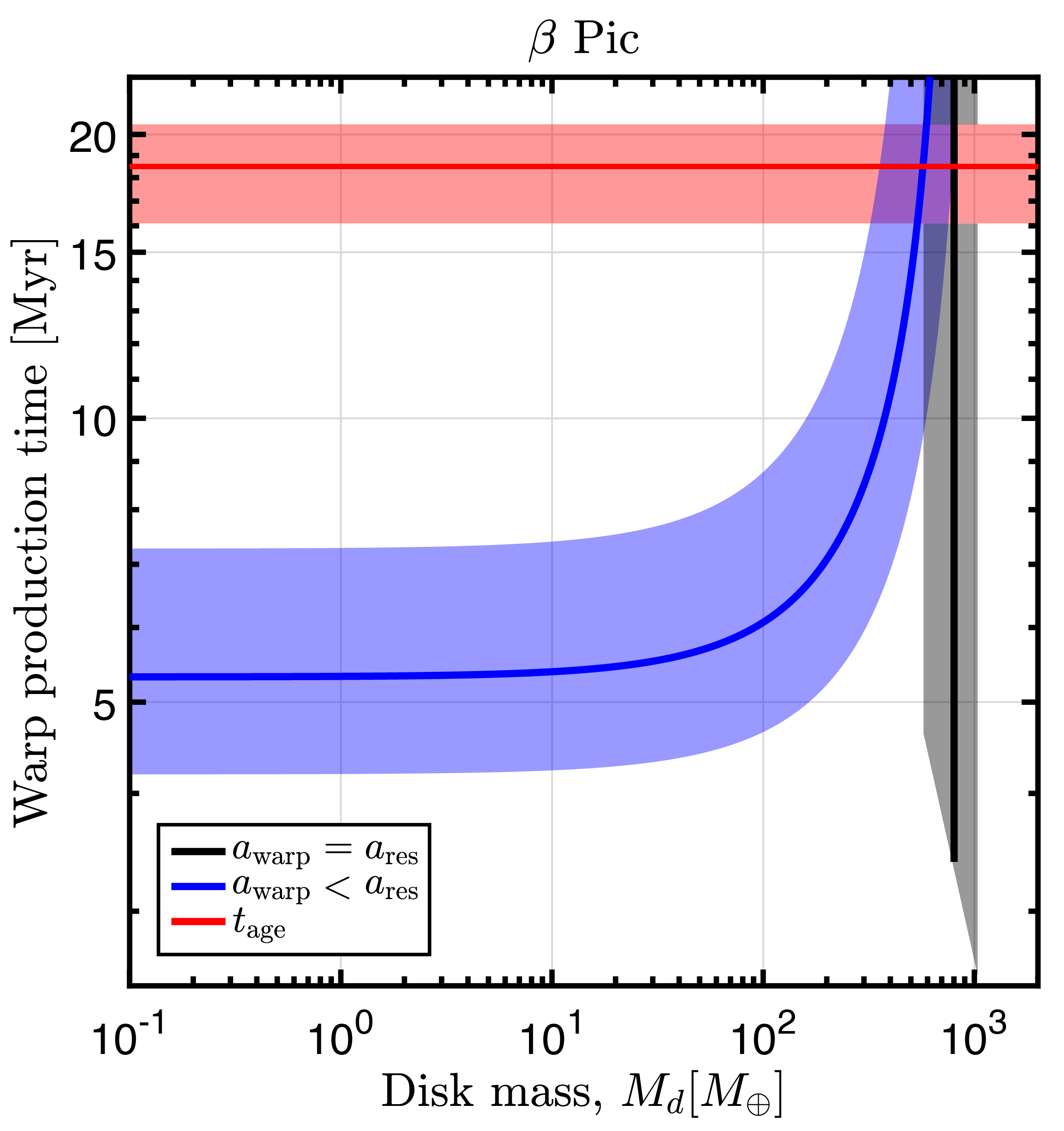}
%%%%%%%%
\vspace{-1.5em}
{\linespread{0.95}\selectfont 
\caption{The time it takes for $\beta$ Pic b to produce a warp in the debris disk at $a\warp \approx 85$ au as a function of disk mass $M_d$. Results are shown for the two possible families of solutions (Section \ref{sec:how_to_use_warp}): the black and blue curves represent ``Family 1'' (i.e., $a\warp = a\res$) and ``Family 2'' (i.e., $a\warp < a\res$), respectively, for the nominal mass of $\beta$ Pic b. The shaded regions around each of these curves result from the $1\sigma$ uncertainty in the mass of $\beta$ Pic b. For reference, the system's age is shown in red, with the shaded area around it enclosing its $1\sigma$ uncertainty. One can see that $\beta$ Pic b, along with a disk of  mass $M_d \lesssim 10^3 M_{\oplus}$, can sculpt the observed warp within the system's lifetime. See the text (Section \ref{sec:beta_pic_luckyQ}) for details.}
\label{fig:beta_pic_t_Md}}
\end{figure}

$\beta$ Pic (HD 39060), with a dynamical mass of $ 1.83 \pm 0.04 M_{\odot}$ \citep{Brandt}  and a dynamical age of $18.5_{-2.4}^{+2.0} $ Myr \citep{Miret2020}, hosts two planets \citep{lagrange2010, lagrange-betapic-c, Lacour2021}. The observed warp in its debris disk  at $a\warp \approx 85$ au \citep[e.g.,][]{Heap2k} is thought to be produced by $\beta$ Pic b ($m_p \approx 9.3 \pm 2.5 M_J$; $a_p = 10.26 \pm 0.10$ au; \citet{Brandt}), with negligible influence from $\beta$ Pic c at $\sim$3 au \citep{Dong2020}. For this scenario to hold, according to Equation (\ref{eq:tau_warp_ppdom}) the warping must have started only {$\approx 4-7$ Myr ago} as shown in Figure \ref{fig:beta_pic_t_Md} \citep[otherwise the warp would be located further out; see also][]{Dawson2011}. Addressing whether this is plausible and indicative of other processes (e.g. $\beta$ Pic b acquiring its current orbit relatively late, gas dispersal timescale, initial conditions of planetesimals, etc.) is beyond the scope of this study.

Instead, we ask: what can be inferred when massive disk models are used? That is, given the system parameters, are there solutions with reasonable values of $M_d \neq 0$ that can reproduce the warp? This is addressed in Figure \ref{fig:beta_pic_t_Md}, where we show the warp production time as a function of disk mass $M_d$, considering the two possible families of solutions with $M_d \neq 0$ (Section \ref{sec:how_to_use_warp}).  Calculations are done for $a\inn \approx 50 $ au and $a\out \approx 150$ au, consistent with ALMA observations \citep{Dent2014, matra2019}, using $p=3/2$ and accounting for the uncertainties in $m_p$ (but not in $a_p$ and $M_c$, which are small enough to be ignored for our purposes). Note that the dust mass inferred from sub-mm observations is $\sim 0.1 M_{\earth}$ \citep{matra2019, krivovwyatt21}, with the total disk mass necessarily being larger than this.

%%%%%%%%%%%%%%%%%%%%%%%%%%%%%%
%%%%%%%%%%%%%%%%%%%%%%%%%%%%%%%%%%

Figure \ref{fig:beta_pic_t_Md} shows that the warp in $\beta$ Pic can be sculpted by planet b  for a broad range of disk masses, $M_d \lesssim 10^{3} M_{\earth}$.
More specifically, the ``Family 2'' solutions (i.e., $a\warp < a\res$; shown in blue) result in longer warp production timescales compared to the $M_d=0$ scenario, while remaining within the system's age limit for up to $M_d \approx 600 M_{\earth}$. These solutions converge to the massless disk scenario as $M_d \rightarrow 0$, such that the warp production time is nearly constant for $M_d \lesssim 10 M_{\earth}$. 
By contrast, the ``Family 1'' solutions (i.e., $a\warp = a\res$; shown in black) can be achieved only with relatively more massive disks: namely, with $M_d/m_p \approx 0.27$, so that $600  \lesssim M_d / M_{\earth} \lesssim 1000$ given the mass of $\beta$ Pic b and its $1\sigma$ uncertainty. In this case, we recall that the warp would remain localized after being carved within $t =T_{\dot{\Omega}_p}/\pi $; thus, the vertical black line in Figure \ref{fig:beta_pic_t_Md}. Given Figure \ref{fig:beta_pic_t_Md}, it is perhaps after all not that surprising that $\beta$ Pic b was discovered with parameters apparently consistent with  expectations based on massless disk models. Before moving on, we highlight that this observation may not necessarily be unique to $\beta$ Pic; the same underlying behavior should be considered when interpreting existing or future debris disks that exhibit warps (see also Section \ref{sec:HD110058}).

Can the disk mass be further constrained? Future improvements in the planetary mass and detailed characterization of the disk's substructure could make this possible. While several observational features speak in favor of our findings with a massive disk -- e.g., the tentatively distance-decreasing aspect ratio, the coexistence of cold and hot planetesimal populations, and the recently imaged ``cat's tail'' in thermal emission  \citep{matra2019, isa2024beta} -- we do not explore these details further. This is so not least because of our model limitations (Section \ref{subsec:limit_future_work}) but also the lack of reliable non-parameteric observational fits (e.g., as recently done for other disks by \citet{Han25}).  For now, we note that the upper limit of $M_d \approx 10^{3} M_{\earth}$ is consistent with collisional model estimates, which suggest $M_d \approx 1200 M_{\earth}$ if the largest colliding planetesimals are $200$ km in size \citep{krivovwyatt21}. A further indication of a massive disk is that $\beta$ Pic b is misaligned with the inner disk, contrary to expectations from massless disk models. The exponential damping of $I_p$ due to resonant friction (Appendix \ref{app:res_friction}) could naturally lead to such a misalignment,\footnote{ \citet{Dawson2011} propose inclination damping as a possible resolution to this conundrum, though in their work, it was artificially imposed.} which we shall explore in a future work. Nevertheless, Figure \ref{fig:beta_pic_t_Md} provides a case in point: the detection of a planet with parameters seemingly compatible with massless disk models does not necessarily imply the disk is massless; rather, it may be a coincidence given the asymptotic behavior of massive disk solutions.

%%%%%%%%%%%%%%%%%%%%%%%%
\begin{figure*} 
%%%%%%%%
\centering
\includegraphics[width=\textwidth]{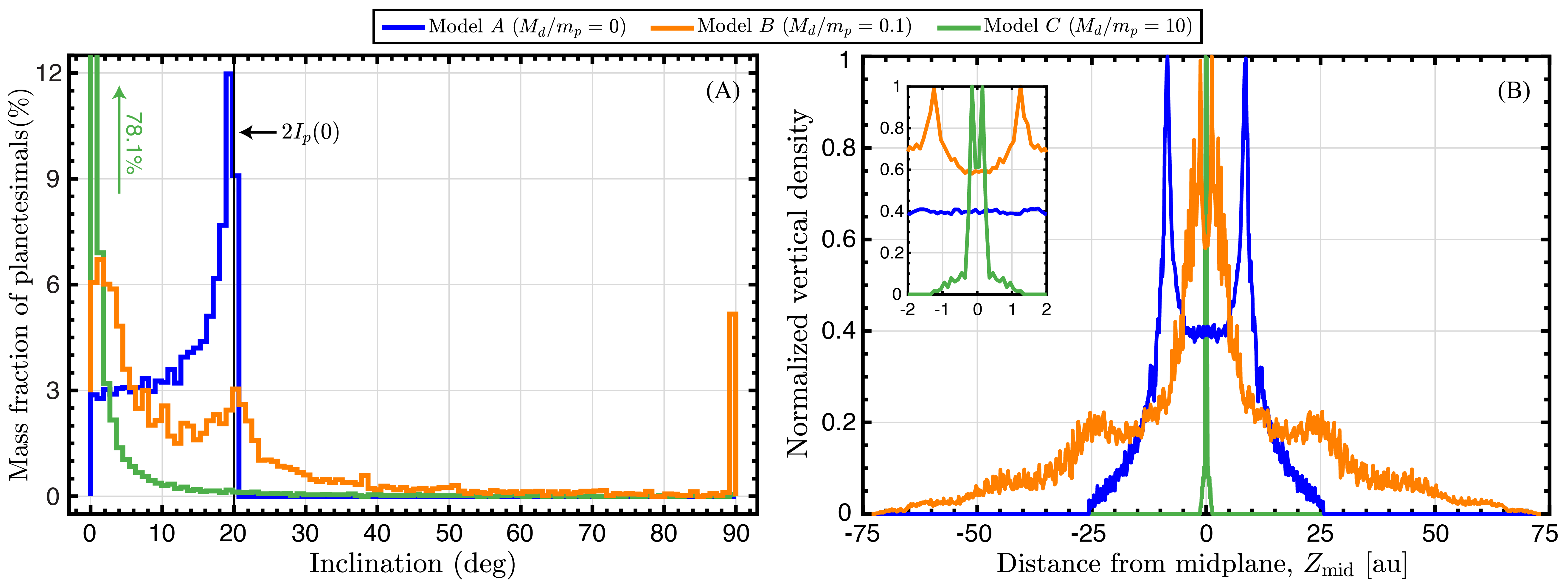}
%%%%%%%%
% Caption   
\vspace{-1.5em}
{\linespread{0.95}\selectfont \caption{Vertical characteristics of the debris disks in each of models \texttt{A}, \texttt{B}, and \texttt{C} (Table~\ref{table:models}). For each model, results correspond to the final snapshot time shown in Figure~\ref{fig:master_orb_el}.
{Panel (A):} Mass-weighted distributions of planetesimal inclinations, computed using Equation (\ref{eq:Id_sol}) and $p = 3/2$, shown as histograms with $100$ equal-sized bins. For reference, the black vertical line represents twice the planetary inclination $2 I_p(0)$. Note that in Model \texttt{C}, nearly $78 \%$ of planetesimals have inclinations of $1^{\circ}$ or less.
{Panel (B):} Normalized vertical density profiles as a function of distance from the disk midplane $Z_{\rm mid}$. Results are averaged over a radial range of $\pm 2.5$ au centered on the reference projected radius $R_0 \approx 49.1$ au (equal to $\sqrt{a\inn a\res}$ in Model \texttt{B}), using a resolution of $dZ_{\rm mid} = 0.1$ au. The inset presents a zoomed view around the midplane. Regardless of disk mass, the non-Rayleigh nature of the inclination distributions and the non-Gaussian shape of the density profiles are evident. For $M_d/m_p = 0.1$, the inclination distribution is bimodal (apart from the peak at $I_d = 90^{\circ}$), and the density profile  resembles a sum of two Gaussians with different widths, indicative  of dynamically cold and hot populations. See the text (Section \ref{sec:cold_hot_rayleigh}) for details.}
% Label
\label{fig:vert_dens_and_histogram}}
\end{figure*}
%%%%%%%%%%%%%%%%%%%%%%%%

In conclusion,  the analysis herein (Section \ref{sec:disc1_warps}) highlights that planetary inferences can vary significantly depending on the disk mass. This underscores the need for caution when interpreting warped debris disks using massless disk models.

%%%%%%%%%%%%%%%%%%%%%%%%%%%%%%%
\subsection{Implications for vertical profiles and aspect ratios}
\label{sec:scale_height_discussion}
%%%%%%%%%%%%%%%%%%%%%%%%%%%%%%%

We next analyze the implications of our results for inclination distributions and vertical density profiles (Section \ref{sec:cold_hot_rayleigh}), followed by a discussion of scale heights  (Section \ref{subsec:aspect_discussion_Sec}).

%%%%%%%%%%%%%%%%%%%%%%%%%%%%%%%%%%%%%%%%%%%%%%%%%%%%%%%%%%%%%%%
\subsubsection{Non-Rayleigh inclinations and non-Gaussian vertical profiles as the norm?}
\label{sec:cold_hot_rayleigh}
%%%%%%%%%%%%%%%%%%%%%%%%%%%%%%%%%%%%%%%%%%%%%%%%%%%%%%%%%%%%%%%

Vertical densities of debris disks are an important observable, as they provide insights into the distributions of planetesimal inclinations and the processes shaping them. Typically, observed debris disks  are modeled using simple Gaussian vertical density profiles \citep[e.g.,][]{grant18, Daley2019, Terrill2023, reasons_matra}. Part of the reasoning for this choice -- beyond its simplicity -- is that it follows from a single Rayleigh distribution of inclinations \citep{matra2019}, which may naturally arise due to self-stirring processes \citep{ida92}. This, however, raises the following  questions: (i) Can the inclination distribution of planetesimals, when perturbed by a planet, be well represented by a single Rayleigh distribution? (ii) If not, how is the resulting vertical density profile best described?

Figure \ref{fig:vert_dens_and_histogram} addresses both of these questions. Namely, panel (A) presents histograms of the mass-weighted inclination distributions of planetesimals at late times in Models \texttt{A}–\texttt{C} (Table \ref{table:models}),  while panel (B) shows the corresponding vertical profiles derived from the density maps of Section \ref{sec:vert_struc_num}. The density profiles are averaged over a $5$ au radial bin centered at a reference projected radius of $R_0 \approx 49.1$ au (equal to the geometric mean of $a\inn$ and $a\res$ in Model \texttt{B}). This is done in the ``disk midplane'', defined as the plane perpendicular to the third principal component of the weighted particle distribution, which serves as a geometric proxy for the disk's axis of symmetry. For $M_d = 0$, the disk midplane coincides with the planetary orbit; for $M_d \neq 0$, it is misaligned by $\approx I_p(0)$. We highlight that similar results are obtained when the disk's midplane is instead defined as perpendicular to its total angular momentum vector (see also Fig. \ref{fig:misalignment_ABC}).

Looking at Figure \ref{fig:vert_dens_and_histogram}, it is evident that the inclinations do not follow a single Rayleigh distribution,  and the vertical density profiles exhibit clear non-Gaussianity. This is so even when the disk mass is ignored. Instead, we find that:

\noindent $\bullet$ In Model \texttt{A} ($M_d =0$), the inclination distribution features a skewed peak at $I_m^{\rm n/disk} = 2 I_p(0)$, beyond which it drops abruptly to zero. Below the peak, the distribution is relatively flat with minor fluctuations. We find that this feature is persistent for $t \gtrsim 0.5 T\sec(a\out)$, akin to the case of massless disks perturbed by \textit{external} planets  \citep{farhat-sefilian-23}.
In terms of vertical density, the profile features a local minimum in the disk midplane, $Z_{\rm{mid}} = 0$,  and is surrounded by two, symmetrical wings that drop off abruptly. The peaks correspond to the maximum vertical extent of planetesimals, $Z_{\rm{peak}} \approx \pm R_0 \tan[I_p(0)]$, defining the ``X''-pattern in Figure \ref{fig:master_Md0_maps}. This behavior has previously been identified in $N$-body simulations of debris disks \citep{Brady2023}, and also underlies the interpretation of the ``dust bands'' in the Zodiacal cloud detected by IRAS \citep{Low1984, Dermott1984, zodiacal97}. We note, however, that the peaks evident in Figure \ref{fig:vert_dens_and_histogram}(B) do not imply distinct dynamically ``cold'' (low-inclination) and ``hot'' (high-inclination) populations; the inclination distribution is not bimodal (Figure \ref{fig:vert_dens_and_histogram}(A)).

\noindent $\bullet$ In Model \texttt{B} ($M_d/m_p = 0.1$), the inclination distribution exhibits three distinct populations: a cold component with a Rayleigh-like peak at $I_d \sim 1^{\circ}$, a hot component with a broader Rayleigh-like peak around $2I_p(0)$, and a sharp peak at $I_d = 90^{\circ}$ driven by the secular resonance. Unlike in massless disks, it develops a high-$I_d$ tail beyond $2 I_p(0)$, reflecting enhanced dynamical excitation. These features persist for $t \gtrsim T_{\dot{\Omega}_p}/\pi$, with the third peak gradually shifting outward to $I_d = 90^{\circ}$ during the early stages.
The corresponding vertical profile exhibits a sharply peaked, symmetric distribution centered at $Z_{\rm{mid}} = 0$, reflecting the presence of a dynamically cold population near the midplane. Notably, a slight dip at $Z_{\rm{mid}} =0$ arises due to the disk midplane being nearly -- but not exactly -- misaligned with the planet by $I_p(0)$. Beyond the central region, the profile falls of steeply and exhibits broader shoulders on both sides -- indicative of dynamically hot population -- that peak at $Z_{\rm{peak}} \approx \pm R_0 \tan[I_m(R_0)]$. Farther from the shoulders, the density tapers gradually up to $\approx Z_{\rm hot}$ (Equation \ref{eq:Z_hot_res}), beyond which it drops sharply. Overall, the profile can be well described as the sum of two Gaussians with different widths or as a Voigt profile -- the convolution of a Gaussian and a Lorentzian.

\noindent $\bullet$  In Model \texttt{C} ($M_d/m_p = 10$), the inclination distribution is sharply peaked at small inclinations and declines rapidly, with over $\approx 78\%$ of planetesimals confined within $I_d \lesssim 1^{\circ}$. The distribution's tail does not follow a simple Rayleigh form, but instead shows a slower decay. We find that the distribution is better approximated either by a superposition of two Rayleigh components with comparable widths or by a Weibull distribution, which interpolates between Rayleigh and exponential forms. 
%%%%%%%
The vertical density profile in this case is sharply peaked and narrowly confined around the midplane, with small but non-negligible wings further out. As in Model \texttt{B}, the profile features a clear dip at $Z_{\rm mid} = 0$, flanked by two prominent peaks. The overall shape can be well approximated by a Lorentzian profile (though fine central features may deviate from this form).

The main takeaway from Figure \ref{fig:vert_dens_and_histogram} is that, regardless of disk mass, non-Rayleigh inclination distributions and non-Gaussian vertical density profiles may not be exceptions  but rather the norm in long-term planet-disk interactions. Additionally, for $M_d/m_p \lesssim 1$, the emergence of both dynamically cold and hot populations is a natural outcome of secular-inclination resonances. This is somewhat akin to  the dual cold and hot population in the Kuiper belt \citep{Brown2001, Kavelaars2009, Gulbis2010},   which is  best modelled by two overlapping Gaussians or a Lorentzian \citep[see][and references therein]{Renu2023} and linked to Neptune's early migration \citep{malhotra95, Han2005_Malhotra}. Our findings do not refute this dynamical interpretation but offer an alternative,  Occam’s razor-style explanation for debris disks -- one that does not rely on planetary migration or primordial (``birth'') inclinations of planetesimals \citep{matra2019}. If planets are present and $M_d/m_p \lesssim 1$, this alone may suffice to explain dual populations, provided the planet--debris disk system is initially misaligned.

Interestingly, to date, only $\beta$ Pic's debris disk has been modeled with non-Gaussian profiles \citep{Apai15, Golimowski_2006, Lagrange_2012}, with the best fit found to be two overlapping Gaussians representing dual populations \citep{matra2019}. Future modelling of debris disks using non-Gaussian vertical profiles could reveal whether this is a unique case or a more widespread feature at the population level. Within our framework, such results could serve as  an indirect proxy for the debris disk's total mass, provided a planet  is detected.  Conversely, assuming a maximum ``allowed'' disk mass, the parameters of an inferred planet may be constrained. Relatedly, we note that fitting observations with single-component Gaussians may bias inferred disk aspect ratios toward larger values, particularly when the true profile is more akin to two Gaussians, as in cases with $M_d/m_p \lesssim 1$ (see also Figure \ref{fig:scaleheight_ABC}). Altogether, these findings emphasize the need for caution when interpreting model fits to observed debris disks -- particularly when inferring the dynamical state of planetesimals and the processes shaping them, including planetary companions.

Finally, we note that the density profiles may be modified by additional effects not considered here, including self-stirring and planetesimal collisions (see also Section \ref{subsec:limit_future_work}). Such processes may alter the relative amplitudes of density peaks (e.g., due to more frequent collisions in denser regions; \citet{nesvold15})  and could even reshape the overall profile. From an observational perspective, on the other hand, limited instrumental resolution would affect the observed surface \textit{brightness}, which in turn may distort the  inferred density profile, particularly on scales smaller than the instrument's point-spread function (PSF). A detailed treatment of such effects is beyond the scope of this work.

%%%%%%%%%%%%%%%%%%%%%%%%%%%%%%%%%%%%%%%%%%%%%%%%%%%%%%%%%%%%%%%
\subsubsection{Aspect ratios: radial dependence and implications}
\label{subsec:aspect_discussion_Sec}
%%%%%%%%%%%%%%%%%%%%%%%%%%%%%%%%%%%%%%%%%%%%%%%%%%%%%%%%%%%%%%%

The disk aspect ratio $\mathcal{H}$ serves as a measure of the dynamical excitation of planetesimals. However, to our knowledge, no analytical formulae exist for $\mathcal{H}$ when a planet perturbs the disk from a few Hill radii away; existing equations apply only to embedded embryos \citep{ida93, matra2019}. We now derive an approximate equation for $\mathcal{H}$.

We assume that planetesimals have evolved significantly such that their relative velocities $v_{\rm rel}$ are isotropic. In this case, 
%%%%%%%%%%%%%%%
\begin{equation}
    v_{\rm rel} =  c I_d v_K  , 
    \label{eq:v_rel_discussion_sec}
\end{equation}
%%%%%%%%%%%%%%%
where $v_K = n_d a_d$ is the Keplerian velocity and  $c$ is a constant \citep{lissauer}. For $c = 1$ with $I_d = \langle I_d(t) \rangle_t$, $v_{\rm rel}$ can be understood as the \textit{mean} relative velocities. This allows us to define the time-averaged vertical height as $\langle \mathfrak{h}_d \rangle = \langle v_{\rm rel}\rangle/n_d$. Accordingly, the mean aspect ratio is:
%%%%%%%%%%%%%%%%%%%%%%
\begin{equation}
  \langle \mathcal{H} \rangle = \frac{\langle \mathfrak{h}_d \rangle}{a_d} = \langle I_d(t) \rangle_t = \frac{1}{2} I_m(a_d) , 
  \label{eq:aspect_ratio_analytical_H}
\end{equation}
%%%%%%%%%%%%%%%%%%%%%%
see also Eq.  (\ref{eq:I_forced_gen}). Note that Equation (\ref{eq:aspect_ratio_analytical_H}) is valid for small values of $I_p(0)$ and $I_d(a_d)$; that is, to the same degree of precision as the disturbing functions in Section \ref{sec:analytical-model}.

According to Equation (\ref{eq:aspect_ratio_analytical_H}), the aspect ratio in the planet-dominated regime, i.e., with $I_m = I_m^{\rm n/disk}$ (Eq. \ref{eq:Im_no_disk}), is:
%%%%%%%%%%%%%%%%%%%%%
\begin{equation}
    \langle \mathcal{H} \rangle^{\rm n/disk} = I_p(0) . 
    \label{eq:H_nodisk_analytical}
\end{equation}
%%%%%%%%%%%%%%%%%%%%%
By contrast, in the disk-dominated regime, i.e., with $I_m = I_m^{\rm disk}$ (Eq. \ref{eq:Im_disk}), we instead have:
%%%%%%%%%%%%%%%%%%%%%
\begin{equation}
\frac{    \langle \mathcal{H} \rangle^{\rm disk} }  { \langle \mathcal{H} \rangle^{\rm n/disk}  }  = 
\frac{I_m^{\rm disk}(a_d)}{I_m^{\rm n/disk}}
=
\bigg|\frac{A_p(a_d)}{A_{d,p}} \bigg| \lesssim 1 ,
    \label{eq:H_disk_analytical22}
\end{equation}
%%%%%%%%%%%%%%%%%%%%%
or equivalently, for the fiducial parameters (Section \ref{sec:modelsystem}), 
%%%%%%%%%%%%%%%%%%%%%
\begin{equation}
{\langle \mathcal{H} \rangle^{\rm disk} }  = 
  9 \times 10^{-2}    ~  \langle \mathcal{H} \rangle^{\rm n/disk} \bigg( \frac{1.85}{\phi_1^c  }\bigg) \bigg( \frac{m_p}{M_d} \bigg)  
     \frac{a_{p,20}^{1/2}}{a_{d,80}^{7/2}}   .
    \label{eq:H_disk_analytical}
\end{equation}
%%%%%%%%%%%%%%%%%%%%%
Equations (\ref{eq:H_nodisk_analytical})--(\ref{eq:H_disk_analytical}) accurately approximate the numerically obtained aspect ratios at $t \gg 1$ in Figure \ref{fig:scaleheight_ABC} (derived using the method of Appendix \ref{app:scale_height}). Naturally, the exception to this occurs interior to the secular resonance (if any), as planetesimals with $I_d(a\res) \rightarrow 90^{\circ}$ spend more time at $R < a\res$; an effect not captured by the analytical approach herein.

More importantly, Equations (\ref{eq:H_nodisk_analytical})--(\ref{eq:H_disk_analytical}) show that, for a wide range of parameters, the disk's vertical aspect ratio $\mathcal{H}$ can be (i) a decreasing function of radius, following $ \propto a_d^{-7/2}$, and  (ii) significantly smaller than that expected from planet-disk interaction models with $M_d = 0$, so that $\langle \mathcal{H} \rangle^{\rm disk} \ll \langle \mathcal{H} \rangle^{\rm n/disk}$. These effects arise in the disk-dominated regime (Section \ref{sec:dyn_reg}), and so may be confined to the outer regions or extend across the disk, depending on system parameters (see also Section \ref{subsec:variations}). Regardless, these findings have two obvious implications which we discuss below.

First, most debris disks appear vertically thin, with aspect ratios typically $\lesssim 0.1$, some as low as $\sim 0.01$, and others reaching up to $\sim 0.30$ \citep{Terrill2023, reasons_matra}. In planet-disk interaction studies, small aspect ratios are generally interpreted as a gravitational tell-tale sign of a near-coplanar planet \citep{pearcewyatt2014, Brady2023}. However, this interpretation may be compromised by neglecting the disk gravity. On the contrary, even in the presence of a highly inclined planet, the disk can maintain a relatively small aspect ratio, or even remain almost razor-thin ($\mathcal{H} \ll 1$) if $M_d/m_p \gtrsim 1$. Thus, caution is needed when interpreting observed scale heights in planet-disk interaction models that assume $M_d =0$ \citep[see also,][]{Pedro2024}.

Second,  the radial profile of a disk's aspect ratio -- whether constant or decreasing with $R$ --  can provide insights into the effects of disk gravity. In principle, this  could constrain the parameters of a potential planet as a function of $M_d$ by making use of Equations (\ref{eq:H_nodisk_analytical})--(\ref{eq:H_disk_analytical}), or conversely, offer a dynamical estimate of the disk mass if a planet is already known. However, by and large for simplicity, most model fits to observations typically assume a Gaussian vertical profile with a distance-independent aspect ratio, $\mathcal{H}(R) =  {\rm const}$ \citep[e.g.,][]{grant18, Daley2019, vizgan2022, Terrill2023, reasons_matra}. As a result, this approach cannot currently be pursued, and future modelling efforts should aim to account and fit for the radial dependence of vertical scale heights.

We note that a recent study by \citet{Han25} explored this avenue using a non-parametric fitting procedure. They found that 8 out of 9 considered disks are consistent with a distance-independent aspect ratio or an $\mathcal{H}(R)$ which increases with radius in their millimeter dust continuum. Interestingly, the debris disk around \HDdec~\citep[][not known to have companions]{MacGregor2019} stands out due to its decreasing radial profile of aspect ratio (see figure 9 in \citet{Han25}). We find that an aspect ratio with a radial profile of $\mathcal{H} \propto R^{-7/2}$ -- as predicted by Equation (\ref{eq:H_disk_analytical}) -- is consistent with the fitting results of \citet{Han25}. However, due to the relatively large uncertainties reported in that study, we refrain from further discussion of this system, instead we simply note that it is a promising target for future modeling and planet-disk interaction studies once higher-resolution data become available.\footnote{\HDdec~is one of the systems that will be targeted by the first ALMA large program dedicated to debris disks, ``The ALMA survey to Resolve exoKuiper belt Substructures'' (ARKS; PI: S. Marino, {ID: 2022.1.00338.L}).} Conversely, for the disks with $\mathcal{H}(R) \approx {\rm const}$ according to \citet{Han25}, one may derive constraints on the maximum $M_d/m_p$ as a function of $a_p$ using Equation (\ref{eq:Mdmp_pl_dom}), i.e., for the system to be in the planet-dominated regime (Section \ref{sec:dyn_reg}).

Finally, we note that the suppression of the disk aspect ratio (equivalently, relative velocities of planetesimals, Eq. \ref{eq:aspect_ratio_analytical_H}) due to the disk gravity may also result in scenarios where the disk remains dynamically cold and unstirred \citep[see also][]{Sefilian2024}. If this occurs only in the outer regions of the disk, then the stirred, observable region would exhibit a constant aspect ratio, as if the disk were massless (Figure \ref{fig:scaleheight_ABC}), which complicates the interpretation. In principle, however, definitive evidence for an aspect ratio consistent with the predictions herein could be translated to constraints on planetary parameters and disk masses.

%%%%%%%%%%%%%%%%%%%%%%%%%%%%%%%
\subsection{Implications for disk mass estimates}
\label{sec:debrisdiskmass_problem_sec}
%%%%%%%%%%%%%%%%%%%%%%%%%%%%%%%

A key takeaway from this study is that a non-zero disk mass can significantly affect the vertical structure of debris disks, with important repercussions for interpreting observed disks and inferring the masses and orbits of yet-undetected planets. Crucially, the resulting observational signatures are a function of the total disk mass, $M_d$, a parameter that remains poorly constrained by conventional techniques. The challenge partly stems from the fact that converting observed thermal or scattered-light emission into total mass estimates relies heavily on collisional cascade models \citep{dohnanyi1969, dd2003, wyattdent2002, krivovwyatt21}.
%%%%%%
These models are sensitive to several poorly constrained parameters, including the size distribution and maximum size of debris, as well as their internal strength and velocity dispersion \citep[][and references therein]{wyatt08collisionsreview}. Indeed, depending on assumptions, total disk mass estimates can reach up to $\sim 10^{3}$--$10^{4} M_{\earth}$, values that appear to exceed the total solid mass available in typical protoplanetary disks \citep{Williams2011}, a discrepancy known as the ``debris disk mass problem'' \citep{krivovwyatt21}. Attempts to resolve this issue include improving collisional models and related physics \citep[e.g.,][]{krivov2006, Thebault2007col, lohne2008, Pan2012, krivovwyatt21}. In parallel, however, studying the structures of debris disks shaped by planet--\textit{massive} disk interactions offers an alternative, microphysics-independent pathway to constraining $M_d$. Depending on the orbit and mass of the planet discovered (or ruled out), this dynamical approach can help bridge the gap between observed dust populations and the total mass budget in debris disks. This further underscores the importance of accounting for disk gravity in studies of planet-debris disk interactions -- an aspect that is often overlooked (Section \ref{sec:intro}).

%%%%%%%%%%%%%%%%%%%%%%%%%%%%%%%
\subsection{Limitations and Future Work} 
\label{subsec:limit_future_work}
%%%%%%%%%%%%%%%%%%%%%%%%%%%%%%%

While this work provides important insights into the role of disk gravity in planet-debris disk interactions, it is not free of limitations and several factors may influence the results.

First, we assumed gas-free debris disks, whereas several -- though not all -- are known to contain significant amounts of gas \citep[e.g.,][]{hughes2018review}. In such cases, gas drag could  affect e.g. the disk's scale height, particularly for small and even mm-sized dust grains \citep[][]{johan22}. Second, by considering only second-order inclination terms in the disturbing function (Section \ref{sec:dist_fn_d_p}), we limited the reliability of the results at large inclinations. In reality, higher-order terms could de-tune the secular resonance by modifying the precession rate at high inclinations, limiting the growth of inclinations at and around that location \citep[as is the case for secular-eccentricity resonances;][]{Malhotra1998}. 
Third, the assumption of a circular planet  prevented us from exploring the coupled eccentricity-inclination dynamics \citep{pearcewyatt2014, ST19, farhat-sefilian-23}, including the possibility of  von Zeipel-Lidov-Kozai dynamics \citep{Terquem2010, Jean2013, Naoz2016, nesvold17, Ito19, Pedro2024}. 
%%%%
One intriguing possibility for future investigation is the establishment of both secular eccentricity and inclination resonances within the disk \citep{Paper1, Paper2}.
%%%%
Another assumption that can be relaxed is the restriction to single-planet systems: the presence of additional planets can alter, for example, the location of resonances, and potentially even their number \citep{wyattetal99, Dong2020}.
%%%%
Furthermore, we ignored self-stirring processes \citep{ida93, Kenyon2008}, which could thicken the disk over time. 
Finally, in line with most debris disk studies, we assumed initially razor-thin disks. This, however, may not necessarily be the case, and the dynamics of an initially puffed up disk and a zero-inclination planet should be considered \citep[][]{marco2023}.

The above  points certainly warrant further investigations. However, the major limitation of this work is the omission of the full effects of the disk gravity. Namely, for the sake of analytical progress, only the axisymmetric component of the disk's back-reaction was considered, while both its non-axisymmetric counterpart and the \textit{self-gravity} were neglected (Sections \ref{sec:analytical-model} and \ref{sec:analytical_analysis}). We shall address these limitations in an upcoming work. For now,  we note that the framework outlined in Section \ref{sec:analytical-model} is well-suited for studying the effects of the disk's full back-reacting potential (but not the disk self-gravity).
%%%% 
%%%% 

Our preliminary numerical calculations using the general form of Equation (\ref{eq:matrix_eom}) show that the primary effect of the disk's non-axisymmetric torque on the planet is to mediate the exchange of angular momentum with the planet. Indeed, as we show in Appendix \ref{app:res_friction}, provided the system features a secular resonance, the interactions may lead to the exponential damping of the planetary inclination due to what is known as \textit{resonant friction} or \textit{secular resonant damping} \citep{tre98, wardhahn2003, hahn07bending}. This happens  such that $I_p(t) \propto \mathrm{exp}(-D_{\rm inc}t/2)$; see Equations (\ref{eq:dIpdt_resfric})--(\ref{eq:Ip_decay_res_fric_sol}).  In such cases, the forced planetesimal inclinations will no longer be time-independent (contrary to this work), and will also decay over time. Indeed, if, at best, $D_{\rm inc} t \gg 1$ (Equation \ref{eq:D_inc_app_gen}), planetesimal inclinations will undergo damped sinusoidal oscillations that asymptotically converge to their free values, i.e., $I_d(t) \rightarrow I_m/2 = \langle I_d(t) \rangle_t$ given by Equation (\ref{eq:I_forced_gen}). As a result, the distribution of $\Delta \Omega(a_d)$ may eventually span the full $[-\pi, \pi]$ range in both planet- and disk-dominated regimes.\footnote{A similar behavior, though in the context of eccentricity dynamics, has been studied by \citet{Paper2}.}  In such cases, for instance, the convergence of $\mathcal{H}(R) \rightarrow \langle I_{d}(t) \rangle_t$ would happen faster, damping out the oscillatory behavior evident in Figure \ref{fig:scaleheight_ABC}. Relatedly, the disk's non-axisymmetric perturbations may also force the planetary nodal precession rate to deviate from that given by Equation (\ref{eq:planetary_node_rate}); our preliminary calculations suggest that the deviation could be by a factor of a few, thus potentially affecting inferences of, e.g., disk masses by a similar factor. Nevertheless, we find that the dynamical regimes and solutions analyzed here  remain  generally valid, giving us confidence that the  simplified analytical framework -- together with Appendix \ref{app:res_friction} --  adequately capture the basic effects of a fully back-reacting disk.

This said, we note that the influence of disk self-gravity remains an open question requiring further investigation. \citet{silsbeekepler} investigated the in-plane dynamics of planetesimals around eccentric stellar binaries by modeling the massive protoplanetary disk as a rigid, precessing slab and demonstrated the existence of up to four distinct dynamical regimes.  These regimes arise depending on whether the disk or the inner companion dominates the precession rates and eccentricity excitation of planetesimals, with all combinations theoretically feasible \citep[see also][]{ST19}. Analogous dynamical complexity for inclinations may also manifest in gravitating debris disk--planet systems. We hope that our analytical results will provide a foundation and starting point for future investigations of these phenomena, whether through direct $N$-body simulations or secular ``$N$-ring" codes \citep[e.g.,][]{hahn2003,  Touma2009, Paper2}.

Finally, we note that the assumption of collisionless planetesimals warrants future consideration. Collisional grinding would naturally deplete the disk mass over time \citep{wyatt08collisionsreview}, attenuating the effects of disk gravity.  This could have important consequences, as it would, for example, open the door to sweeping the secular resonance \citep{hep80, ward81scan}. Within our  setup, as $M_d$ decreases, the resonance would sweep outward through the disk (Figure \ref{fig:I_forced_a}), dynamical effects of which would depend on the mass depletion details and be encoded in the resulting disk structure.

Relatedly, in the disk-dominated regime (Section \ref{sec:dyn_reg}), the suppressed planetesimal inclinations could hinder the planet-induced secular stirring expected based on massless-disk models \citep{Mustill2009}. This is because the resulting relative velocities of planetesimals (Equation \ref{eq:v_rel_discussion_sec}) may remain below the $\sim 50-100$ m/s threshold required for destructive collisions \citep{wyatt08collisionsreview}.
%%%%%
This effect has been demonstrated in the context of debris disks perturbed by coplanar, eccentric planets \citep{Sefilian2024}.
%%%%%
Such effects could be considered in tandem with processes affecting small ($\lesssim \mu m$-sized) particles, such as radiation pressure, which may further affect the disk's vertical structure \citep{thebault2009}, introducing dependencies on the observational wavelength. Clearly, there is still much to hammer out when it comes to gravitating planet--debris disk systems!

%%%%%%%%%%%%%%%%%%%%%%%%%%%%%%%%%%%%%%%%
\section{Summary}
\label{sec:summary}
%%%%%%%%%%%%%%%%%%%%%%%%%%%%%%%%%%%%%%%%

In this work, we investigated the long-term inclination dynamics of debris disks in single-planet systems, with a focus on the often-overlooked role of the disk's gravity. To this end, we developed a simplified analytical model based on the classical Laplace--Lagrange secular theory, valid to second order in inclinations. The model captures perturbations from an inner planet on an inclined (but circular) orbit and includes the disk's gravitational back-reaction on the planet, while neglecting the disk's self-gravity. To flesh out the fundamental physical effects, we focused on the axisymmetric component of the disk's potential. Despite its simplicity, this study represents -- to the best of our knowledge -- the first detailed analytical investigation of how a debris disk's gravity affects its vertical structure. In particular, we examined how the disk-to-planet mass ratio $M_d/m_p$, among other parameters, governs the orbital evolution of planetesimals and the resulting large-scale disk morphology, such as  warps and scale heights. Our main findings are summarized below:

%%%%%%%%%%%%%%%%
%%%%%%%%%%%%%%%%
\begin{enumerate}

\item We identified two distinct regimes of planetesimal dynamics:  planet-dominated and disk-dominated. These regimes arise depending on which agent -- planet or disk -- dominates the nodal precession rate of the planetesimals relative to the planet. The transition between them is marked by a secular-inclination resonance, with their presence and extent determined primarily by  $M_d/m_p$. 

\item The planet-dominated regime applies across the entire disk only when $M_d/m_p \ll 1$. Initially, the disk undergoes a transient phase during which a warp propagates outward. Ultimately, the disk settles into a symmetrical, boxy structure centered on the planet's orbital plane, with  a characteristic ``X''-pattern when viewed \mbox{edge-on}.

\item The disk-dominated regime applies across the entire disk when $M_d/m_p \gtrsim 1$. The disk remains dynamically rigid, staying vertically thin with  suppressed planetesimal inclinations -- often over an order of magnitude smaller than in models assuming $M_d =0$. The disk-induced planetary nodal precession prevents the planet from torquing the disk into alignment with its orbital plane.

\item A secular-inclination resonance is established within the disk when $M_d/m_p \lesssim 1$. The disk develops an outward-propagating warp, initially launched at its inner edge, which ultimately stalls at the resonance location as a long-lived feature. With its inner regions aligned with the planetary orbital plane and its outer regions remaining flat, the disk exhibits a broken, curly ``X''-pattern.

\item We present simple, disk-mass-dependent equations to infer the properties of a yet-undetected planet capable of producing an observed warp. We show that models with $M_d = 0$ underestimate the range of planetary mass-semimajor axis combinations for realistic disks with non-zero masses. As an example, we applied our results to \HDone, showing that a $\gtrsim 0.1 M_J$ planet at $\lesssim 16$ au can sculpt the observed warp, provided that $M_d \lesssim 300 M_{\earth}$. 

\item Our results can be used to constrain the total masses of warped debris disks. For example, we show that the well-known warp in the $\beta$ Pic debris disk is consistent with being carved by $\beta$ Pic b, provided that  the disk mass is $M_d \lesssim 10^{3} M_{\earth}$.

\item The disk's aspect ratio is primarily dictated by $M_d/m_p$ and the planetary inclination $I_p(0)$. In planet-dominated regions, the aspect ratio is distance-independent at late times,  $\mathcal{H}(R) \approx I_p(0)$, while in disk-dominated regions, it sharply decreases with distance with a strongly suppressed magnitude, $\mathcal{H}(R)/I_p(0) \propto R^{-7/2} \ll 1$.

\item Regardless of $M_d$, the inclination distribution of planetesimals does not follow a Rayleigh profile, and the vertical density profile differs from a Gaussian. In the presence of a secular resonance, the disk harbors both dynamically cold and hot populations, rendering the inclination distribution bimodal and the vertical density profile akin to the sum of two Gaussians with different widths.

\end{enumerate}
%%%%%%%%%%%%%%%%
%%%%%%%%%%%%%%%%

More generally, the current state of affairs in the modelling of planet--debris disk interactions warrants a cautious approach. Our results show that the disk's gravity -- often neglected without justification -- plays an important dynamical role,  even when the disk is less massive than the planet. At best, this oversight can compromise the accuracy of inferred planetary parameters; at worst, it leads to misleading interpretations of planetary system formation and evolution. A similar oversight persists in observational modelling, where Gaussian assumptions with constant $\mathcal{H}(R)$ are common for vertical profiles; we advocate instead for fitting non-Gaussian profiles and allowing for radial variations in aspect ratios.

In closing, we note that this work serves as a primer, offering a simplified analytical framework for exploring the role of disk gravity in misaligned planet--debris disk systems. Acknowledging its strengths and limitations, it provides a useful guide for identifying which planet--disc parameters and dynamical regimes merit the computational investment of more comprehensive studies, whether via N-body simulations or specialized secular ``N-ring'' approaches \citep[e.g.,][]{hahn2003, Touma2009, Paper2}.

%%%%%%%%%%%%%%%%%%%%%%%%%%%%%%%%%%%%%%%
\section*{Acknowledgments}
%%%%%%%%%%%%%%%%%%%%%%%%%%%%%%%%%%%%%%%
\noindent  The authors thank Joshua Lovell and Marija Jankovic for their engaging discussions, Natasha Simonian  for her assistance in enhancing the artistic quality of select figures, and the referee, Alexander Mustill, for a prompt report that improved the clarity of the manuscript. 
% Antranik
A.A.S. is supported by the Heising-Simons Foundation through a 51 Pegasi b Fellowship.
% Kaitlin
The results reported herein benefited from collaborations and/or information exchange within NASA’s
Nexus for Exoplanet System Science (NExSS) research coordination network sponsored by NASA’s
Science Mission Directorate and project “Alien Earths” funded under Agreement No. 80NSSC21K0593. K.M.K. acknowledges support under NASA Theoretical and Computational Astrophysical Networks (TCAN) via grant No. 80NSSC21K0497 and NASA XRP via grant No. 80NSSC24K0163.
% Renu 
%%%%%%%RM acknowledges the ``Alien Earths'’ program (supported by the National Aeronautics and Space Administration under Agreement No. 80NSSC21K0593).
%Virginie
V.F.-G. acknowledges funding from the National Aeronautics and Space Administration through the Exoplanet Research Program under Grant No 80NSSC23K0288 (PI: V. Faramaz).

%%%%%%%%%%%%%%%%%%%%%%%%%%%%%%%%%%%%%%%%%
\section*{Data availability}
%%%%%%%%%%%%%%%%%%%%%%%%%%%%%%%%%%%%%%%%%
\noindent The data that support the findings of this study are available from the corresponding author upon reasonable request. To minimize visual distortion, contour plots in this study employ the perceptually-uniform colormap \textit{batlow} \citep{crameri}.

%%%%%%%%%%%%%%%%%   REFERENCES      %%%%%%%%%%%%%%%%%%%%
\bibliography{Sefilian2025Vertical}{}
\bibliographystyle{mnras}
%%%%%%%%%%%%%%%%%%%%%%%%%%%%%%%%%%%%%%%%%%%%%%%%%%%%%%%%

%%%%%%%%%%%%%%%%%%%%%%%%%%%%%%%%%%%%%%%%
%%%%%%%%%%%%%% APPENDIX %%%%%%%%%%%%%%%%
%%%%%%%%%%%%%%%%%%%%%%%%%%%%%%%%%%%%%%%%
\appendix

%%%%%%%%%%%%%%%%%%%%%%
\section{The Disk's Back-reaction on the Planet: The Continuum Limit}
\label{sec:app_A}
%%%%%%%%%%%%%%%%%%%%%%

Here, we calculate the disturbing function $R_{d,p}$ of an inclined planet due to an external, massive debris disk, modelling the latter as a continuous structure (rather than a collection of discrete bodies as in Section \ref{sec:modelsystem}). To achieve this, we replace the summations in Equation (\ref{eq:R_disc_planet}) with integrations. In doing so, individual planetesimal masses $m_j$ are replaced by $dm_d = 2\pi a_d \Sigma_d^{t=0}(a_d) da_d$ \citep{statler99, irina18}, where $\Sigma_d^{t=0}(a_d) \propto a_d^{-p}$ is given by Equation (\ref{eq:Sigma_d}). For completeness and future use, we also assume that the disk's inclination profile follows a power-law with an index $q$:
%%%%%%%%%%%
\begin{equation}
    I_d(a_d) = I_0 \bigg( \frac{a\out}{a_d} \bigg)^{q} ~~~ {\rm for} ~~ a\inn \leq a_d \leq a\out ,  
\end{equation}
%%%%%%%%%%%%%%%
and that the disk elements share a common longitude of ascending node with  $\Omega_d(a_d) = {\rm const}$. This approach and its assumptions are analogous to those in \citet{Paper1}, originally applied to coplanar setups (see their Appendix A). 

With these ingredients, and after some straightforward algebra,  $R_{d,p}$ can be expressed in the following form,
%%%%%%%%%%%%%%%%%%%%%%%%
\begin{equation}
    R_{d,p} = n_p a_p^2 \bigg[
    \frac{1}{2} A_{d,p} I_p^2 + B_{d,p} I_p \cos( \Omega_p - \Omega_d) 
    \bigg], 
\end{equation}
%%%%%%%%%%%%%%%%%%%%%%%%
where $A_{d,p}$ and $B_{d,p}$ are given by: 
%%%%%%%%%%%%%%%%%%%%%%%%
\begin{eqnarray}
A_{d,p} &=& - 2 \pi \frac{G \Sigma_d^{t=0}(a\inn)}{n_p a_p} \frac{a\inn}{a_p} \phi_1 < 0 , 
\label{eq:Adp_contn}
\\
B_{d,p} &=&  \pi \frac{G \Sigma_d^{t=0}(a\inn)}{n_p a_p} \frac{a\inn}{a_p} I_d(a\inn) \phi_2  > 0. 
\label{eq:Bdp_contn}
\end{eqnarray}
%%%%%%%%%%%%%%%%%%%%%%%%
The coefficients $\phi_1$ and $\phi_2$ appearing in Equations (\ref{eq:Adp_contn}) and (\ref{eq:Bdp_contn}), respectively, are given by (recall that $\delta \equiv a\out/a\inn$): 
%%%%%%%%%%%%%%%%%%%%
\begin{eqnarray}
    \phi_1 &=& 
    \frac{1}{4} \bigg( \frac{a_p}{a\inn} \bigg)^{1-p} \int_{a_p/a\out}^{a_p/a\inn} \alpha^{p-1} b_{3/2}^{(1)}(\alpha)  d\alpha  ,  
    \nonumber
    \\
    & & \qquad\qquad\qquad = \frac{3}{4} \bigg( \frac{a_p}{a\inn} \bigg)^2 \frac{1-\delta^{-1-p}}{p+1} \phi_1^c , 
    \label{eq:phi1_app}
    \\
    \phi_2 &=&    \frac{1}{2} \bigg( \frac{a_p}{a\inn} \bigg)^{1-p-q} \int_{a_p/a\out}^{a_p/a\inn} \alpha^{p+q-1} b_{3/2}^{(1)}(\alpha) d\alpha , 
    \nonumber
    \\
    & & \qquad\qquad\qquad =   \frac{3}{2} \bigg( \frac{a_p}{a\inn} \bigg)^2 \frac{1-\delta^{-p-q-1}}{p+q+1}  \phi_2^c  . 
    \label{eq:phi2_app}
\end{eqnarray}
%%%%%%%%%%%%%%%%%%%%
At this point, we note that by substituting  Equation (\ref{eq:phi1_app}) into Equation (\ref{eq:Adp_contn}), one arrives at Equation (\ref{eq:Adp_Bdp_approx}).

In Equations (\ref{eq:phi1_app}) and (\ref{eq:phi2_app}),  the coefficients $\phi_k^c$ ($k=1, 2$) are ``corrective'' factors that account for the deviation of the Laplace coefficients from their first-order approximations, which were used to derive the second lines in those equations. The expression of $\phi_1^c$ is given by Equation (A7) of \citet{Paper1} and is not reiterated here (more on this below). The coefficient $\phi_2^c$, on the other hand, is given by:
%%%%%%%%%%%%%%%%%%%%
\begin{equation}
    \phi_2^c =  \frac{1}{3} \frac{p+q+1}{1-\delta^{-p-q-1}} \frac{a\inn}{a_p} \int_{1}^{\delta} u^{-p-q-1} b_{3/2}^{(1)}\bigg(\frac{1}{u}\frac{a_p}{a\inn}\bigg) du . 
    \label{eq:phi2_c_app}
\end{equation}
%%%%%%%%%%%%%%%%%%%%
Interestingly, we note that the expression of $\phi_2^c$ reduces to that of $\phi_1^c$ upon replacing $p+q$ by $p$ in Equation (\ref{eq:phi2_c_app}). Accordingly, the function $\phi_2^c$ has the same behavior as $\phi_1^c$, which, as shown in \citet{Paper1}, depends strongly on $a_p/a\inn$, but only weakly on $p$ and $\delta$ \citep[see also][]{petrovich19}. Thus, regardless of ($p, q, \delta$), one has $\phi_1^c(p, \delta) \approx \phi_2^c(p+q, \delta)$, such that both functions approach unity as $a_p/a\inn \rightarrow 0$, and diverge as $a_p/a\inn \rightarrow 1$;  see Figure 13(A) in \citet{Paper1}. To leading order, the divergence occurs such that $\phi_k^c \sim (1-a_p/a\inn)^{-1}$, obtained by noting that $b_{3/2}^{(1)}(\alpha) \sim (1-\alpha)^{-2}$ when  $\alpha \rightarrow 1$.

%%%%%%%%%%%%%%%%%%%%%%%%%%%%%%%%%%
\section{Damping of planetary inclinations due to (secular) resonant friction}
\label{app:res_friction}
%%%%%%%%%%%%%%%%%%%%%%%%%%%%%%%%%%

We now derive an analytical expression for the evolution of the planetary inclination resulting from its secular interactions with a massive external disk. Unlike in Section \ref{sec:analytical_analysis}, we include the disk's full back-reacting potential, not just its axisymmetric component. We show that this may cause the planetary inclination to damp exponentially due to a process known as \textit{resonant friction} \citep[][and references therein]{tre98, wardhahn2003, hahn07bending, Paper2}. A similar calculation has been carried out for coplanar, eccentricity dynamics by \citet{tre98}; here, we adapt their procedure to address the inclination dynamics.

To this end, we first consider the equations of motion given by Equations (\ref{eq:zdot_j}) and (\ref{eq:zdot_p}) and look for precessing modal solutions of the form ${\zeta}_p(t) , {\zeta}_j(t) \propto {\rm exp}(i \Omega_c t )$, where $\Omega_c = \beta + i \gamma$ is some complex quantity ($\beta, \gamma \in \mathbb{R}$). Doing so, one arrives at the following dispersion relation: 
%%%%%%%%%%%%%%%%%%%%%
\begin{equation}
    A_{d,p} - \Omega_c = \sum_{j=1}^{N} \frac{ \delta B_{j,p} B_p(a_j) }{ A_p(a_j) - \Omega_c} . 
    \label{eq:disp_rel_tr}
\end{equation}
%%%%%%%%%%%%%%%%%%%%%
We next consider the continuum limit ($N \rightarrow \infty$). Replacing the sum over $N$ in Equation (\ref{eq:disp_rel_tr}) by an integral and using the fact that $dm_d/da_d = 2 \pi a_d \Sigma_d^{t=0}(a_d)$ \citep{statler99}, we arrive at:
%%%%%%%%%%%%%%%%%%%%%
\begin{equation}
    A_{d,p} - \Omega_c = 
    \frac{\pi}{8} n_p \frac{m_p}{M_c}
    \int_{L}  F(a_d) da_d  , 
    \label{eq:disp_rel_tr_cont}
\end{equation}
%%%%%%%%%%%%%%%%%%%%%
where, for brevity, we have defined $\alpha \equiv a_p/a_d$ and 
%%%%%%%%%%%%%%%%%%%%%
\begin{equation}
    F(a_d) = \frac{a_d \Sigma_d^{t=0}(a_d)}{M_c}
    \frac{n_d(a_d) \alpha^3}{A_p(a_d)-\Omega_c} \bigg[ b_{3/2}^{(1)}(\alpha) \bigg]^2 .
\end{equation}
%%%%%%%%%%%%%%%%%%%%%
In Equation (\ref{eq:disp_rel_tr_cont}), the integral with subscript ``L'' indicates a Landau integral \citep[as is common in plasma physics and galactic dynamics;][]{landau46, lyndenbell62}, introduced to account for the singularity at $A_p(a_d) = \Omega_c$ such that the integral for ${\rm Im}[\Omega_c] = \gamma > 0$ is the continuation of the one for ${\rm Im}[\Omega_c] = \gamma < 0$. Assuming that the planet-planetesimal coupling is weak enough so that ${\rm Re}[\Omega_c] = \beta \approx A_{d,p}$, i.e., $d\Omega_p/dt \approx A_{d,p}$ (Eq. \ref{eq:Omegap_sol}),  it follows that
%%%%%%%%%%%%%%%%%%%%%
\begin{equation}
 \frac{dI_p^2(t)}{dt} 
 = - 2 \gamma I_p^2(t)
= I_p^2(t)
\frac{\pi}{4} n_p \frac{m_p}{M_c} \cdot {\rm Im}  \bigg[ \int_{L} F(a_d) da_d \bigg] .
\label{eq:int_mda}
\end{equation}
%%%%%%%%%%%%%%%%%%%%%
If we further assume that ${\rm Im}[\Omega_c] = \gamma$ is small and negative (i.e., weak coupling), then using the relationship $\lim_{\epsilon \to 0 } {\epsilon}/{[\epsilon^2 + x^2]} =    \pi {\rm sgn}(\epsilon)  \delta(x)$ \citep{Gradshteyn}, where $\delta(x)$ is the Dirac-delta function, we may write 
%%%%%%%%%%%%%%%%%%%%%
\begin{eqnarray}
    {\rm Im}\bigg[ \frac{1}{A_p(a_d)-\Omega_c} \bigg] 
    &=& \frac{\gamma}{ (A_p - A_{d,p})^2 + \gamma^2},  
    \nonumber
    \\ 
    &\approx& -\pi \delta\bigg(A_p(a_d)  - A_{d,p}\bigg) .    
    \label{eq:Im_Omega_c_limit}
\end{eqnarray}
%%%%%%%%%%%%%%%%%%%%%
Inserting Equation (\ref{eq:Im_Omega_c_limit}) into Equation (\ref{eq:int_mda}) and noting that the Dirac-delta function in Equation (\ref{eq:Im_Omega_c_limit}) corresponds to the secular resonance condition (Section \ref{sec:analytical_analysis}), such that 
%%%%%%%%%%%%%%%%%%%%%
\begin{equation}
    \delta \bigg(A_p(a_d)-A_{d,p}\bigg) = \frac{\delta(a_d-a\res)}{ |dA_p/da_d|_{a_d = a\res} }, 
    \label{eq:delta_fn_dirac}
\end{equation}
%%%%%%%%%%%%%%%%%%%%%
we finally arrive at the following expression: 
%%%%%%%%%%%%%%%%%%%%%
\begin{equation}
    \frac{1}{I_p^2(t)} \frac{dI_p^2(t)}{dt} = \frac{2}{I_p(t)} \frac{d I_p(t)}{dt} = 
    - D_{\rm inc} < 0 . 
    \label{eq:dIpdt_resfric}
\end{equation}
%%%%%%%%%%%%%%%%%%%%%
Note that the analytical continuation required by Landau's prescription ensures that Equation (\ref{eq:dIpdt_resfric}) is valid for either sign of ${\rm Im}[\Omega_c] = \gamma \approx 0$. In Equation (\ref{eq:dIpdt_resfric}), $D_{\rm inc} > 0 $ is a positive coefficient given by:
%%%%%%%%%%%%%%%%%%%%%
\begin{eqnarray}
    D_{\rm inc} &=& \frac{\pi^2}{4} \frac{m_p}{M_c} \frac{a\res^2 \Sigma_d^{t=0}(a\res)}{M_c} 
    \alpha\res^3 
    \label{eq:D_inc_app_gen}
    \\
  & & \qquad \qquad  \times
  \big[b_{3/2}^{(1)}(\alpha\res) \big]^2 \frac{n_p n_d(a\res)}{|dA_p/d\log a_d|_{a_d = a\res}}, 
  \nonumber 
\end{eqnarray}
%%%%%%%%%%%%%%%%%%%%%
where $\alpha\res \equiv a_p/a\res$. Thus, the planetary inclination damps exponentially at a rate specified by $D_{\rm inc}$ following
%%%%%%%%%%%%%%%%%%%%%
\begin{equation}
    I_p(t) = I_p(0) \cdot {\rm exp}( - D_{\rm inc} t / 2 ). 
    \label{eq:Ip_decay_res_fric_sol}
\end{equation}
%%%%%%%%%%%%%%%%%%%%%
This completes our derivation. In physical terms, the fact that  $D_{\rm inc}$ (Eq. \ref{eq:D_inc_app_gen}) is evaluated at the resonance location $a\res$ implies that the damping of the planetary inclination is primarily dictated by the torque exerted on the planet by the resonant planetesimals. This makes intuitive sense, as the planetesimals at (and around) the resonance remain coupled to the planet such that $\Delta \Omega(t) \approx \pi/2$ (see, e.g., Section \ref{subsec:orbit-evol-massive-less-disks} and Figure \ref{fig:master_orb_el}(b)), producing a time-independent torque.

In closing, we point out that the damping rate given by Equation (\ref{eq:D_inc_app_gen}) matches that derived by \citet[][see their equation (17)]{wardhahn2003} using an alternative approach based on wave solutions in the tight-winding approximation. This is not surprising, as in the secular limit (i.e., when the pattern speeds are much smaller than the planetesimal's mean motion), these results shall naturally follow from Toomre's stellar dispersion relation \citep{Toomre69}. Nevertheless, it is academically satisfying to see them derived from Lagrange's planetary equations, as demonstrated here. Finally, we note that secular resonant friction arises specifically from orbit-averaged torques and can be understood as a specialized form of the well-known dynamical friction process \citep{BT}. Interestingly, a similar effect has been identified in the context of intermediate-mass black holes embedded in stellar disks around supermassive black holes \citep{sz21, Ginat2023}.

%%%%%%%%%%%%%%%%%%%%%%%%%%%%%%%%%%
\section{Laplace Coefficients}
\label{app:laplace_coeff_geom}
%%%%%%%%%%%%%%%%%%%%%%%%%%%%%%%%%%

The standard Laplace coefficients  can be efficiently computed using their relations to, e.g., Gauss hypergeometric functions, avoiding direct numerical integrations. Indeed, following \citet{Izsak63}, $b_s^{(m)}(\alpha)$ of Equation (\ref{eq:bsm_normal}) reads as:
%%%%%%%%%%%%%%%%%%%%%%%%
\begin{equation}
    b_s^{(m)}(\alpha) = \frac{2 \alpha^m \Gamma(s+m)}{\Gamma(s) \Gamma(m+1)} F(s, m+s; m+1; \alpha^2) , 
    \label{eq:bsm_gauss_geom}
\end{equation}
%%%%%%%%%%%%%%%%%%%%%%%%%%
where $\Gamma(x)$ is the gamma function (i.e., the analogue of factorial for non-integers), and $F(a,b;c;z)$ is the Gauss hypergeometric function which can be calculated using existing effective algorithms \citep{PressNR}. To lowest order, the function $F$ can be approximated as:
%%%%%%%%%%%%%%%%%%
\begin{equation}
    F(s, m+s; m+1; \alpha^2) = 1  + s \frac{m+s}{m+1} \alpha^2 + \mathcal{O}(\alpha^4) .
\end{equation}
%%%%%%%%%%%%%%%%%%
Equation (\ref{eq:bsm_gauss_geom}) is valid for $0 \leq \alpha \leq 1$, and we note that
%%%%%%%%%%%%%%%%%%%%%%%%%%
\begin{equation}
b_{s}^{(m)}(\alpha^{-1})= \alpha^{2s} b_{s}^{(m)}(\alpha) 
\end{equation}
%%%%%%%%%%%%%%%%%%%%%%%%%%
by definition; see Equation (\ref{eq:bsm_normal}).

%%%%%%%%%%%%%%%%%%%%%%%%%%%%%%%%%%
\section{Tools for analyzing disk structures}
\label{app:disk_sigma_analysis}
%%%%%%%%%%%%%%%%%%%%%%%%%%%%%%%%%%

In this Appendix, we first briefly outline the process of converting the orbital element distribution of planetesimals into maps of disk surface density (App. \ref{app:map_construction}), followed by technical details on how we measure disk scale heights (App. \ref{app:scale_height}) and quantify planet-disk (mis)alignment (App. \ref{app:ang_mom}).

%%%%%%%%%%%%%%%%%%%%%
\subsection{Construction of Disk Surface Density Maps}
\label{app:map_construction}
%%%%%%%%%%%%%%%%%%%%%

We begin by outlining the procedure used to construct maps of disk surface density.  The procedure is largely adapted from the one described in \citet[][]{Paper1}, but it is generalized to account for orbital inclinations and viewing angles. For similar calculations incorporating sky-plane viewing angles, see \citet{farhat-sefilian-23}.

To begin with, at a given time $t$, the orbit of each planetesimal (labeled $j = 1,.., N$) is populated with $N_{np}$ new particles. Each new particle is assigned the same orbital elements as its parent planetesimal $j$ but with a mass $m_j/N_{np}$ and  a mean anomaly $l$ randomly sampled from a uniform distribution between  $0$ and $2\pi$. This approach effectively imitates Gauss' orbit-averaging technique \citep[valid for secular dynamics;][]{tremaine2023}, while simultaneously enhancing the resolution to $N_t \equiv N \times N_{np} \gg 1$ planetesimals.

Next, the instantaneous position ($X, Y, Z$) of each planetesimal  is calculated in a Cartesian frame centered at the host star. This is done using the standard  formulae:
%%%%%%%%%%%%%%%%%%%%%%
\begin{equation}
\begin{pmatrix}
X \\ Y \\ Z 
\end{pmatrix}
= r
\begin{pmatrix}
\cos \Omega \cos(\omega+f) - \sin \Omega \sin(\omega+f) \cos I 
\\
\sin \Omega \cos(\omega+f) + \cos \Omega \sin(\omega+f) \cos  I 
\\
\sin(\omega + f) \sin I
\end{pmatrix}    
,
\label{eq:XYZ-planetesimals}
\end{equation}
%%%%%%%%%%%%%%%%%%%%%%%%%%%%
where, as usual, $r = a (1-e\cos E)$ is the radial distance from the origin, and $E$ and $f$ are the eccentric and true anomalies, respectively. Note that for circular but inclined orbits (as in this work), one has $r = a$,   $f = E = l $,  and,  without loss of generality, the argument of periapsis can be set to $\omega = 0$.

Last but not least, to obtain the  disk surface density distribution, $\Sigma_d(t)$, at a given time: (i) the instantaneous positions of all particles  are binned onto a 3-D grid with the desired resolution,  (ii) the masses within each grid cell are aggregated, and (iii) the resulting values are projected onto the desired plane of view -- i.e., $(X,Y)$, $(X,Z)$ or $(Y,Z)$ -- by summing the contributions along the orthogonal axis and normalizing the result by the grid cell area in the projected plane.

In this work, unless otherwise stated, we adopt $N_{np} = 10^4$ and a grid defined by at least $400^3$ equal-sized cubic cells. In closing, we note that the resulting surface density maps can also be convolved with a Gaussian filter of a specified  size and shape, if desired. 
%%%%
The same process can also be applied to obtain an approximate surface brightness distribution of the disk: namely, by weighting the individual masses with an additional factor of $T(r) \propto r^{-1/2}$ (prior to binning the data), appropriate for thermal emission in the Rayleigh--Jeans limit from large ($\gtrsim$ mm) dust grains. This assumes optically thin, blackbody emission, which may not hold in all regimes.

%%%%%%%%%%%%%%%%%%%%%%%%
\subsection{Scale Heights and Aspect Ratios}
\label{app:scale_height}
%%%%%%%%%%%%%%%%%%%%%%%%

To calculate the disk's scale height $\mathfrak{h}_d$ (or aspect ratio $\mathcal{H}$) at a given time $t$, we bin the $N_t = N \times N_{np}$ particles radially and analyze the vertical mass distribution in each bin. To be specific, we first define the cylindrical radius $R = \sqrt{X^2+Y^2}$, and bin the data into $N_r$ evenly spaced radial bins (with $dR= 0.25$ au, unless otherwise stated). Next, for each radial bin, we compute the mass-weighted mean vertical position $Z_0(R)$ using the distribution's first moment:  
%%%%%%%%%%%%%%%%%
\begin{equation}
    Z_0(R) = \frac{\sum_{i} m_i(R) Z_i(R)}{\sum_{i} m_i(R)} .
    \label{eq:Z_0_R}
\end{equation}
%%%%%%%%%%%%%%%%%
Here, the index $i$ counts the particles within a given radial bin centered at $R$, each with a vertical position $Z_i$ and mass $m_i$.  
Note that we compute $Z_0(R)$ since the vertical mass distribution of particles in the disk may not necessarily be symmetric around $Z = 0$. In practice, $Z_0(R) \approx 0$ in most cases, but we use Eq. (\ref{eq:Z_0_R}) to maintain full generality; in particular it can account for offsets when resonant particles are excited to $R \rightarrow 0$.

Next, we calculate the scale height $\mathfrak{h}_d(R)$ in a given radial bin  as the square root of the mass-weighted second moment of the vertical positions about $Z_0(R)$, that is:
%%%%%%%%%%%%%%%%%
\begin{equation}
\mathfrak{h}_d(R) =  \sqrt{\frac{\sum_i m_i(R) [Z_i(R) - Z_0(R) ]^2}{\sum_i m_i(R)}} .
\end{equation}
%%%%%%%%%%%%%%%%%
Finally, we obtain the disk aspect ratio as follows:
%%%%%%%%%%%%%%%%%
\begin{equation}
\mathcal{H}(R) \equiv \frac{\mathfrak{h}_d(R)}{R} .
\label{eq:H_appD2}
\end{equation}
%%%%%%%%%%%%%%%%%
This approach is equivalent to azimuthal averaging  over the nodal positions while accounting for the particle mass distribution, thus providing a reliable measure of the disk's vertical thickness. In closing, we note that using the surface density (i.e., dividing the particle masses by the grid cell area) instead of the masses themselves does not affect the result, as the grid cell area would cancel out in the moment calculations.

Finally, we emphasize that if, at a given $R$, the majority of the mass is in planetesimals with $Z>R$, then  Equations (\ref{eq:Z_0_R})--(\ref{eq:H_appD2}) lose their physical meaning, indicating a transition to a non-disk-like structure. Relatedly, the definition of $\mathcal{H}(R)$ is better suited for non-bimodal distributions of $Z(R)$; otherwise, the resulting aspect ratio would be inflated and may not represent the true thickness of the disk.

%%%%%%%%%%%%%%%%%%%%%%%%
\subsection{Planet-disk misalignment}
\label{app:ang_mom}
%%%%%%%%%%%%%%%%%%%%%%%%

At a given time $t$, we assess the (mis)alignment between the planet and the disk using their angular momentum vectors. The angular momentum vector  $\vec{L}_k$ of an object labeled $k$ (e.g., the planet $p$ or planetesimals $j = 1, ..., N$)  reads as: 
%%%%%%%%%%%%%%%%%%%%%%
\begin{equation}
\vec{L}_k = m_k n_k a_k^2 \sqrt{1-e_k^2} 
\begin{pmatrix}
\sin I_k \sin\Omega_k
\\
-\sin I_k  \cos\Omega_k
\\
\cos I_k
\end{pmatrix} 
,
\end{equation}
%%%%%%%%%%%%%%%%%%%%%%
where $e_k$ is the eccentricity (which, in this study, is taken to be zero). We thus define the planet-disk misalignment as 
%%%%%%%%%%%%%%
\begin{equation}
   \theta_{{\rm al}} =  \arccos \bigg( \frac{\vec{L}_p \cdot \vec{L}_d}{||\vec{L}_p|| \cdot ||\vec{L}_d||} \bigg) ,
   \label{eq:theta_al_app_D}
\end{equation}
%%%%%%%%%%%%%%
where $\vec{L}_p$ and $\vec{L}_d = \sum_{j=1}^{N} \vec{L}_j$ represent the angular momentum vectors of the planet and the disk, respectively. Note that $\vec{L}_k$ does not depend on the mean anomalies, which is why we use the $N$ simulated planetesimals to calculate $\vec{L}_d$, i.e., without spawning $N_{np}$ new particles as in Appendix \ref{app:map_construction}.

%%%%%%%%%%%%%%%%%%%%%%%%%%%%%%%%%%%%%%%%%%%%%%%%%%%%%%%%
% Don't change these lines
\bsp	% typesetting comment
\label{lastpage}
\end{document}